\documentclass[twocolumn]{aastex631}

\usepackage{amsmath}
\usepackage{grffile}
\usepackage{dcolumn}
\usepackage{bm}
\usepackage{url}
\usepackage{epsfig}
\usepackage{mathrsfs}  
\usepackage{subfigure}
\usepackage{multirow}
\usepackage{algorithm}
\usepackage{algorithmicx}
\usepackage{algpseudocode}
\usepackage{amssymb}
\usepackage{tensor}

\usepackage{amsfonts}
\usepackage{amssymb}
\usepackage{graphicx}
\graphicspath{{plots/}}
\usepackage{xcolor}

\usepackage{fancyhdr}

\begin{document}

\title{Modeling Solids in Nuclear Astrophysics with Smoothed Particle Hydrodynamics}

\author[0000-0002-1487-0360]{I.~Sagert}
\affiliation{Los Alamos National Laboratory, Los Alamos, New Mexico 87545}

\author[0000-0003-4156-5342]{O.~Korobkin}
\affiliation{Los Alamos National Laboratory, Los Alamos, New Mexico 87545}

\author[0000-0003-2656-6355]{I.~Tews}
\affiliation{Los Alamos National Laboratory, Los Alamos, New Mexico 87545}

\author[0000-0003-4614-0378]{B.-J.~Tsao}
\affiliation{Los Alamos National Laboratory, Los Alamos, New Mexico 87545}

\author[0000-0002-8435-9533]{H.~Lim}
\affiliation{Los Alamos National Laboratory, Los Alamos, New Mexico 87545}

\author[0000-0002-4510-7325]{M.~Falato}
\affiliation{Los Alamos National Laboratory, Los Alamos, New Mexico 87545}

\author[0000-0002-2116-2493]{J. Loiseau}
\affiliation{Los Alamos National Laboratory, Los Alamos, New Mexico 87545}

\begin{abstract}
Smoothed Particle Hydrodynamics (SPH) is a frequently applied tool in computational astrophysics to solve the fluid dynamics equations governing the systems under study.
For some problems, for example when involving asteroids and asteroid impacts, the additional inclusion of material strength is necessary in order to accurately describe the dynamics.
In compact stars, that is white dwarfs and neutron stars, solid components are also present. 
Neutron stars have a solid crust which is the strongest material known in nature. 
However, their dynamical evolution, when modeled via SPH or other computational fluid dynamics codes, is usually described as a purely fluid dynamics problem.
Here, we present the first 3D simulations of neutron-star crustal toroidal oscillations including material strength with the Los Alamos National Laboratory SPH code FleCSPH. 
In the first half of the paper, we present the numerical implementation of solid material modeling together with standard tests.  
The second half is on the simulation of crustal oscillations in the fundamental toroidal mode.
Here, we dedicate a large fraction of the paper to approaches which can suppress numerical noise in the solid.
If not minimized, the latter can dominate the crustal motion in the simulations.
\end{abstract}
\section{Introduction}
Since its introduction in 1977 \citep{Lucy1977,Gingold1977}, Smoothed Particle Hydrodynamics (SPH) has become a numerical method that is frequently applied to terrestrial problems like modeling fluid dynamics and solid material problems in science, industry, and engineering \citep{Vacondio2021, Lind2020, Liu2010, Liu2002}. 
The first solid material application of SPH was for the simulation of high-velocity impacts \citep{Libersky1991}, followed by an extension to model brittle solids \citep{Benz1995}.  
However, SPH methods can also be used to study computational nuclear astrophysics problems such as core-collapse supernovae \citep{Becerra2019, Fryer2004}, compact-star mergers \citep{Bauswein2013, Dan2014, Rosswog2015, Rosswog2022}, and planetary impacts \citep{Reinhardt2017, Stickle2017, Schaefer2016, Owen2015, Schaefer2007}.
These studies benefit from the excellent conservation properties of SPH and its capability to model large material deformations including matter ejection into vacuum. 
The Los Alamos National Laboratory (LANL) SPH code FleCSPH has been developed as part of the Next Generation Code Ristra Project with the goal to utilize the LANL FleCSI computational framework for exascale computing \citep{Loiseau2020}. 
It has been used for the numerical study of compact stars, including single-star oscillations \citep{Tsao2021}, expansion of neutron star merger ejecta \citep{Stewart2022}, and binary mergers \citep{Kaltenborn2022}.
The code includes different analytical and tabulated astrophysical and material equations of state (EoSs), 
Newtonian gravity via the Fast Multipole Method (FMM) \citep{Korobkin2021}, and external potentials for boundary conditions and particle relaxation \citep{Kaltenborn2022}. 
Recent additions include a set of fixed general-relativistic (GR) background metrics for relativistic single compact stars and material strength modeling \citep{Tsao2021}. 
The latter and its application to the modeling of solid neutron-star crust dynamics are the focus of this paper.\\
In astrophysics, solid materials are usually associated with planetary physics.  
However, solids can also be found in white dwarfs \citep{Montgomery1999} and neutron stars (NSTs) \citep{Chamel2008} when densities are sufficiently high for ions to crystallize and form a Coulomb crystal. 
For white dwarfs, this happens in the core where the solid is formed from carbon and oxygen ions.
In NSTs, the Coulomb crystal is found on the surface, in form of a crust that is made of iron-like atomic nuclei. 
Microphysical studies have found that the NST crust is the strongest material known in nature~\citep{Caplan:2016uvu}. 
Hence, the crust can support oscillation modes which would not be present in a fluid.
One example are toroidal oscillations which are associated with low-frequency quasi-periodic patterns in giant X-ray flares~\citep{Israel:2005av,Strohmayer:2005ks,Watts:2005ue,Strohmayer:2006py,Hambaryan:2010av}.
Oscillations between the fluid core and the solid crust via the so-called interface-mode might occur during the inspiral phase of a NST merger, resulting in a possible modification of the gravitational wave (GW) signal \citep{Pan2020} and even the shattering of the crust~\citep{ Tsang2012,Neill:2020szr}.
Ideally, these effects should be examined in fully dynamical simulations that include the constitutive and breaking model of the crust.  
However, current state-of-the-art simulation approaches to study the dynamics of NSTs and their binary mergers usually do not explicitly model the crust as a solid but assume the entire star to be fluid \citep{Rosswog2021, Bernuzzi2020,Dietrich2017, Baiotti2017, Bauswein2013}. \\
In this work, it is our aim to enable such studies of NST crust dynamics using SPH.
With that, we will first give an overview of the constitutive equations and their implementation in the LANL code FleCSPH. 
We will then present standard tests for solid material modeling and compare to analytic solutions as well as other simulation codes where applicable. 
We will then turn to an astrophysical setup and model oscillations of the NST crust in the fundamental toroidal mode \citep{Steiner:2009yg}. 
Here, we will focus on the numerical challenges to model the gelatin-like behavior of the crust in the presence of numerical noise and artefacts. 
Finally, we will present results for the toridal oscillations and compare the extracted frequencies to values from analytic estimates in the literature.
In the equations below, Greek indices label spatial coordinates, and Einstein summation convention is used.
\section{Solid Material Modeling in SPH}
\subsection{General Equations}
SPH solves the conservation equations of mass, momentum and specific internal energy in the Lagrangian frame as given by
\begin{align}
    &\frac{d \rho^{} }{d t} = - \rho \frac{\partial}{\partial x^\alpha} v^\alpha \,, \label{eq::mass_conservation}\\
    &\frac{d v^\alpha}{dt} = \frac{1}{\rho} \frac{\partial}{\partial x^\beta} \sigma^{\alpha \beta} \,,
    \label{eq::momentum_conservation}\\
    &\frac{du^{}}{dt} = \frac{1}{\rho} \sigma^{\alpha \beta} \frac{\partial}{\partial x^\beta} v^\alpha
    \label{eq::energy_conservation}\,,
\end{align}
with mass density $\rho$, velocity $\mathbf{v}$, and the specific internal energy $u$. 
For solids, the stress tensor $\sigma^{\alpha \beta}$ can be written as $\sigma^{\alpha \beta} = - P \delta^{\alpha \beta} + S^{\alpha \beta}$ with the bulk pressure $P$ and deviatoric stress tensor $S^{\alpha \beta}$. 
While the pressure is usually determined from an EoS, $S^{\alpha \beta}$ requires a constitutive model that can describe material strength and breaking \citep[for an introduction, see e.g.,][]{Schaefer2016, Owen2010, Howell2002, Benz1995, Libersky1991}.\\
Hooke's Law for elasticity connects $S^{\alpha \beta}$ to the strain tensor $\epsilon^{\alpha \beta}$ and the shear modulus $\mu$ via
\begin{align}
    S^{\alpha \beta} = 2 \mu \: ( \epsilon^{\alpha\beta} - \frac{1}{3} \delta^{\alpha\beta} \epsilon^{\gamma \gamma})\,.
    \label{eq::epp_elastic}
\end{align}
For large strains, many materials may also undergo plastic deformations where Hooke's Law no longer applies. 
Here, the evolution of $S^{\alpha \beta}$ is described by a yield criterion and flow model.  
The von-Mises yield criterion uses the second invariant $J_2$ of the stress tensor,
\begin{align}
 \sigma_v = \left(3 \: J_2\right)^{1/2} &= \left(\frac{3}{2} \sigma^{\alpha \beta} \sigma^{\alpha \beta}\right)^{1/2} \nonumber\\
&= \left( \frac{1}{2} \left[ \left(S^{11} - S^{22}\right)^2 + \left(S^{11} - S^{33}\right)^2 \right. \right. \nonumber\\
 & \left. \left. + \left(S^{22} - S^{33}\right)^2 \right] + 3 \left(S^{12}\right)^2 \right. \nonumber\\
 &+ \left. 3 \left(S^{23}\right)^2 + 3 \left(S^{31}\right)^2 \right)^{1/2} \,,
 \end{align}
to define a scalar measure of the stress.
Together with a material-specific yield stress $Y_0$, $\sigma_v$ is then used to limit the growth of the deviatoric stress via:
\begin{align}
    S^{\alpha \beta}  = f S^{\alpha \beta}, \:\: f = \mathrm{min} \left[ \frac{Y_0}{\sigma_v}, 1 \right] \,.
    \label{eq::epp_plastic}
\end{align}
Equations (\ref{eq::epp_elastic}) and (\ref{eq::epp_plastic}) form the so-called elastic--perfectly plastic (EPP) strength model. 
More advanced models are available for, e.g., metals at large strains \citep{Preston2003, Johnson1983, Steinberg1980} as well as porous materials \citep{Jutzi2009}. 
A simple extension to the EPP model uses linear isotropic hardening in the plastic phase \citep{Greto2019, Burton2015}. 
Here, the yield stress increases with plastic strain $\epsilon_p$ as
\begin{align}
    Y\left(\epsilon_p + \Delta \epsilon_p \right) &= Y_0 + H\left(\epsilon_p + \Delta \epsilon_p \right)\,, \\
    \Delta \epsilon_p &= \frac{\sigma_v - Y\left(\epsilon_p\right)}{3 \mu + H}\,,
    \label{eq::plastic_strain}
\end{align}
given a material-dependent hardening parameter $H$. \\
Damage models describe the evolution of flaws or cracks in materials which can grow up to a specific threshold, releasing the local stress, until the corresponding region (or SPH particle) is fully damaged and does not respond to strain anymore. 
After that, the fully damaged material behaves like a fluid \citep{Schaefer2016, Benz1995}.
For the NST crust, it has been shown that cracks do not form due to the large pressures and material failure occurs collectively under large strains \citep{Horowitz2009}. 
With that, we will focus on the maximum-strain damage model for breaking. 
Here, instead of modeling the formation and propagation of cracks, material failure occurs as soon as the breaking strain is reached. 
To calculate the local scalar strain $\epsilon_{loc}$, we use the maximum tensile stress $\sigma_\mathrm{max} = \mathrm{max}\left[\sigma_1, \sigma_2, \sigma_3 \right]$ where $\sigma_\gamma$ are the principal stresses, determined by a principal axis transformation on the stress tensor $\sigma^{\alpha \beta}$ \citep{Schaefer2016} . 
The local strain is then given by
\begin{align}
    \epsilon_{loc} = \frac{\sigma_\mathrm{max} }{E}\,, \:\:\: E = \frac{9K\mu}{(3K + \mu)}\,,
\end{align}
and can be compared to the material's breaking strain with failure for $\epsilon_{lock} \geq \epsilon_{break}$. 
\subsection{SPH Discretization}
\label{sed:discretization}
We begin with a brief description of the SPH method. 
For a comprehensive overview, see e.g., \cite{Rosswog2015, Liu2010, Monaghan2005}, and \cite{Liu2003}. 
SPH is a meshless Lagrangian method that applies particles which are advected with the material flow. 
A physical quantity $A$ at the location of particle $i$ is determined from a weighted sum over its neighboring particles $j$ via
\begin{align}
    A(\vec{r}_i) &= \sum_j \frac{m_j}{\rho_j} \: A(\vec{r}_j) \: W_{ij}\,, \\
    W_{ij} &= W(|\vec{r}_i - \vec{r}_j|, h)\,,
\end{align}
using a smoothing function $W_{ij}$. The latter depends on the distance of particles $i$ and $j$ and vanishes beyond a support radius that is characterized by the smoothing length $h$.
The spatial derivatives of $A$, including gradient, divergence, and curl, can be calculated via the gradient of $W_{ij}$ which allows the discretization of the conservation equations (\ref{eq::mass_conservation})-(\ref{eq::energy_conservation}) \citep{Rosswog2015a}. \\
The continuity equation in SPH is given by 
\begin{align}
\frac{d \rho_i}{dt} = \rho_i \sum_j \frac{m_j}{\rho_j} \left( v^\alpha_i - v^\alpha_j \right) \frac{\partial W_{ij}}{\partial x_i^\alpha}\,.
\label{eq::continuity}
\end{align}
An alternative is to determine the density directly via a summation over particle masses, i.e. $\rho_i = \sum_j m_j W_{ij}$.
This method explicitly ensures mass conservation, however, Eq.~(\ref{eq::continuity}) is often preferred for solids as it e.g. can better represent the density at object boundaries. 
Momentum conservation can be expressed as
\begin{align}
    \frac{d v_i^\alpha}{dt} &= - \sum_b m_j \left( \frac{\sigma_i^{\alpha \beta}}{\rho_i^2} + \frac{\sigma_j^{\alpha \beta}}{\rho_j^2} \right) \frac{\partial W_{ij}}{\partial x^\beta_i} \\
    &= - \sum_j m_j \left[\left( \frac{P_i}{\rho_i^2} + \frac{P_j}{\rho_j^2} + \Pi_{ij} \right) \frac{\partial W_{ij}}{\partial x^\alpha_i} \right. \nonumber\\
    &- \left. \left( \frac{S_i^{\alpha \beta}}{\rho_i^2} + \frac{S_j^{\alpha \beta}}{\rho_j^2} \right) \frac{\partial W_{ij}}{\partial x^\beta_i}\right] 
    \label{eq::mom_conservation}\,
\end{align}
and the conservation of specific internal energy is given by
    \begin{align}
        \frac{d u_i}{dt} &= \sum_j \left[m_j \left( \frac{P_i}{\rho_i^2} + \frac{\Pi_{ij}}{2} \right) \left( v_i^\alpha - v_j^\alpha \right) \frac{\partial W_{ij}}{\partial x_i^\alpha} \right] \nonumber\\
        &+ \frac{1}{\rho_i} S_i^{\alpha \beta} \: \frac{d{ \epsilon}_i^{\alpha \beta}}{dt} \,.
        \label{eq::en_conservation}
    \end{align}
In both, $\Pi_{ij}$ is the artificial viscosity tensor which we describe following \cite{Monaghan1983}. 
For solids, we also have to evolve the deviatoric stress tensor 
\begin{align}
    \frac{d S^{\alpha \beta}}{dt} &= 2 \mu\left( \frac{d \epsilon^{\alpha \beta}}{dt} - \frac{\delta^{\alpha \beta}}{3} \: \frac{d \epsilon^{\gamma \gamma}}{dt} \right) \nonumber\\
    &+ S^{\alpha \gamma} \: \frac{dR^{\gamma \beta}}{dt} - \frac{dR^{\alpha \gamma}}{dt} S^{\gamma \beta}\,,
\end{align}
with the strain rate and rotational rate tensors
\begin{align}
    \frac{d \epsilon_i^{\alpha \beta}}{dt} &= \sum_j \frac{m_j}{2\rho_i}\left[ \left(v^\alpha_j - v^\alpha_i \right) \frac{\partial W_{ij}}{\partial x_i^\beta} \right. \nonumber\\
    &\left. + \left(v^\beta_j - v^\beta_i \right) \frac{\partial W_{ij}}{\partial x_i^\alpha} \right] ,  \\
    \frac{dR_i^{\alpha \beta}}{dt} &= \sum_j \frac{m_j}{2\rho_i} \left[ \left(v^\alpha_j - v^\alpha_i \right) \frac{\partial W_{ij}}{\partial x_i^\beta} \right. \nonumber\\
    &\left. - \left(v^\beta_j - v^\beta_i \right) \frac{\partial W_{ij}}{\partial x_i^\alpha} \right]\,.
\end{align}
The latter transforms the stresses from the reference frame associated with the material to the laboratory reference frame. \\
Equations (\ref{eq::mom_conservation}) and (\ref{eq::en_conservation}) can be extended by the inclusion of gravitational forces and external potentials \citep{Tsao2021, Kaltenborn2022}. 
The latter are used in FleCSPH for particle relaxation and to impose boundary conditions. 
We will take advantage of this capability when describing NSTs. 
In a nutshell, an external potential can confine particles in a given density profile. 
In equilibrium, the momentum equation gives 
\begin{align}
    \mathbf{g}_{\rm ext} = \nabla P \: \rho^{-1}\,. 
\end{align}
Using a polytropic EoS as described in Eq.~(\ref{eq::polytropic}), this results in 
\begin{align}
    \mathbf{g}_{\rm ext} = K \: \Gamma \: \rho^{\Gamma -2} \: \nabla \rho\,,
\end{align}
and with $-\nabla \phi_\mathrm{ext} = \mathbf{g}_\mathrm{ext}$ leads to
\begin{align}
    \phi_\mathrm{ext} = - K \: \Gamma \: \rho^{\Gamma -1} \left(\Gamma - 1\right)^{-1}. 
    \label{eq::phi_ext}
\end{align}
With that, adding $\mathbf{g}_\mathrm{ext}$ to the momentum equations will result in particles distributing according to the desired density profile while they are trapped in a potential $\phi_\mathrm{ext}$. 
FleCSPH contains different smoothing functions $W$ including cubic and quintic splines, gaussian and super gaussian, as well as the Wendland C2, C4, and C6 kernels. 
For all simulations in this paper, except the first standard test on colliding rubber rings, we will apply the Wendland C6 kernel as it is not subject to the so-called pairing instability in SPH \citep{Dehnen2012} and leads to low velocity noise in dynamical simulations \citep{Rosswog2015a}.
In addition, for all simulations except the colliding rings, we allow the smoothing length to change according to   
\begin{align}
    h_i = \eta \left(\frac{m_i}{\rho_i}\right)^{1/D}
    \label{eq::adaptive_h}
\end{align}
where $D$ is the dimensionality of the physics problem and $\eta$ is a user-specified number of order unity. The adjustment of the smoothing length ensures an approximately constant number of particle neighbors for changing density. 

One frequently observed numerical artefact when modeling solids with SPH is the so-called tensile instability. 
It occurs in the presence of negative pressure or tension and leads to particle clumping. 
Different techniques are availabe to cure this instability, including the usage of a Moving-Least-Square interpolant \citep{Monaghan2005, Dilts2000, Dilts1999}.
The simplest approach is the addition of a small artificial repulsive stress ~\citep{Gray2001}. 
This stress $\xi^{\alpha \beta}$ is calculated by first diagonalizing the stress tensor into $\tilde{\sigma}^{\alpha \beta}_i$.
For each positive diagonal component an artificial stress term
\begin{align}
 \tilde{\xi}^{\alpha \beta}_i = -\epsilon_s \frac{\tilde{\sigma}_i^{\alpha \beta}}{\rho_i^2}
 \label{eq:art_stress}
\end{align}
is calculated with $\epsilon_s$ being a free parameter often set to 0.3. 
The artificial repulsive stress is then rotated back into the original coordinate system to $\xi^{\alpha \beta}_i$ and added to the conservation of momentum and conservation of specific internal energy equations via
\begin{align}
    \frac{d v_i^\alpha}{dt} 
     &= - \sum_j m_j \left[\left( \frac{P_i}{\rho_i^2} + \frac{P_j}{\rho_j^2} + \Pi_{ij} \right) \delta^{\alpha\beta} \right. \nonumber\\
     &- \left. \left( \frac{S_i^{\alpha \beta}}{\rho_i^2} + \frac{S_j^{\alpha \beta}}{\rho_j^2} + \left(\xi^{\alpha \beta}_i + \xi^{\alpha \beta}_j \right) f_{ij}^n \right) \right] \frac{\partial W_{ij}}{\partial x^\beta_i}\,,\\
    \frac{d u_i}{dt} 
    &= \sum_j m_j \left[\left(\frac{P_i}{\rho_i^2} + \frac{\Pi_{ij}}{2} \right) \delta^{\alpha \beta} + \frac{\xi_{ij}^{\alpha \beta}}{2} f_{ij}^n \right] \nonumber\\
    &\times \left( v_i^\alpha - v_j^\alpha \right) \frac{\partial W_{ij}}{\partial x_i^\beta} 
    + \frac{S_i^{\alpha \beta}}{\rho_i} \dot{\epsilon}_i^{\alpha \beta}\, .
\end{align}
Here 
\begin{align}
    f_{ij}^n = \left(\frac{W_{ij}}{W_{\Delta x}}\right)^n
\end{align}
with $\Delta x$ being the particle spacing and $n$ frequently set to $4$. 
This ensures that the repulsive stress is only applied to the nearest neighbors.  
Finally, to prevent velocity fluctuations, FleCSPH includes the xSPH-velocity method \citep{Monaghan1989} where particles are moved with a velocity  
\begin{align}
    \frac{d\mathbf{x}_i}{dt} = \mathbf{v}_i + x_{\mathrm{sph}} \sum_j \frac{2\:m_j}{\rho_i \rho_j} \left( \mathbf{v}_j - \mathbf{v}_i \right) W_{ij}\,,
\end{align}
that is smoothed over the neighboring particles with the factor $x_\mathrm{sph} = 0.5$. 
\subsection{Material Equations of State}
FleCSPH has several implemented analytic and tabulated EoSs for terrestrial materials and nuclear matter. 
For the benchmark problems discussed in the next chapters, we apply the Tait-Murnaghan~\citep{Monaghan2005}, Mie-Gr\"uneisen~\citep{Rice1958}, and Osborne~\citep{Howell2002, Riney1970} EoSs.
Another frequently used EoS model for solids was suggested by \cite{Tillotson1962}.
While it is implemented in FleCSPH, it is not applied in this paper and will therefore not be discussed here.\\
The Tait-Murnaghan or liquid EoS gives the pressure as a function of mass density $\rho$: 
\begin{align}
    P(\rho) = \frac{\rho_0 \: c_0^2}{\gamma} \: \left[\left(\frac{\rho}{\rho_0}\right)^\gamma - 1 \right] + P_0\,.
\end{align}
Here, $c_0$ is the sound speed, $\rho_0$ the initial mass density or density at the initial pressure $P_0$, and $\gamma$ is a material-dependent constant.
This EoS is easy to implement and popular for modelling weakly compressible fluids.
Note, that for relativistic matter, the sound speed is determined via 
\begin{align}
    c_0 = \sqrt{\frac{dP}{d\epsilon}}
\end{align}
where $\epsilon$ is the energy density. 
For solid material experiments on Earth, however, matter is usually in the non-relativistic regime with $\epsilon \sim \rho$.
As a consequence, in order to determine $c_0$, the derivative of the pressure is often taken with respect to the mass density. \\
In the Mie-Gr\"uneisen EoS the pressure is a function of density $\rho$ and internal energy $u$ as
\begin{align}
    P(\rho,u) &= \rho_0 \: c_0^2 \: \chi \left[ 1 - \frac{\Gamma_0}{2} \chi \right] \left(1 - s \chi \right)^{-2} + \Gamma_0 \rho u\,, 
    \label{eq::MG_EoS}
    \\ 
    \chi &= 1 - \frac{\rho_0}{\rho}\,. \nonumber
\end{align}
Here, $\rho_0$ is again the initial density, $c_0$ the sound speed, $\Gamma_0$ the Mie-Gr\"uneisen Gamma, and $s$ the linear Hugoniot slope coefficient. 
A different form can be found in \cite{Wilkins1998}:
\begin{align}
    P (\rho, u) = k_1 \chi + k_2 \chi^2 + k_3 \chi^3 + \Gamma_0 \rho_0 u\,,
\end{align}
where $\Gamma_0$, $\rho_0$, and $\chi$ are defined as for Eq.~(\ref{eq::MG_EoS}) and $k_1 - k_3$ are material-specific parameters.\\
Finally, the Osborne EoS has the form: 
\begin{align}
    P(\rho,u) &= \frac{a_1 \mu + \bar{a} \mu^2 + u\rho_0 \left( b_0 + b_1 \mu + b_2 \mu^2 \right)}{u \rho_0 + e_0} \nonumber\\
    &+ \frac{\left(u \: \rho_0\right)^2 \left(c_0 + c_1 \mu \right) }{u \rho_0 + e_0}\,, \\
    \mu &= \frac{\rho}{\rho_0} - 1, \:\:\:
    \bar{a} =\begin{cases}
		a_2, & \mu \geq 0\\
        a_2^\star, & \text{otherwise}
	\end{cases}\,,
\end{align}
with $a_1$, $a_2$, $a^\star_2$, $b_0 - b_2$, $c_0$, $c_1$, and $e_0$ being material-dependent parameters while $\rho$ is again the mass density and $u$ the specific internal energy. 
\section{Standard Material Tests}
While some of the fluid dynamics capabilities of FleCSPH have been demonstrated in previous works \citep{Loiseau2020}, here, we focus on the verification of its material strength modeling. 
We choose three standard setups - the colliding rubber rings, the Verney implosion, and the Taylor anvil impact. 
There is a large abundance of other different test problems such as the oscillating plate \citep{Gray2001} or high-velocity impacts and crater formation \citep{Caldwell2018}. 
Also, neither of the chosen setups tests material failure. 
However, in this paper, our astrophysical application solely relies on FleCSPH's implementation of elasticity.
With that, the three presented test problems should be sufficient.   
\subsection{Colliding Rubber Rings}
The colliding rubber ring test is a benchmark problem for the numerical implementation of elasticity \citep{Gray2001, Schaefer2016} and artificial numerical fragmentation due to the tensile instability.
Two rubber rings with $\rho_0 = 1.01\:\mathrm{g/cm^3}$, $c_0 = 8.54 \times 10^4\:\mathrm{cm/s}$, and $\mu = 0.22 \: \rho_0 \: c_0^2$ collide at $0.118\:c_0$ (each ring has a velocity of $0.059\:c_0$). 
The rings have an inner and outer radius of $3\:\mathrm{cm}$ and $4\:\mathrm{cm}$, respectively.
The density is evolved via Eq.~(\ref{eq::continuity}) and particle positions are updated with xSPH velocities.  
Numerical repulsive stress with $\epsilon_s = 0.3$ is added to prevent particle pairing. 
The pressure is given by the Tait-Murnaghan EoS with $\gamma = 1$ and $P_0 = 0$. 
This problem has no analytic solution but we can compare our simulation outcomes to published studies. 
For this, we closely follow the setup in \cite{Schaefer2016, Gray2001} and use 4480 particles in a rectangular lattice with 0.1\:cm spacing. 
We use a cubic spline kernel with a constant support radius of $0.3\:\mathrm{cm}$. 
The post-collision time $t$ is given with respect to the moment when the pressure of the SPH particles that are closest to each other but belong to different rings begins to change. 

\begin{figure}[!htbp]
    \centering
    \includegraphics[width = 0.4\textwidth]{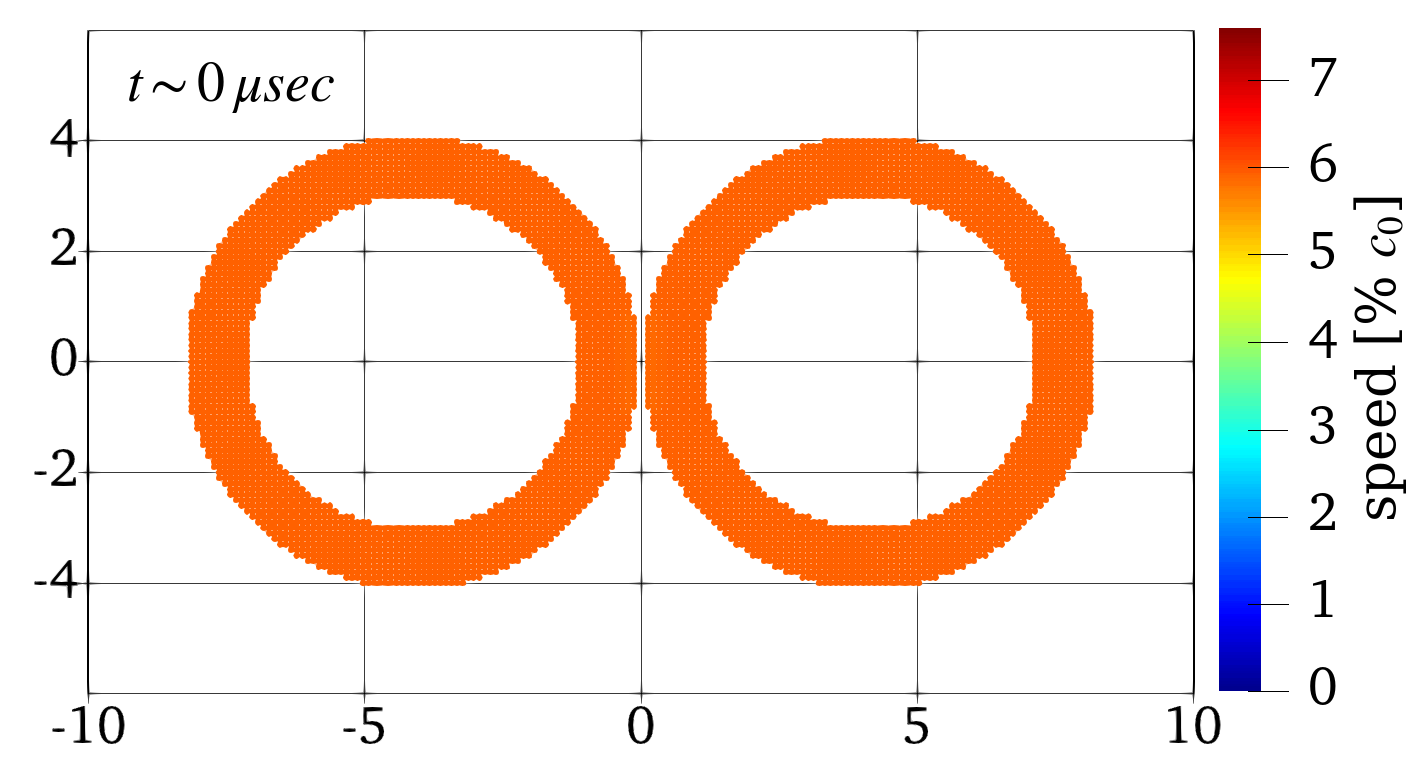}
    \hspace{0.05cm}
    \includegraphics[width = 0.4\textwidth]{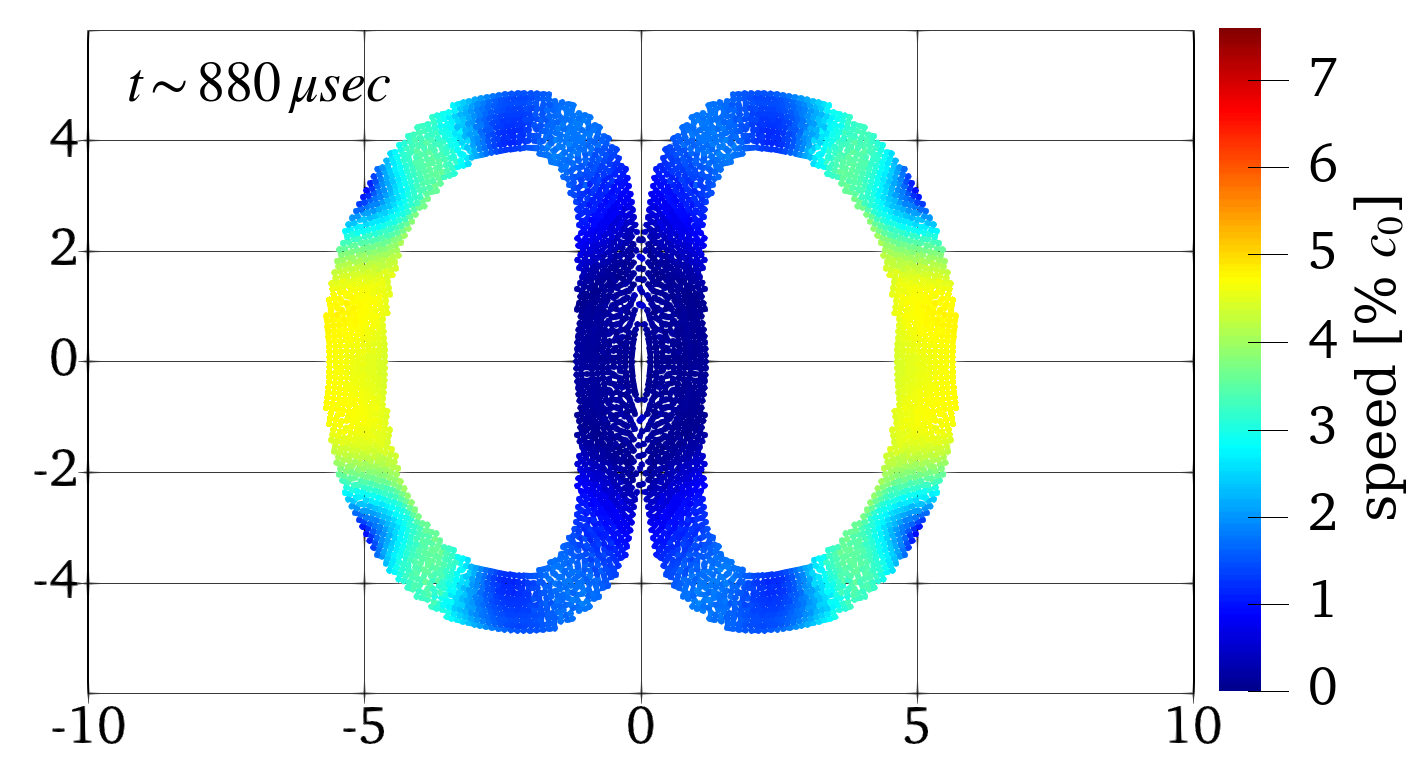}
    \includegraphics[width = 0.4\textwidth]{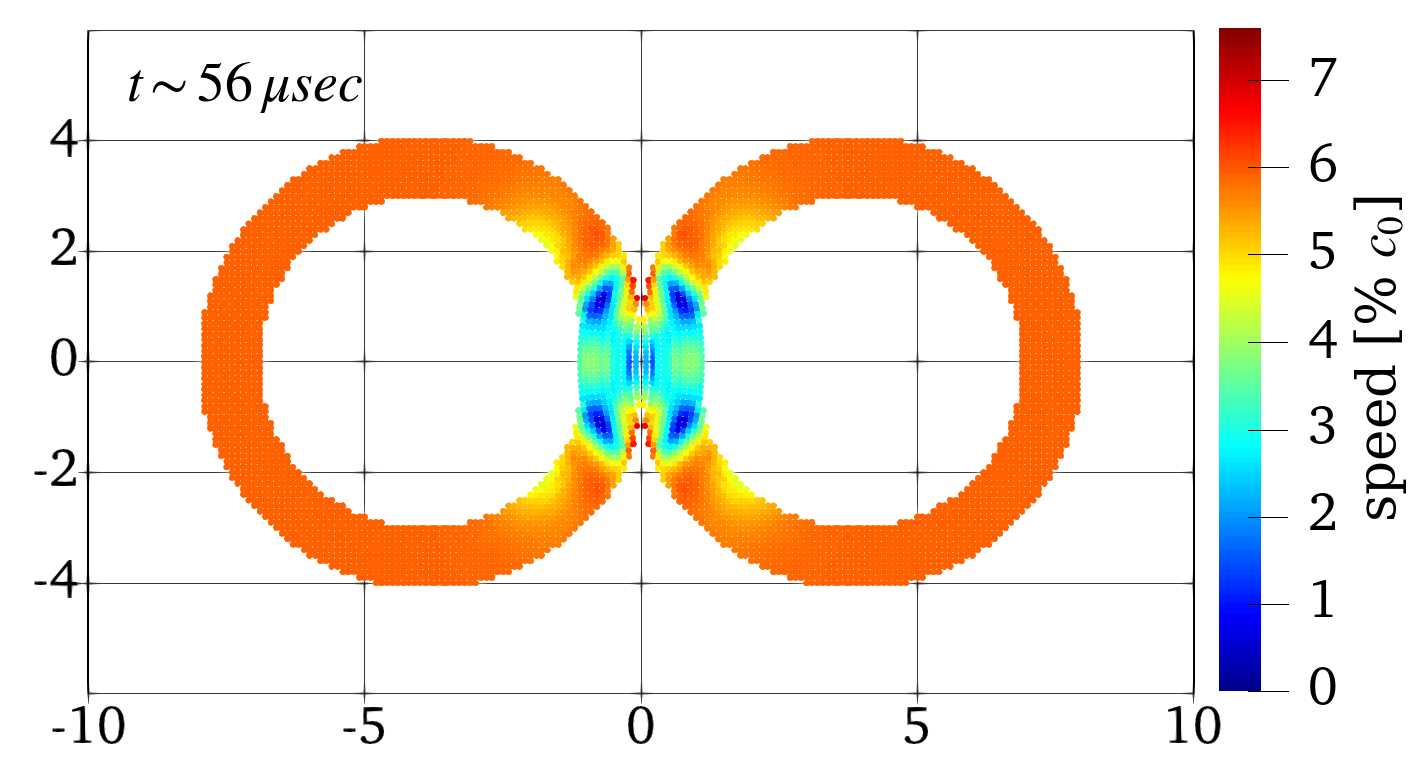}
    \hspace{0.05cm}
    \includegraphics[width = 0.4\textwidth]{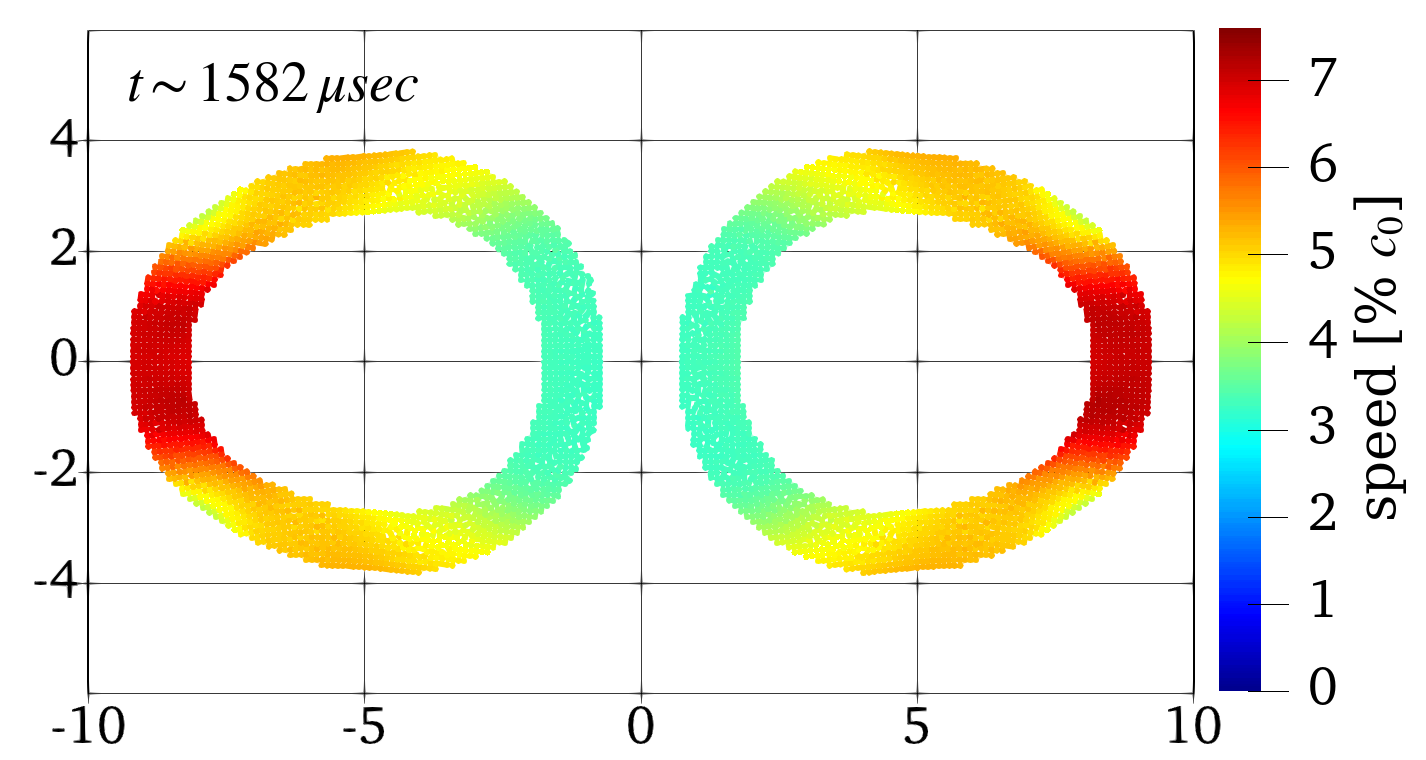}    
    \caption{Snapshots of the colliding rubber rings at different times. Color coded is the particle speed in percentage of the sound speed $c_0$. The speed distributions can be compared to \cite{Schaefer2016} with good agreement.}
    \label{fig:2D_rings}
\end{figure}

In the simulation, we observe that after contact, the rings are gradually compressed, reach maximum compression, bounce off of each other, and move apart while oscillating. 
Figure \ref{fig:2D_rings} shows snapshots of the simulation at different times. 
The color coding corresponds to the particle speed and the units are in percentage of $c_0$. 
The speed can be compared to results in \cite{Schaefer2016} showing good agreement.
Subtle differences in the particle speed, especially at $t = 880\:\mathrm{\mu sec}$, are visible but are, as preliminary tests showed, most likely due to differences in the numerical viscosity treatment.
While \cite{Schaefer2016} seems to use the kernel support radius in the implemented viscosity model, we are using the smoothing length $h$. 
\begin{table}
\begin{center}
 \begin{tabular}{|l c c c|} 
 \hline
Time [$\mathrm{\mu sec}$] & Width [cm] & Height [cm] & Distance [cm] \\ [0.5ex] 
 \hline
0 & 8.0 & 8.0 & 0.0 \\ 
500 & 5.1 & 9.6 & 0.0 \\
1000 & 6.1 & 9.4 & 0.0\\
1500 & 8.2 & 7.9 & 1.0 \\
2000 & 8.9 & 7.1 & 3.9 \\
\hline 
\end{tabular}
\end{center}
\caption{\label{table_2D_rings} Width, height, and distance of the colliding rubber rings at selected times as obtained from the FleCSPH simulation in Fig.~\ref{fig:2D_rings}.}
\end{table}
\begin{figure}
    \centering
    \includegraphics[width = 0.45\textwidth]{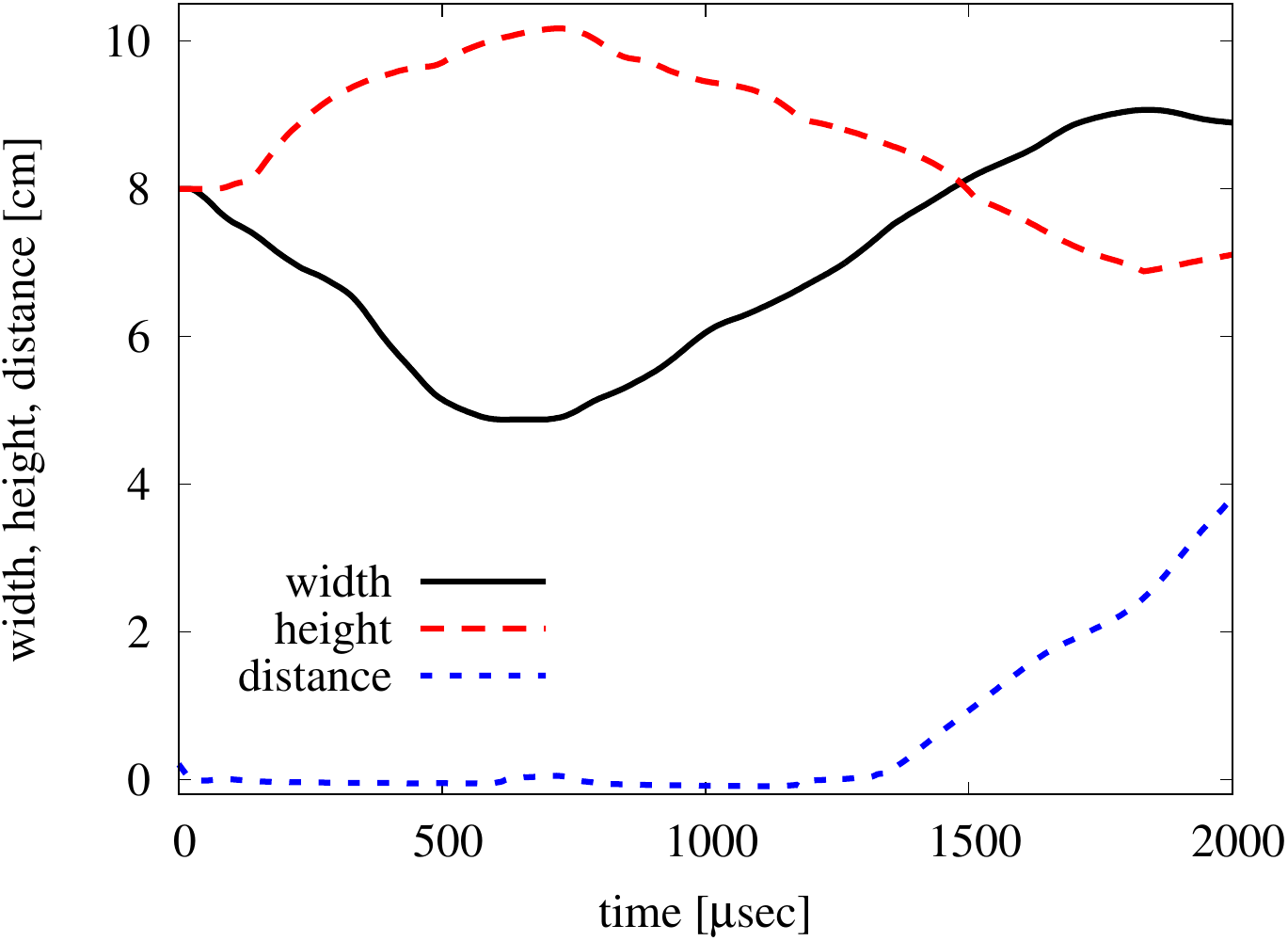}
    \caption{Width, height, and distance of the two rubber rings in Fig.~\ref{fig:2D_rings} as a function of time.}
    \label{fig:2d_rings_comparison}
\end{figure}
Throughout the simulation, the rings are stable.
This is expected when numerical repulsive stress is used. 
Otherwise, fragmentation is observed during ring collision \citep{Schaefer2016} which we have also seen in our simulations. 

Figure \ref{fig:2d_rings_comparison} gives a more detailed time evolution of the rings' width, height, and distance while Table~\ref{table_2D_rings} provides values for selected times.  
For both, the width is given by the distance in the $x$-direction between the innermost and outermost particles that belong to the same ring and have a height of $0\:\mathrm{cm} \leq y \leq 0.12\:\mathrm{cm}$. 
The ring distance is determined by taking twice the value of the $x$-coordinate of the innermost SPH particle. 
Finally, the height is calculated as the $y$-distance between two particles with maximum and minimum $y$ values. 
\subsection{Verney Implosion}
In the Verney test, a spherical metal shell implodes after receiving a radial inward velocity \citep{Verney1968}. 
During the convergence, the shell thickens. 
Its kinetic energy is converted into internal energy through plastic work until the implosion terminates at a stopping radius. 
Assuming incompressible and EPP behavior, a given shell setup, material strength, and final stopping radius, the implosion has an analytic solution given an initial radial velocity distribution of the form
\begin{align}
    v_r (r) = v_0 \: \left(\frac{R_i}{r}\right)^\alpha .
    \label{eq::verney_vel}
\end{align}
Since simulations usually use compressible materials, there are deviations from the analytic solution including small-amplitude oscillations after the stopping radius has been reached. 

In addition to the original 3D spherical implosion, \cite{Howell2002} introduced a 2D cylindrical version of the test. 
Here, we will simulate both, following the setups of \cite{Burton2015} and \cite{Howell2002} for a beryllium shell.
For their work, \cite{Burton2015} use a finite volume cell-centered Lagrangian hydrodynamics approach while \cite{Howell2002} apply a finite volume free-Lagrange Godunov scheme. 

We will use the EPP strength model and show the results of the Osborne and Mie-Gr\"uneisen EoSs. 
The parameters for both as well as the shear modulus and yield stress for beryllium are given in Table~\ref{tab::be_eos}. 
\begin{table}
\centering
 \begin{tabular}{|l c | l c|} 
 \hline
 Osborne & [g cm $\mathrm{\mu}$s] & Strength & [Ba] \\ [0.5ex] 
 \hline
 $\rho_0$ & 1.845 & $\mu$ & $1.5111 \times 10^{12}$ \\
 $a_1$ & 0.9512 & $Y$ & $3.30 \times 10^9$ \\
  $a_2^\star$ & -0.3453 & & \\
 \cline{3-4}
 $a_2$ & 0.3453 & Mie-Grueneisen& [g cm s]\\
 \cline{3-4}
 $b_0$ & 0.9269 & $\rho_0$& 1.85 \\
 $b_1$ & 2.9484 & $c_0$ & $7.998 \times 10^5$ \\
 $b_2$ & 0.5080 & $\Gamma$ & 1.16\\
 $c_0$ & 0.5644 & $s$& 1.124 \\ 
 $c_1$ & 0.6204 & & \\
 $e_0$ & 0.800 & & \\
 \hline
 \hline
\end{tabular}
\caption{Osborne EoS parameters in the gram-centimeter-microsecond unit system, Beryllium strength parameters \citep{Howell2002, Riney1970}, and EoS parameters for the Mie-Gr\"uneisen model \citep{Burton2015}.}
\label{tab::be_eos}
\end{table}
For the 2D simulation, we set up a shell with density $\rho = 1.845 \: \mathrm{g/cm^3}$, inner radius $R_i = 8\:$cm, and thickness of 2\:cm. The smoothing function is the Wendland C6 kernel and is set to adapt to the particle density according to eq.(\ref{eq::adaptive_h}).
\begin{figure}
    \centering
    \includegraphics[width = 0.45\textwidth]{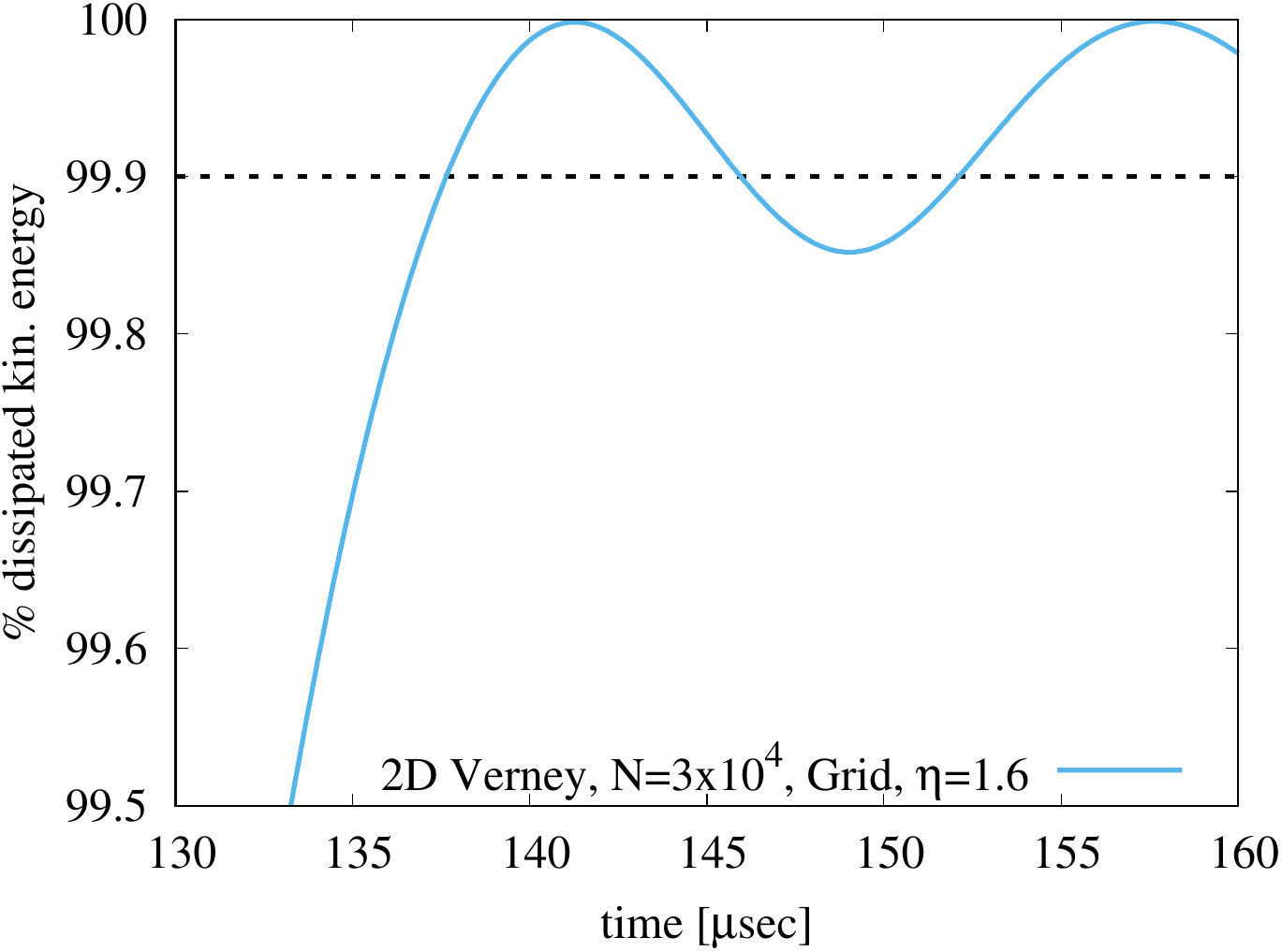}
    \hspace{0.1\textwidth}
    \caption{Percentage of dissipated kinetic energy with respect to its initial value for the 2D Verney implosion.}
    \label{fig:verney_2D_energy}
\end{figure}
The initial radial velocity distribution is given by eq.(\ref{eq::verney_vel}) with $\alpha = 1$ and $v_0 = 4.902 \times 10^4 \:$cm/s. 
The final inner radius of the shell is $4\:$cm.
\cite{Howell2002} quote a stopping time of $137 \: \mu$s when 99.9\% of the initial kinetic energy is converted into internal energy. 
\begin{figure}
    \centering
    \includegraphics[width = 0.48\textwidth]{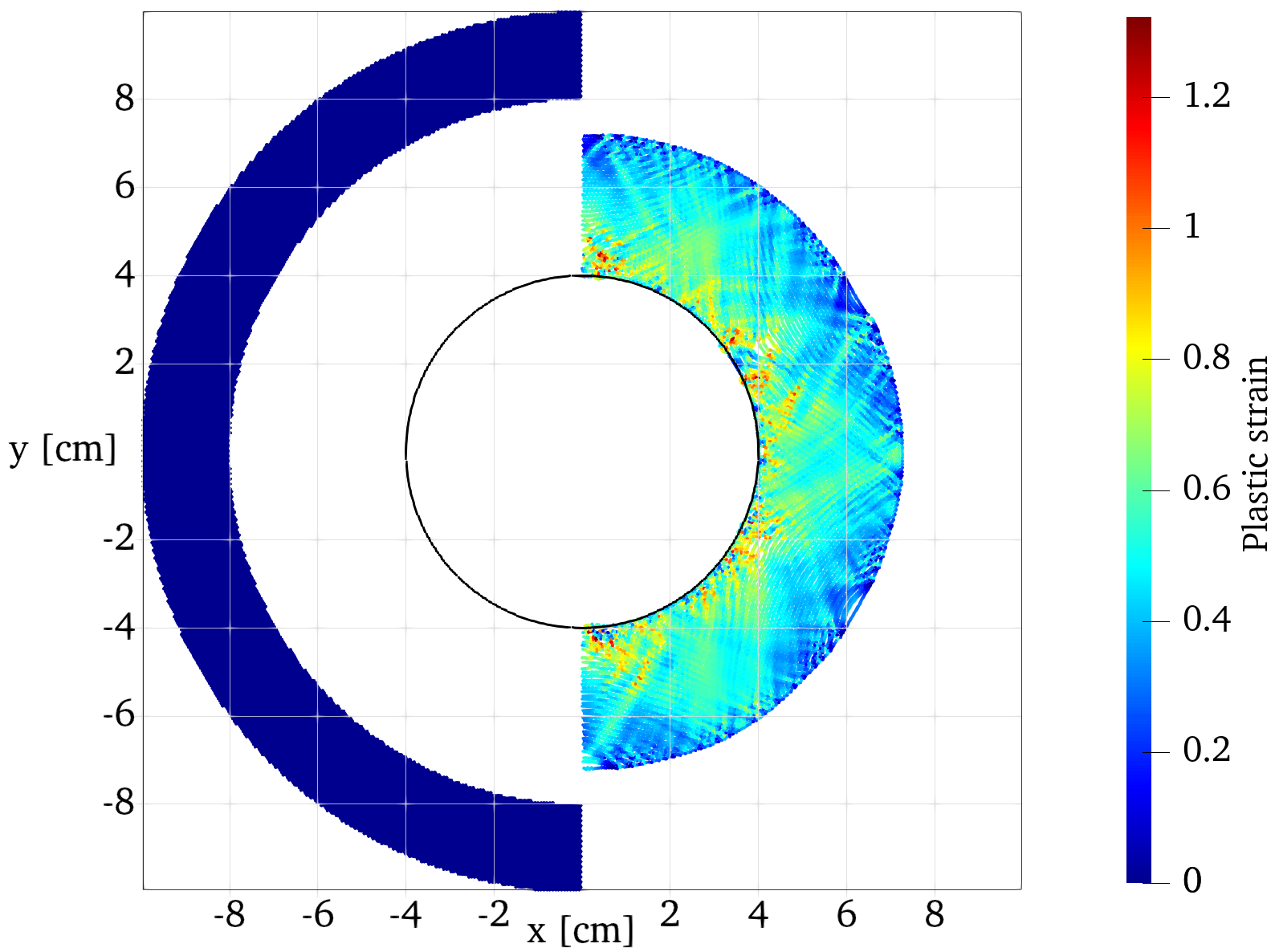}
    \caption{Particle distribution and plastic strain in the 2D Verney implosion simulation. For compactness, only one half of the simulation space is shown. The left side shows the initial setup. The right side shows the particles at $137\: \mu$s, when 99.9\% of the initial kinetic energy has dissipated. The black ring shows the position of the stopping radius.}
    \label{fig:2d_verney}
\end{figure}
We use $\sim 3 \times 10^4$ particles in two different setups.
In the first, the particles are distributed in a triangular lattice while in the second, they are arranged into rings with roughly equal radial and tangential separation.
For both, we use $\eta = 1.6$. 
First, we discuss the results for the triangular distribution.  
Figure~\ref{fig:verney_2D_energy} shows the evolution of the kinetic energy dissipation determined via 
\begin{align}
    100\times \left(1 - \frac{E_\mathrm{kin} (t)}{E_\mathrm{kin} (t=0)}\right)
\end{align}
for the 2D Verney setup with the Osborne EoS.
The 99.9\% mark is reached at about $137\: \mu$s which is in agreement with \cite{Howell2002}, followed by elastic oscillatory behavior. 
About 100 of the $3 \times 10^4$ particles have a radial distance that is between $3.9\:\mathrm{cm}$ and $4\:\mathrm{cm}$. 
Figure~\ref{fig:2d_verney} shows the corresponding particle configurations with plastic strain at 0 and $137\:\mathrm{\mu}$s, respectively. 
For the latter, we find a non-radial pattern which should not be present in the implosion. 
This is due to the initial placement of particles in a lattice and has been seen in other SPH simulations as in e.g. \cite{Owen2015b}.
The latter work also demonstrated that a ring placement can eliminate these effects.
With that, we turn to the second setup.
Here, we initially find that the implosion results in a very smooth distribution of plastic strain. 
However, as it progresses and the shell widens, the particle representation only has two options to maintain constant density. 
Either particles reconfigure drastically or the rings increase their distance to each other. 
We find the latter.  
This becomes a problem when the ring separation exceeds the particle smoothing length which first happens for the innermost ring. 
As a consequence, the latter continues to detach and begins to deform.
As the simulation progresses and the shell width continues to increase, more of the inner rings exhibit the same behavior.
\cite{Owen2015b} pointed out that the Verney problem benefits from the usage of asymmetric smoothing kernels since the radial and azimuthal spacing of particles changes so anisotropically. 
Since FleCSPH does not have that capability, we attempt to prevent the ring detachment by setting $\eta = 3.5$ and thereby generally increasing the smoothing length.
Interestingly, in this case, we also achieve the desired effect. 
Even though the distance between the rings grows as the shell collapses, it never exceeds the smoothing length and we find that the rings stay attached to each other.
We tested the effect of using $\eta = 3.5$ for the triangular particle configuration.
As expected, even though the grid effects were still present, they were much more smoothed out.
\begin{figure}
    \centering
    \includegraphics[width = 0.48\textwidth]{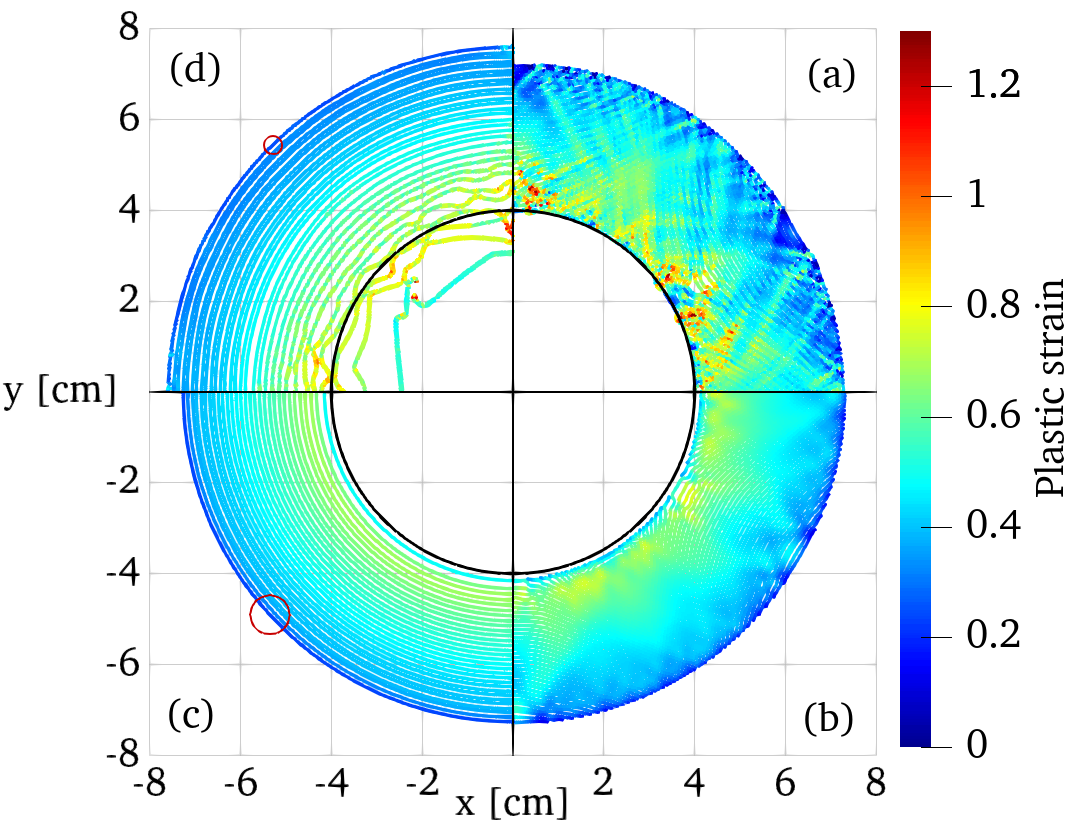}
    \caption{One quadrant of the particle distribution with corresponding plastic strain in the 2D Verney implosion simulation at $137\: \mu$s for a triangular particle initialization with (a) $\eta = 1.6$ and (b) $\eta = 3.5$ as well as a ring initialization with (c) $\eta = 3.5$ and (d) $\eta = 1.6$. The small red circles in figures (c) and (d) visualize the particle's smoothing length. The black ring shows  the position of the stopping radius.}
    \label{fig:2d_verney_all}
\end{figure}
The resulting configurations at 137\:s are given in Fig.~\ref{fig:2d_verney_all} where we only show one quadrant for each setup for better comparison.
For $\eta = 3.5$ particles from neither the lattice nor the ring distribution cross the stopping radius. 
For the lattice initialization, the smallest minimum particle radial distance is about 4.12\:cm while for the ring initialization is is about 4.15\:cm.
\begin{figure}
    \centering
    \includegraphics[width = 0.45\textwidth]{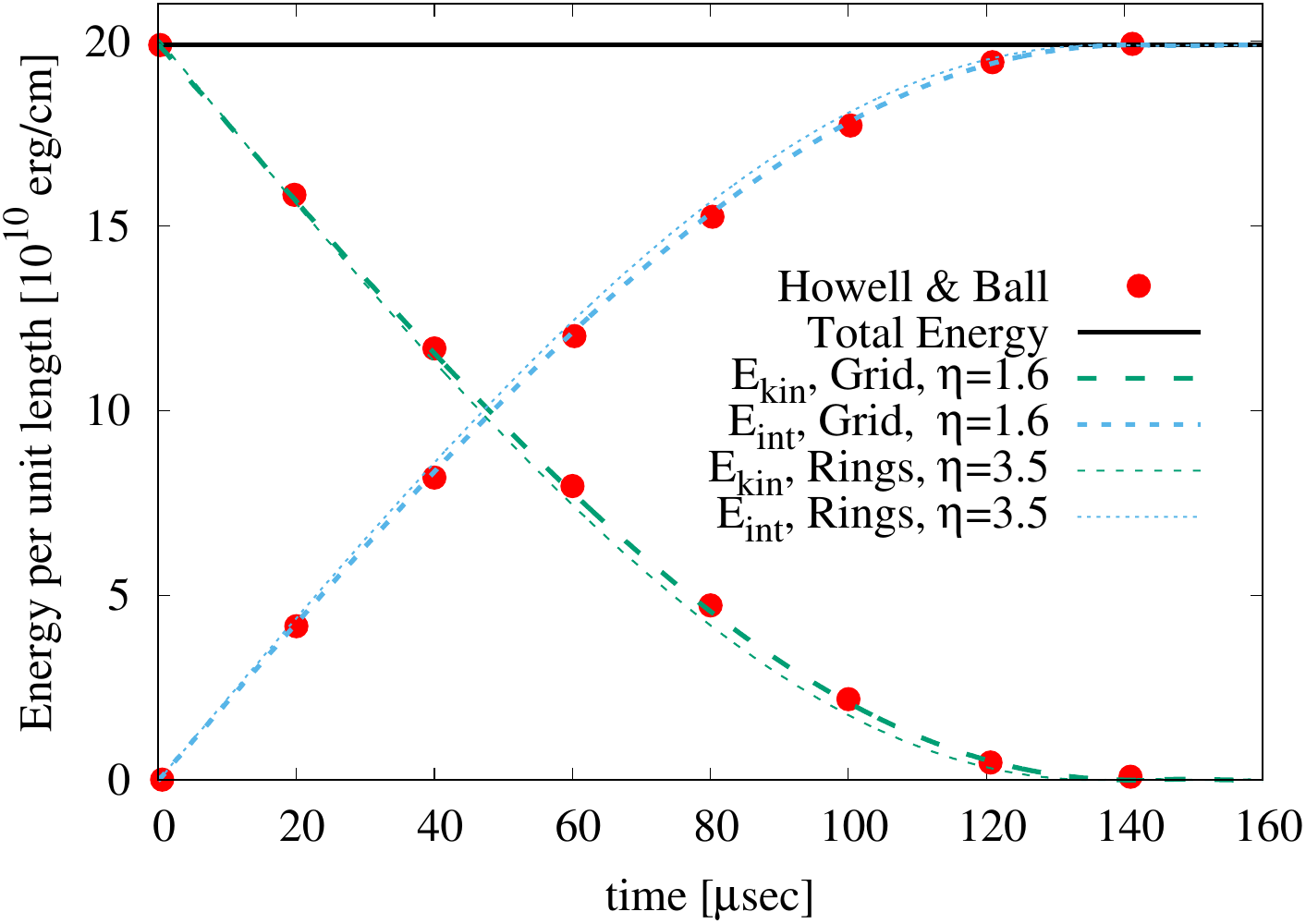} 
    \hspace{0.1\textwidth}
    \caption{Total, kinetic, and internal energy per unit length of the 2D Verney implosion together with the results from 
    \cite{Howell2002}. Thick lines correspond to the triangular lattice particle initialization with $\eta = 1.6$ while thin lines show the energy for the ring initialization with $\eta = 3.5$.}
    \label{fig:verney_2D_energy2}
\end{figure}
Despite the lattice imprint, the overall evolution of total kinetic and internal energies for the lattice particle placement with $\eta = 1.6$ is in very good agreement with the results of \cite{Howell2002} as shown in Fig.~\ref{fig:verney_2D_energy2}.
For the ring distribution with $\eta = 3.5$, we find that the energies still agree fairly well but slightly under and over predict the kinetic and internal energies, respectively. 
We can most likely attribute that to the more dissipative nature of the simulations with larger smoothing length.
\begin{figure}
    \centering
    \includegraphics[width = 0.45\textwidth]{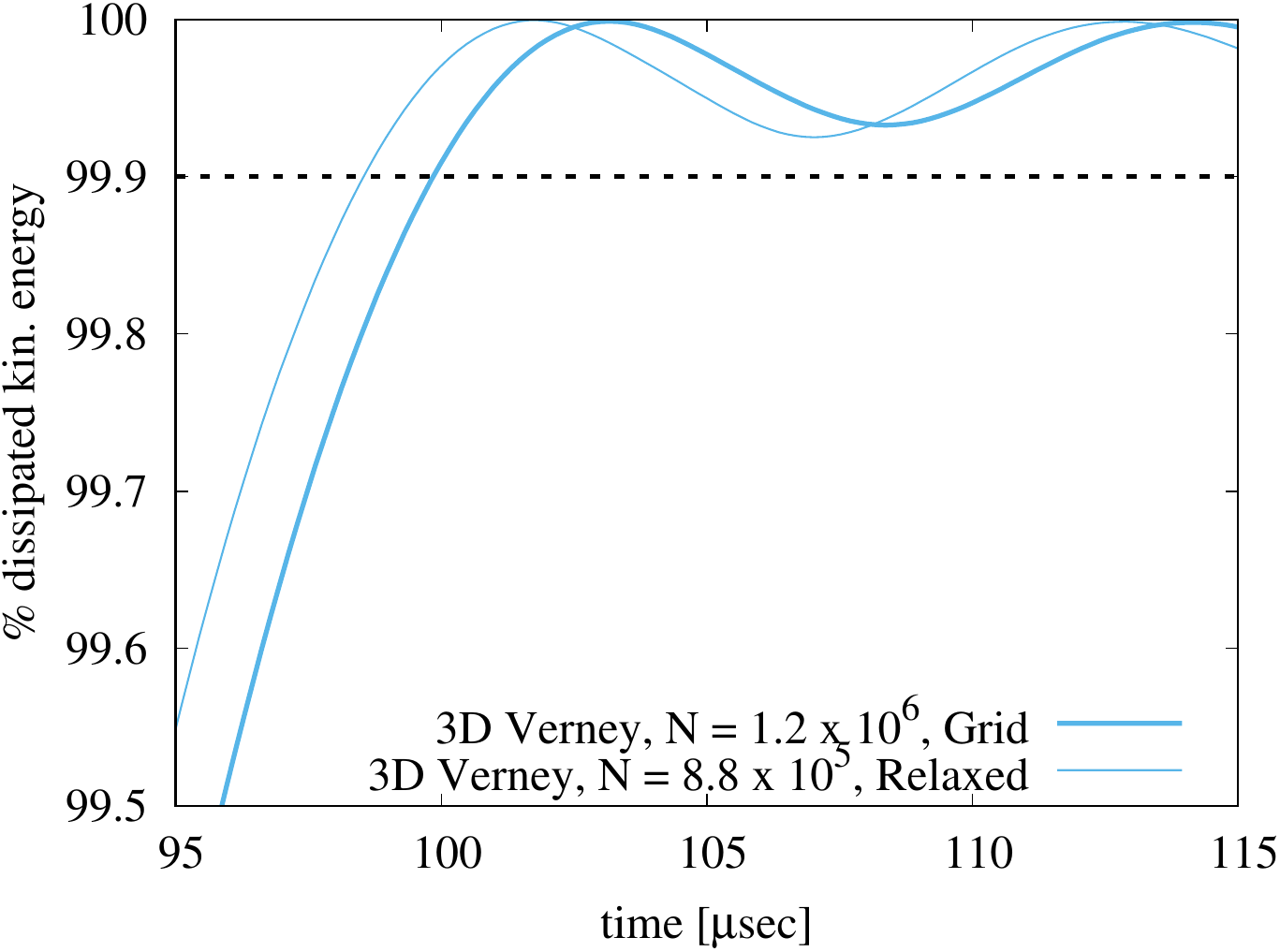}
    \caption{Time evolution of the dissipated kinetic energy for the grid and relaxed particle initializations.}
    \label{fig:verney_3D_energy}
\end{figure}
\begin{figure}[!htbp]
    \centering
    \includegraphics[width = 0.48\textwidth]{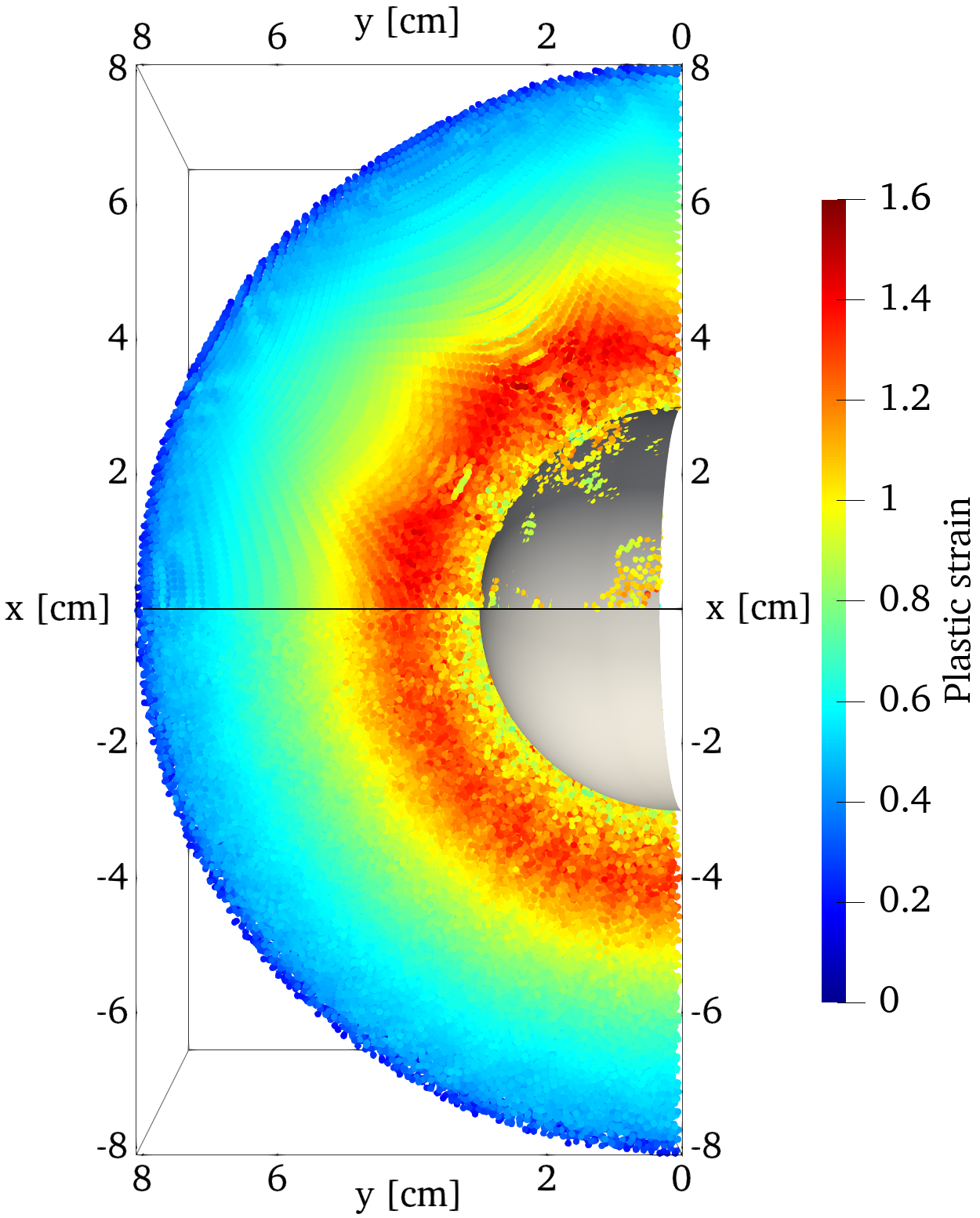}
    \caption{Particle distribution and plastic strain in the 3D Verney implosion simulation at ca. 100\:s. The top half corresponds to 1/8 of a setup with 1.2 million particles distributed in a regular grid while the bottom half are 1/8 of $8.8 \times 10^5$ particles that have first been relaxed in an external potential. We add 1/4 of a grey half sphere with a radius of 3\:cm to show the location of the analytic stopping radius.}
    \label{fig:3d_verney}
\end{figure}
For the simulation in 3D, we again use two different setups. The first has about $1.2 \times 10^6$ particles set in a triangular grid, representing a shell with inner radius of 8\:cm and outer radius of 10\:cm. 
For the second simulation, we initially distribute particles in a sphere with homogeneous density and radius of 11\:cm. The particles are randomly perturbed from their grid positions and then relaxed in an external potential. 
After the relaxation process, we cut a shell with inner and outer radii of 8\:cm and 10\:cm, respectively. This results in a total of $8.8 \times 10^5$ particles.
In both setups, we use $\alpha = 2$ and $v_0 = 6.75 \times 10^4 \:$cm/s. 
The stopping radius of 3\:cm is predicted for a time of 100 $\mathrm{\mu}$s~\citep{Burton2015}. 
We recover it when 99.9\% of the kinetic energy is dissipated using the grid particle initialization. 
For the relaxed configuration, the 99.9\% energy mark is reaches at 98.5\:$\mathrm{\mu}$s, as shown in Fig.~\ref{fig:verney_3D_energy}. 
For both initial conditions, the particle configuration at $102\: \mu$s are shown in Fig.~\ref{fig:3d_verney} where we plot the plastic strain for 1/8 of the shells for clarity.  
We also add 1/8 of a grey sphere with radius 3\:cm to mark the predicted stopping radius.
\begin{figure}
    \centering
    \includegraphics[width = 0.45\textwidth]{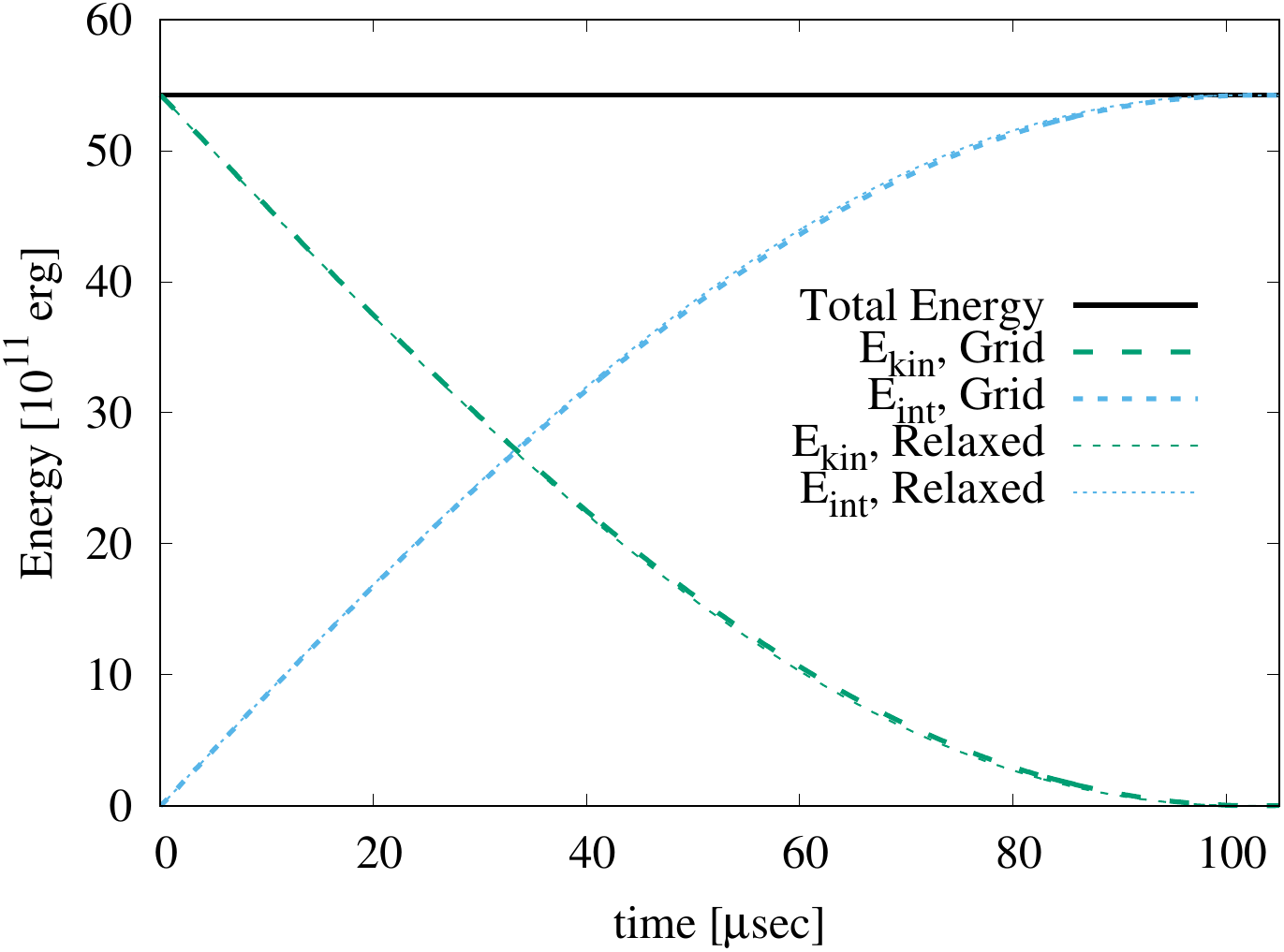} 
    \hspace{0.1\textwidth}   
    \caption{Time evolution of the total, internal, and kinetic energies in the 3D Verney implosion simulation for grid and relaxed initial particle configurations.}
    \label{fig:verney_3D_energy2}
\end{figure}
For the grid initialization, the radial distance of the inner particles at stopping time is ca. 3\:cm, however, about 1240 particles have a smaller radial distance, between 2.8\:cm and 3\:cm. 
We also see an imprint of the lattice structure in the plastic strain. 
For the relaxed particle setup, none of the particles crosses the stopping radius and the minimal radial distance is about 3.05\:cm. 
We find that the plastic strain field is much smoother than in the grid setup while the evolution of the total, kinetic, and internal energies agrees very well, as shown in Fig.~\ref{fig:verney_3D_energy2}.
\subsection{Taylor Anvil}
Here, a cylindrical metal rod is subject to high strain rates after its impact on a rigid surface \citep{Meyers1994}.
The test is motivated by experimental work of \cite{Taylor1948} who studied dynamic non-uniform deformation of cylindrical projectiles which were fired at a steel target.
The deformation is a result of elastic and plastic waves propagating in the cylinder and therefore depends on the constitutive behavior of the underlying material. 
During the impact, the initial kinetic energy is converted into internal energy.
With that, the final shape, e.g. the position of the foot, is dependent on plastic dissipation \citep{Burton2015}.
Here, we use a copper rod with an initial radius of 0.32\:cm, a length of 3.24\:cm, and velocity of $2.27 \times 10^4\:$cm/s.
\begin{figure}[!htbp]
    \centering
    \includegraphics[width = 0.42\textwidth]{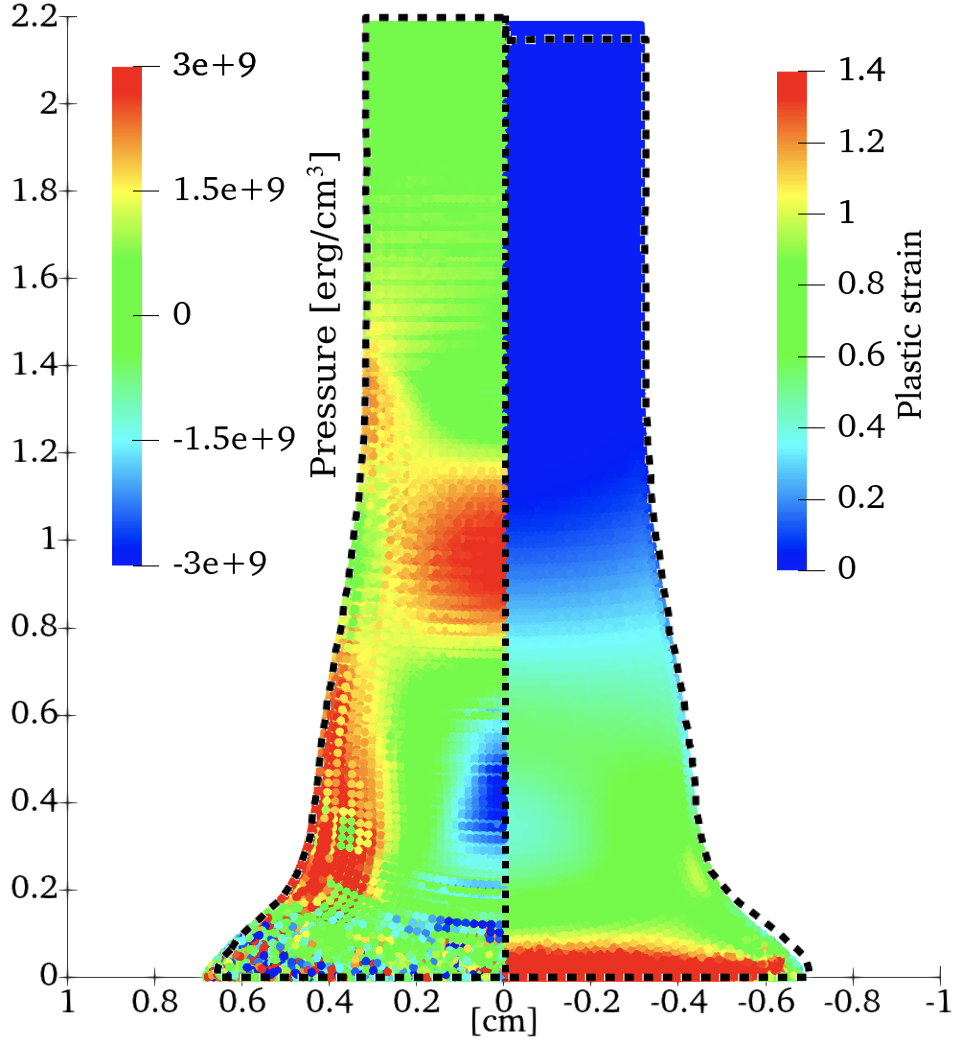}
    \caption{Cross-section of the 3D Taylor anvil, initialized by placing particles in a lattice, with the pressure field on the left and the plastic strain on the right at $80 \: \mathrm{\mu sec}$. The dashed outline shows the shape of the rod from \cite{Burton2015} at the same time for the CCH2 (left) and CGR (right) schemes.}
    \label{fig:3d_anvil}
\end{figure}
\begin{figure}[!htbp]
    \centering
    \includegraphics[width = 0.42\textwidth]{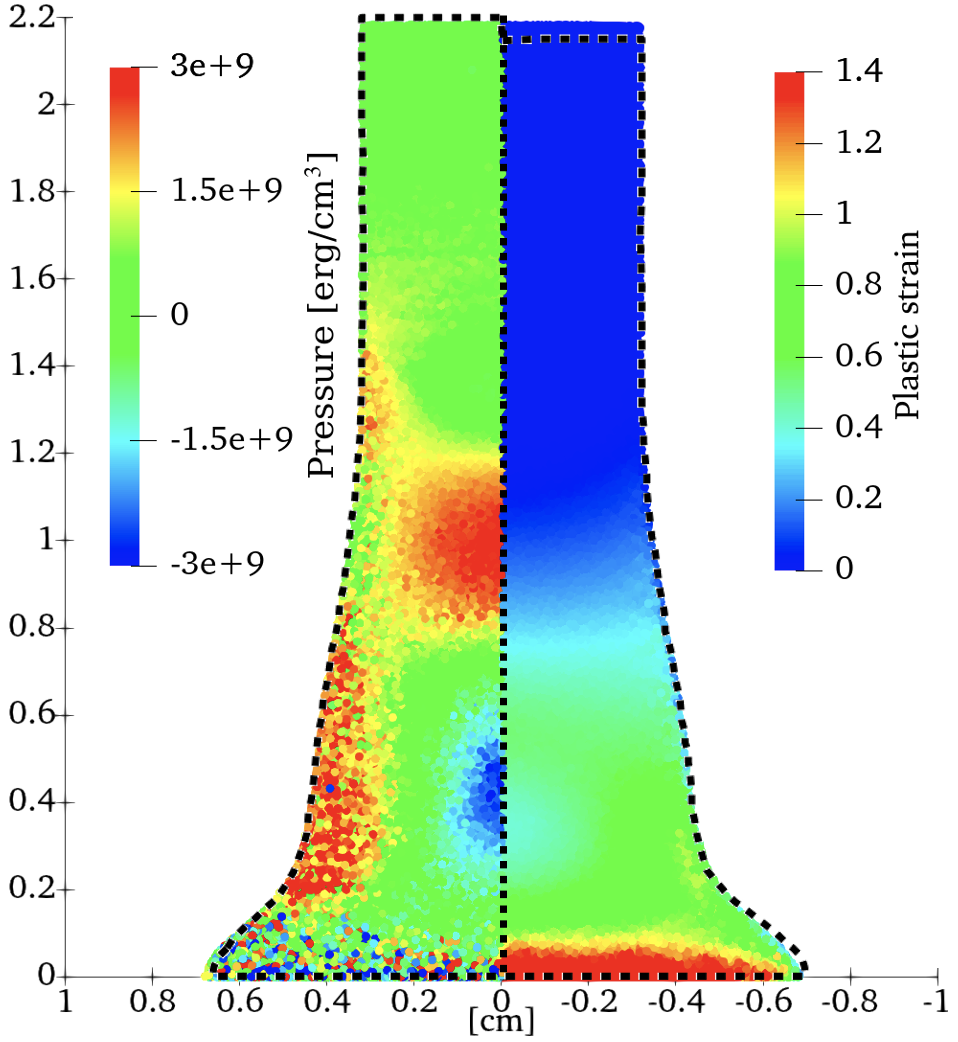}
    \caption{Cross-section of the 3D Taylor anvil, initialized by relaxing particles, simulation with the pressure field on the left and the plastic strain on the right at $80 \: \mathrm{\mu sec}$. The dashed outline shows the shape of the rod from \cite{Burton2015} at the same time for the CCH2 (left) and CGR (right) schemes.}
    \label{fig:3d_anvil_relax}
\end{figure}
\begin{figure}[!htbp]
    \centering
    \includegraphics[width = 0.45\textwidth]{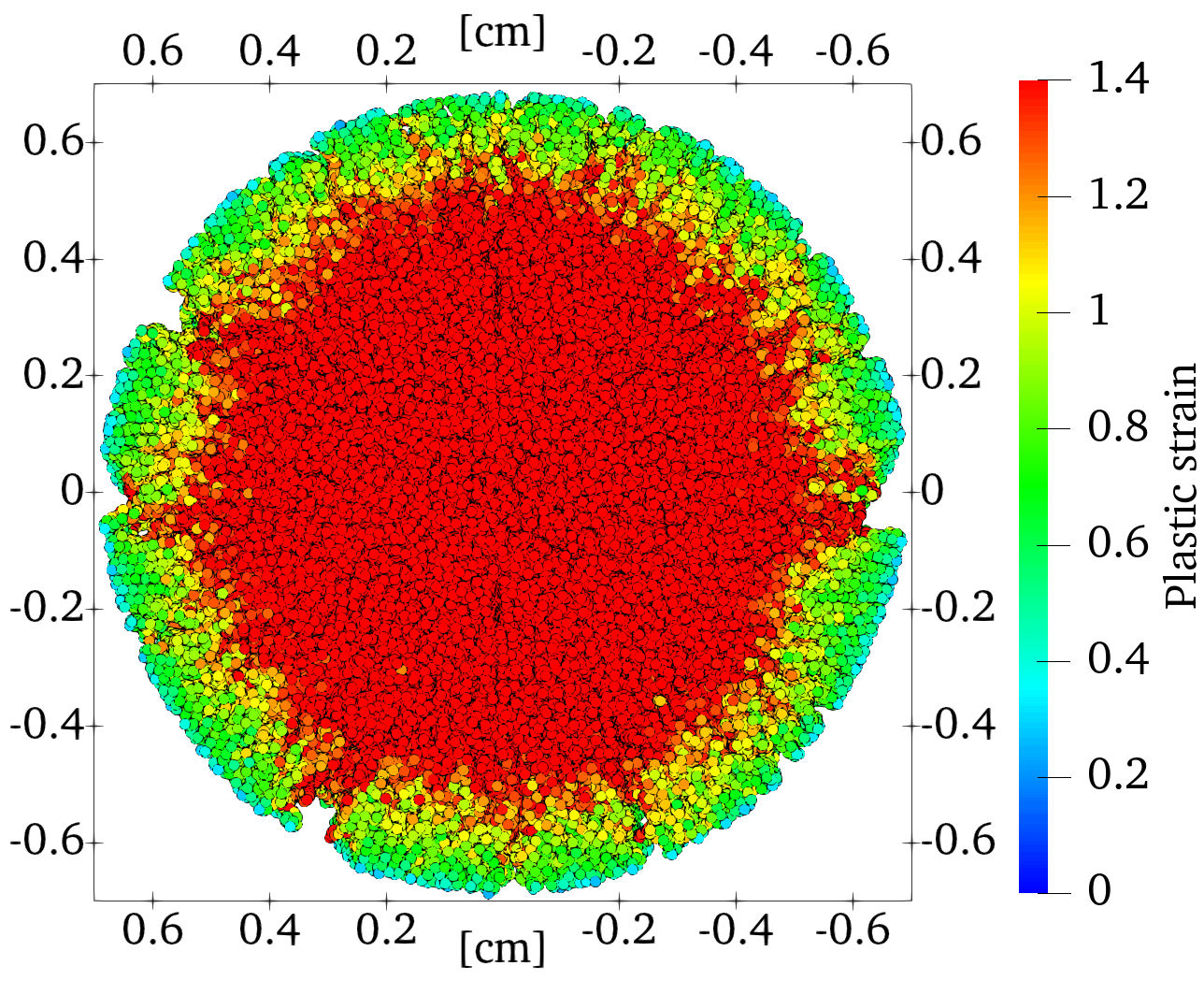}
    \caption{Foot of the Taylor anvil in Fig.~\ref{fig:3d_anvil} as viewed from the bottom. Color-coded is the plastic strain.}
    \label{fig:3d_anvil_2}
\end{figure}
There are no analytical predictions for this test, however, we can compare e.g. the deformation of the rod to published work. 

We apply the Mie-Gr\"uneisen EoS with $\rho_0 = 8.93\: \mathrm{g/cm^3}$, $c_0 = 3.94 \times 10^5 \: \mathrm{cm/s}$, $s = 1.48$, and $\Gamma = 2$. 
The strength model is EPP with linear hardening with $\mu = 4.3333 \times 10^{11} \: \mathrm{Ba}$, $Y_0 = 4 \times 10^9 \: \mathrm{Ba}$, and $H = 10^9 \: \mathrm{Ba}$. 
As in the Verney test, we run the simulation with two different setups. 
One with a particle initialziation in a lattice and one where particles undergo a relaxation process first.
For the lattice, we use $7 \times 10^5$ particles.
For the relaxed particle configuration, we first initialize particles in a sphere with a radius of 2\:cm. 
We perturb the positions of the particles randomly and then relax them in an external spherical potential assuming a constant density. 
After the relaxation process, we extract particles with $-1.6\:\mathrm{cm} \leq z \leq 1.6\:\mathrm{cm}$ and $\sqrt{x^2 + y^2} \leq 0.32\:\mathrm{cm}$
and impose the initial velocity and material properties. 
For both setups, we use the Wendland C6 smoothing kernel with $\eta = 1.6$.
There is some dependence of the outcome on modeling the interaction with the target \citep{Burton2015,Morgan2013}.
Here, we assume simple reflective boundary conditions, where particles that pass the surface of the target are instantaneously reflected back. 
We compare our results to simulations by \cite{Burton2015} who constrained the rod to stick to the surface. 
Figure \ref{fig:3d_anvil} shows the cross-section of the rod for the lattice initialization at $80\:\mathrm{\mu s}$ where the left side gives individual particle pressures and the right side the plastic strain. 
In addition, we plot the outline of rod shapes from \cite{Burton2015} as a dashed line. 
The left side corresponds to their applied CCH2 hydrodynamic scheme while the right side is obtained with the CGR scheme.  
In general, the shapes of the bar agree very well.
Our anvil heel is at $\sim 2.18\:\mathrm{cm}$ while the toe is located at  $\sim 0.68\:\mathrm{cm}$, see Fig.~\ref{fig:3d_anvil_2}. 
For comparison, \cite{Burton2015} obtain a toe position of around $0.70 \:\mathrm{cm}$ and $\sim 2.15\:\mathrm{cm}$ for the heel using CGR while CCH2 gives a toe at $\sim 0.66\:\mathrm{cm}$ and heel at $\sim 2.20 \: \mathrm{cm}$.
We find a lot of noise in the pressure, especially close to the impact surface. 
This is most likely the result of our simple treatment of the interface between the projectile and target. 
It could also be improved by more modern SPH modeling techniques (see e.g. \cite{Yan2022, Mohseni2021, Chalk2020, Molteni2009} and references therein).
As in the Verney test, we see an imprint of the lattice particle placement, here mostly in the pressure. 
It should be noted that our plotting of individual particle pressures is not equivalent to pressure visualizations of grid-based methods. 
For that, a more correct approach is to involve a convolution over the particle's neighbors. 
However, while this could alleviate the pressure noise, it would not cure the pressure stripes which are aligned with the (deformed) lattice used for the initial particle positions. 
The pressure strips are most likely the result of fluid motion reducing particle resolution in one direction while increasing it in the perpendicular direction which is exacerbate by the lattice arrangement.

We can compare the outcome of the impact to a relaxed particle placement. 
The corresponding snapshot is shown in Fig.\ref{fig:3d_anvil_2}.
Here, the noise is the individual particle pressure in a bit more pronounced, however, we also see that the pressure alignment is seemingly gone.
In general, the shape of the compressed rod and the pressure distribution agrees very well with the lattice initialization.
\subsection{Conclusion to Solid Material Tests}
We end the discussion of our solid material benchmark tests here and move to the astrophysical application of FleCSPH in the next section. 
As a concluding remark we can say that FleCSPH passes the presented tests by showing good agreement with analytic solutions and other numerical methods for shape and energy evolution.
The usage of relaxed particle configurations seems to improve grid imprints and particle alignment that are seen when particles are initialized in a lattice. 
This aspect should be explored further.
Future work in solid material modeling with FleCSPH will also include the implementation of more advanced techniques to e.g. suppress particle noise. 
\section{The Neutron-Star Crust}
NSTs are one of the final stages of stellar evolution and are created in supernova explosions of stars with masses in the range of about  $8-25M_{\odot}$~\citep{Janka2007}.
The masses of NSTs range from $\sim 1.2-2.1 M_{\odot}$ with radii of the order of 12~km~\citep{Dietrich:2020efo}.
This leads to central densities that exceed the nuclear saturation density, $n_{\rm sat} \sim 0.16$fm$^{-3}$ (corresponding to a mass density of $\rho_{\rm sat} \sim 2.7\cdot 10^{14}$g cm$^{-3}$) by several times. 
NSTs have a layered structure. 
Although there are still many uncertainties, it is assumed that the core is largely occupied by a fluid composed of a mix of nucleons and leptons.
As the densities in the center reach several times $n_{\rm sat}$, more exotic particles such a hyperons or deconfined quarks can appear ~\citep{Alford:2004pf,Alford:2006vz,Lonardoni:2014bwa,Annala:2019puf}.
The NST core is surrounded by a solid crust, which consists of neutron-rich nuclei arranged in a Coulomb lattice~\citep{Baym1971,Chamel:2008ca}. 
The crust's thickness is of the order of 1~km and contains $\sim1$\% of the total NST mass. 
It can be subdivided into an outer and inner part.  
The outer crust is a Coulomb crystal of neutron-rich nuclei in the iron region surrounded by a degenerate electron gas. 
With larger densities, the neutron chemical potential grows and the nuclei become increasingly neutron-rich. Eventually, the neutron chemical potential is sufficiently large for neutrons to drip out of the nuclei. 
This defines the onset of the inner crust that is formed by a lattice of progressively heavier and more neutron-rich ions, surrounded by both a neutron and an electron gas. 
At the base of the inner crust, when densities approach half of $n_{\rm sat}$, the competition between nuclear and Coulomb forces might lead to the formation of non-spherical nuclei and exotic shapes of nuclear matter, called nuclear pasta~\citep{Newton:2009zz,Caplan:2016uvu}. 
At even higher densities, the nuclei dissociate and form a fluid consisting of neutrons, protons, and electrons in $\beta$ equilibrium. 
This is where the NST core begins.

Understanding the composition of the crust and its properties is important for many different aspects of NST research. 
NST crust material is very interesting because it is a solid and therefore possesses non-zero shear strength~\citep{Horowitz2009}.
This and the crust's location close to the star's surface, could possibly lead to observational signatures of its shear motion which might be detectable as low-frequency quasi-periodic oscillations in giant X-ray flares~\citep{Strohmayer:2006py,Steiner:2009yg}.
The crust could also modify the GW signal from NST mergers.
For example, during the inspiral phase, as two stars move closer, the orbital period of the binary decreases and might pass characteristic frequencies of NST crustal oscillation modes. 
The resulting resonant coupling between a mode and the orbital frequency can draw from the orbit, change the GW phase, and even shatter the crust~\citep{Tsang2012}. 
The latter might result in an electromagnetic precursor signals, seconds before the main short GRB~\citep{Neill:2020szr}. 
The solid nature of the crust might also allow for the existence of mountains on NSTs, which could emit continuous gravitational waves~\citep{Haskell:2015psa}.

As with numerical studies of terrestrial solids, the simulation of the NST crust requires the knowledge of its EoS, the constitutive model, and breaking behavior. 
The shear modulus of inner and outer crust material as well as breaking strains have been studied via analytical calculations and microphysical simulations with Molecular Dynamics~\citep{Horowitz2009, Caplan:2016uvu} and Skyrme-Hartree Fock~\citep{Newton:2009zz}.
The first Molecular Dynamics study of elastic deformations found a linear stress-strain relation, followed by a small region of plasticity before an abrupt breaking with a breaking strain of $\sim 0.1$ and $\sim 0.04$ for shear and tensile deformations, respectively~\citep{Horowitz2009}. 
Being about $10^{10}$ times stronger than terrestrial engineering materials makes the crust the strongest known solid in nature. 
The stress-strain relation of the crust resembles the one of brittle solids on Earth. 
However, different to those, the high pressures in the crust prevents voids from forming and fractures from propagating. 
This absence of localized defects results in abrupt failure once the breaking strain is reached rather than continuous yielding at low strains.\\

We follow the work of \cite{Tews:2016ofv} for values of the shear modulus of the NST crust. 
Here, the inner crust is modeled via the Wigner-Seitz approximation, where we consider a spherical nucleus described by a liquid droplet model in a spherical unit cell.
The nucleus is surrounded by a neutron fluid that is in chemical and mechanical equilibrium with the nucleus. 
We describe the bulk energy of the nucleus as well as of the neutron fluid using calculations from chiral effective field theory (EFT).
It is a systematically improvable theory to describe interactions between nucleons consistent with all symmetries of quantum chromodynamics.
EFT provides an order-by-order framework that enables one to improve the precision and accuracy of nuclear interactions. 
Theoretical uncertainties can be easily estimated from order-by-order calculations~\citep{Epelbaum:2008ga,Machleidt:2011zz}.
Within this framework, two-nucleon and consistent many-nucleon forces are included by construction.
Chiral EFT was successfully applied to studies of atomic nuclei~\citep{Wienholtz:2013nya,GarciaRuiz:2016ohj,Lonardoni:2017hgs,Hu:2021trw} and nuclear matter~\citep{Hebeler:2009iv,Tews:2012fj,Hagen:2013yba,Drischler:2017wtt}.
Using the neutron-matter results from \cite{Lynn:2015jua} and empirical information on symmetric matter, we have used this approach to calculate the EoS of the inner crust, i.e., pressure as a function of energy or number densities, as well as geometrical properties of the crust, i.e., the cell radii and volumes as functions of density.\\
Using the Wigner-Seitz approximation with the assumption of a body-centered cubic Coulomb lattice with electron screening effects, the shear modulus of the crust can be written as
\begin{align}
\mu = 0.1194 \: \left( 1 - 0.01 \: Z^{2/3} \right)^2  \: \frac{Z^2 e^2}{a} \: n_i
\end{align}
where $n_i = V_W^{-1}$ is the volume of the unit cell and $a = R_W$ its radius, while $Z$ is the charge number of the nucleus in the cell. 
We assume $Z=40$ for the charge number of the nuclei. 
However, variations of $Z$ do not lead to large changes in the shear modulus~\citep{Tews:2016ofv}.
We find that the shear modulus can be expected to be in the range of $\sim 10^{28} \: \mathrm{erg/cm^3}$ to $\sim 10^{30} \: \mathrm{erg/cm^3}$ for densities $10^{12} \: \mathrm{g/cm^3}$ to $10^{14} \: \mathrm{g/cm^3}$~\citep{Tews:2016ofv}.
This is in agreement with Molecular Dynamics and Skyrme Hartree-Fock simulations.
\begin{figure}
    \centering
    \includegraphics[width = 0.45\textwidth]{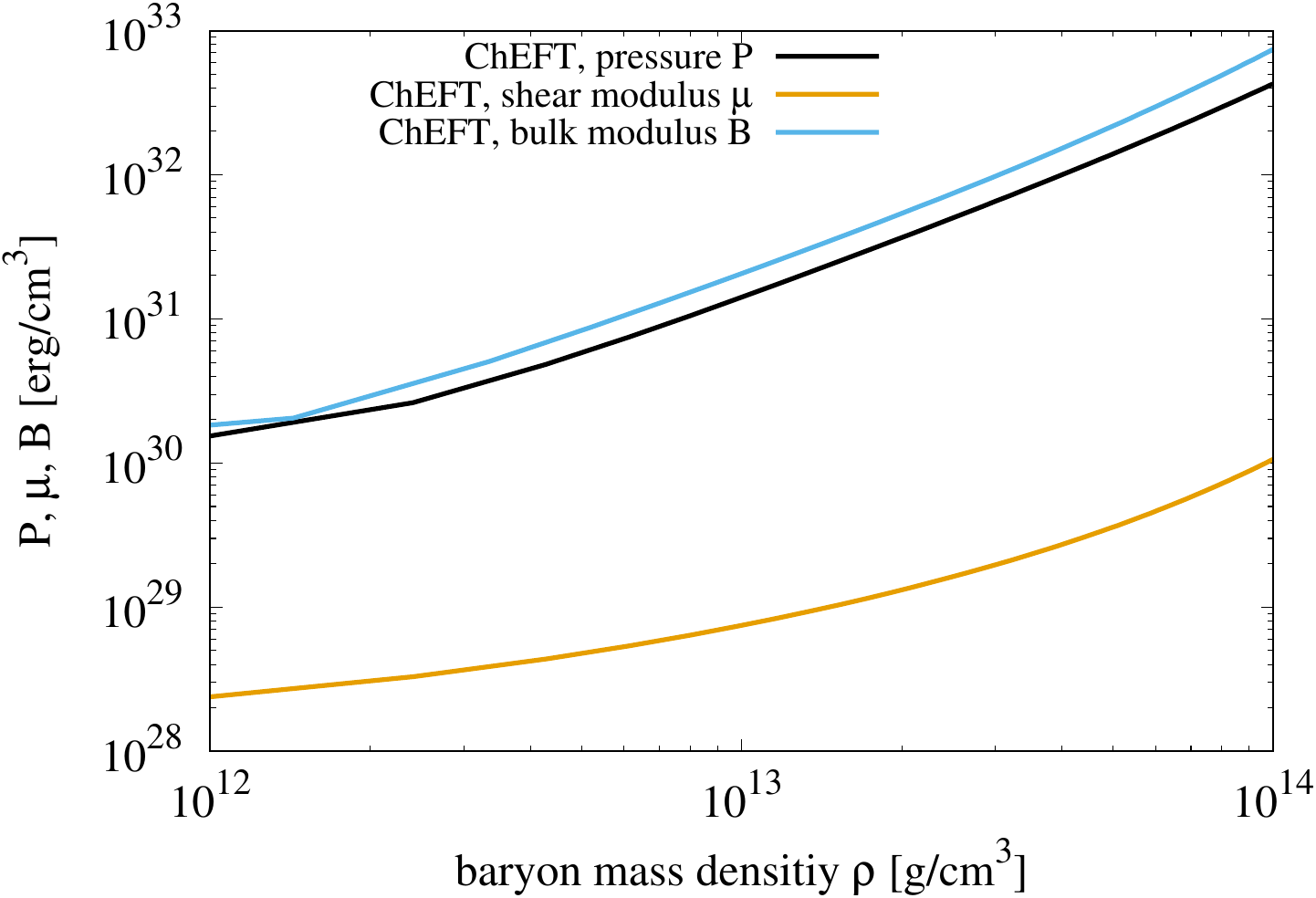}
    \caption{Pressure, bulk modulus, and shear modulus of neutron-star crust material determined via Chrial EFT calculations.}
    \label{fig:eos_mub_reduced}
\end{figure}
In Figure~\ref{fig:eos_mub_reduced}, we plot the resulting EoS, bulk and shear moduli of the NST crust material. 
It can be seen that the bulk modulus is 2-3 orders of magnitude larger than the shear modulus. 
Note that this relation between bulk and shear modulus is found in gelatin-like solid materials while for typical terrestrial solids like rocks and metals the bulk and shear moduli are usually of the same magnitude. 
\section{Crust Simulations with FleCSPH}
In principle, FleCSPH is very well equipped to study the NST crust. 
It includes analytic and tabulated nuclear-matter EoSs at zero and finite temperature, respectively. 
In addition, the code contains Newtonian gravity via the Fast Multipole Method, a set of general-relativistic background metrics for relativistic isolated NSTs, and the capability to use potentials to mimic external forces and self-gravity \citep{Tsao2021}. 
While the code has already been used for compact-star \citep{Kaltenborn2022, Tsao2021, Loiseau2020} and kilonova modeling \citep{Stewart2022}, the addition of constitutive models should allow the study of solids in compact star environments. 
However, as we will describe in the next sections, the fact that the NST crust behaves like a gelatin and not like a traditional solid results in several numerical challenges. 
With that, rather than exploring the full spectrum of possible NST crust solid material behavior, we will focus on modeling a rather simple problem of toroidal crust oscillations in the fundamental mode. 
Such oscillations are also of physical relevance as they might explain low-frequency modes seen in quasi-periodic oscillations in giant X-ray flares~\citep{Strohmayer:2005ks,Watts:2005ue}. 
These are assumed to be caused by rearrangements of the NST magnetic field which is coupled to the crust. 
Analytic estimates of the crustal oscillation modes can be done in Newtonian and relativistic regimes. 
Some applied approximations include the absence of crust-core coupling effects via the magnetic field, the usage of slab instead of a spherical geometries, and the assumption of incompressible matter with a constant shear modulus. 
For the $m=0$ modes, \cite{Piro:2005jf} estimated
\begin{align}
 f_{P,l} = \frac{1}{2\pi} \: \sqrt{\frac{\mu}{\rho}} \frac{\sqrt{l(l+1)}}{R} \,,
 \label{eq::freq_piro}
\end{align}
while \cite{Samuelsson:2006tt} arrived at
\begin{align}
 f_{S,l} = \frac{1}{2\pi} \: \sqrt{\frac{\mu}{\rho}} \sqrt{\frac{(l-1)(l+2)}{2RR_c}} \,
 \label{eq::freq_samuelsson}
\end{align}
where $\sqrt{\mu/\rho}$ is the shear velocity in the crust and $R_c$ and $R$ are the crustal radial distance and the star's radius, respectively.  
Here, both expressions are adjusted for the non-relativistic regime. 
Their small difference in the numerator originates in the way how results are transferred from the planar to a spherical geometry. 
The estimate from \cite{Samuelsson:2006tt} has been found to fit the $l$-scaling of the oscillation frequencies better but it might not describe the frequency of the fundamental $m=0,\:l=2$ mode.
Given the presence of analytical estimates and FleCSPH's solid material and compact-star modeling capabilities, we can attempt to initialize fundamental toroidal oscillation of the NST crust and compare with the resulting frequencies from Eqs.~(\ref{eq::freq_piro}) and (\ref{eq::freq_samuelsson}).
However, potential numerical challenges are: 
\begin{figure*}[htb!]
    \centering
    \includegraphics[width = 1.0\textwidth]{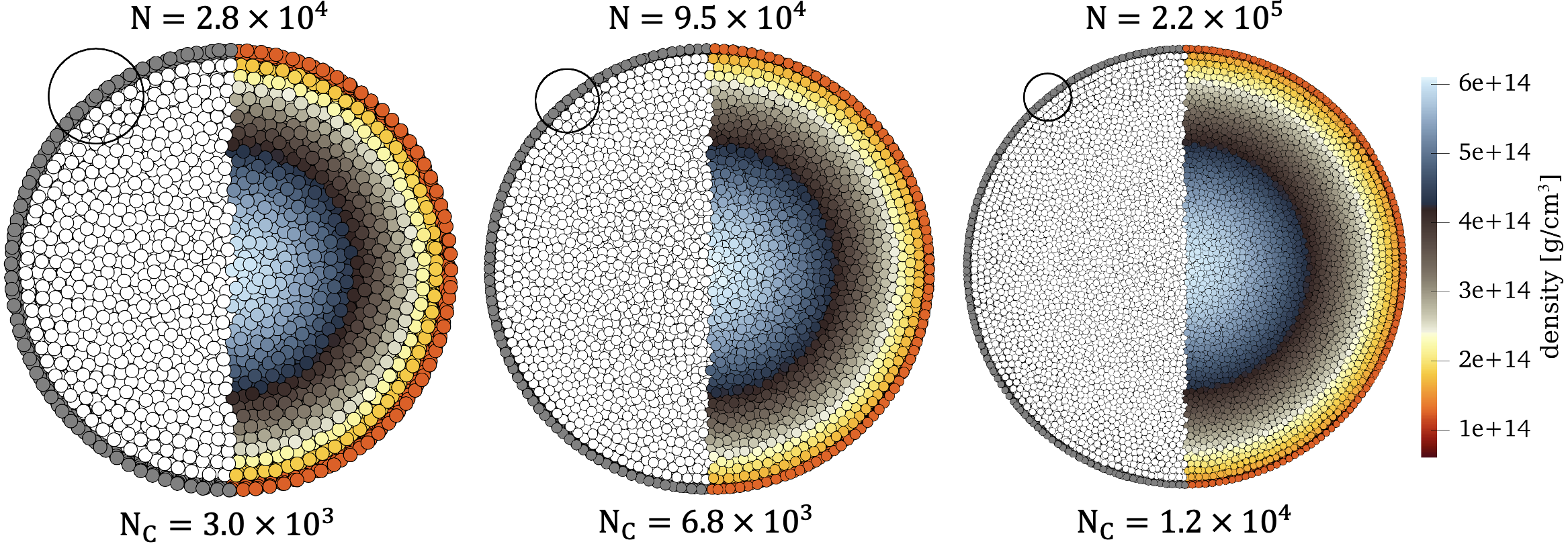}
    \caption{The three NST setups with crust and core, composed of $N = 2.8\times 10^{4}$, $9.5 \times 10^{4}$, and $2.2 \times 10^{5}$ particles in total. $N_C$ gives the number of particles in the crust only. The left side of the stars shows the distinction into crust particles (grey) and core particles (white) while the color-coded right side shows the density. In all three cases, the crust is defined as a layer of particles at a density of $\sim 1.4 \times 10^{14} \mathrm{g/cm^3} \sim 0.5 \rho_{\rm sat}$. The circle in the top left of each star gives the smoothing length of a crust particle, marking the particle's SPH neighbors as the particles located within the circle. The particle distributions are achieved with relaxation steps 1 and 2, as described in section \ref{section::sim_setup}.}
    \label{fig:star_cross}
\end{figure*}
\begin{enumerate}
     \item Due to its lower densities and small radial width, the NST crust is usually represented by much fewer particles than the NST core. 
     This results in the majority of computational resources being spent on the simulation the core dynamics rather than the crust and limits our ability to resolve low crustal densities.   
    \item Assuming the absence of magnetic fields, there should be no physical coupling mechanism between the shear motion of the solid crust and the fluid core. In other words, the crust should undergo toroidal motion without disturbing the NST core. However, in the general SPH scheme, the hydrodynamic equations are solved by summing over all neighboring particles. If a SPH crust particle is located close to the crust-core interface (CCI), some of its neighbors will belong to the core.
    This is visualized in Fig.~\ref{fig:star_cross}. As a consequence, there is communication across the CCI which can numerically couple the shear motion of both NST components.
    \item The gelatine-like solid nature of the crust makes it very sensitive to density fluctuations which can occur in SPH. This, in turn, leads to perturbations in pressure, particle accelerations and velocities.
    The latter, when not minimized, can dominate the motion of the NST crust.
\end{enumerate}
The first issue can be solved by the usage of external potentials to confine SPH particles to a given crustal density profile without the need to include the core \citep{Tsao2021}.
The second and third challenges will be addressed below. 
\subsection{Simulation Setup}
\label{section::sim_setup}
All stars are simulated with the Wendland C6 smoothing kernel, adaptive $h$ and $\eta = 1.5$.
We initialize NSTs with the polytropic EoS
\begin{align}
    P(\rho) = K \rho^\Gamma\,,
    \label{eq::polytropic}
\end{align}
where we use $\Gamma = 2$ and $K = 10^5 \: \mathrm{g^{(\Gamma -1)} \: cm^{(3\Gamma -1)}/s^2}$.
We focus on stars with a mass of $1.4\,M_\odot$ which, given the parameters for Eq.~(\ref{eq::polytropic}), results in a radius of about 15~km. 
We will test the effect of SPH particle number $N$, i.e. resolution, by simulating stars with $N \sim 2.8 \times 10^4$, $9.5 \times 10^4$, and $2.2 \times 10^5$. 
Since particle masses are kept constant, a larger $N$ usually results in the resolution to lower crustal densities and larger radii. 
Given this and Eqs.~(\ref{eq::freq_piro}) and (\ref{eq::freq_samuelsson}), the final oscillations frequencies might vary with $N$. 
To just test the effect of a higher resolution on the oscillation of the crust at a given density, we therefore modify the NST density profile to have a steep falloff shortly after the crust-core transition. 
\begin{figure}
    \centering
    \includegraphics[width = 0.35\textwidth]{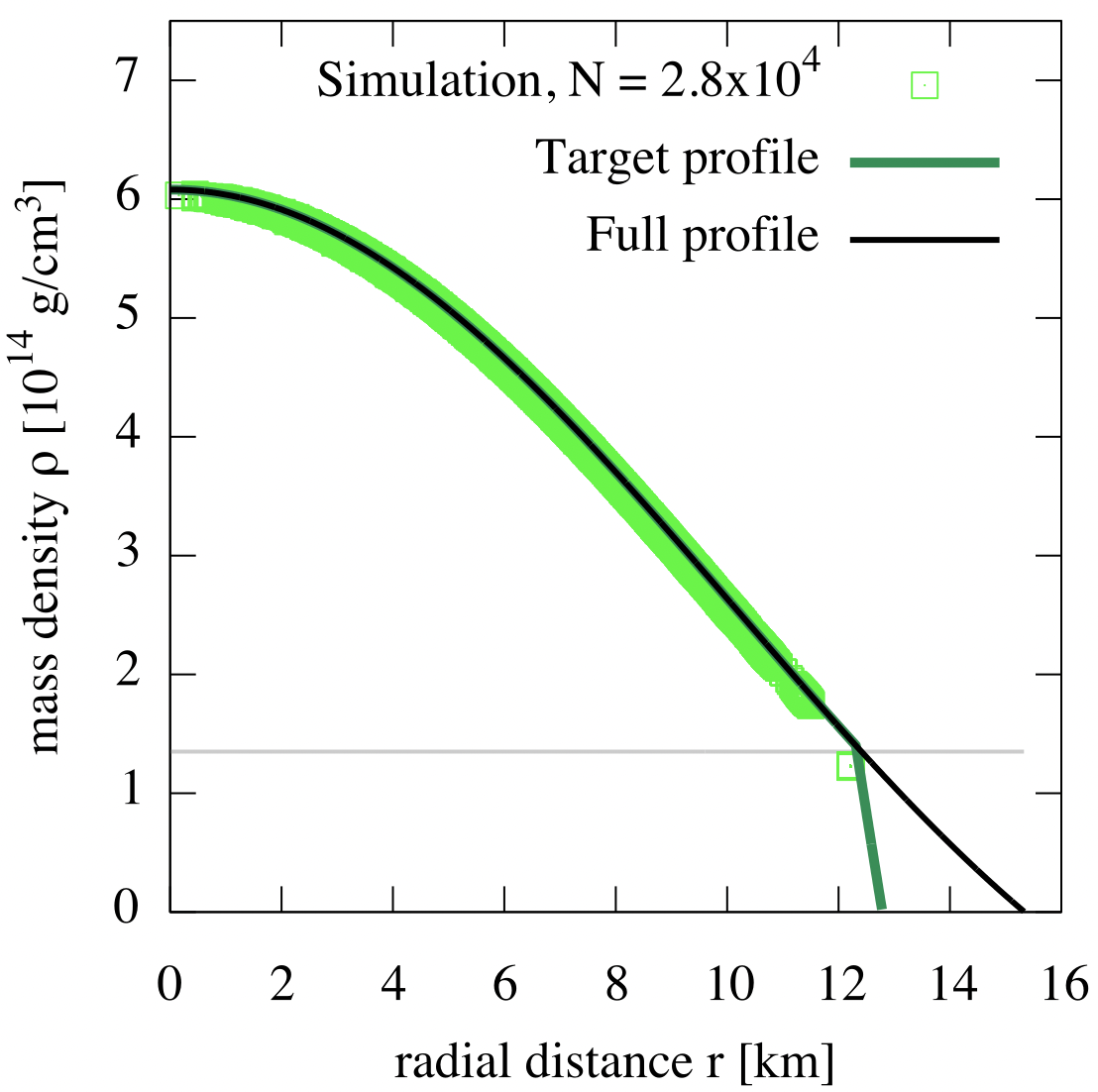}\\
    \vspace{0.4cm}
    \includegraphics[width = 0.35\textwidth]{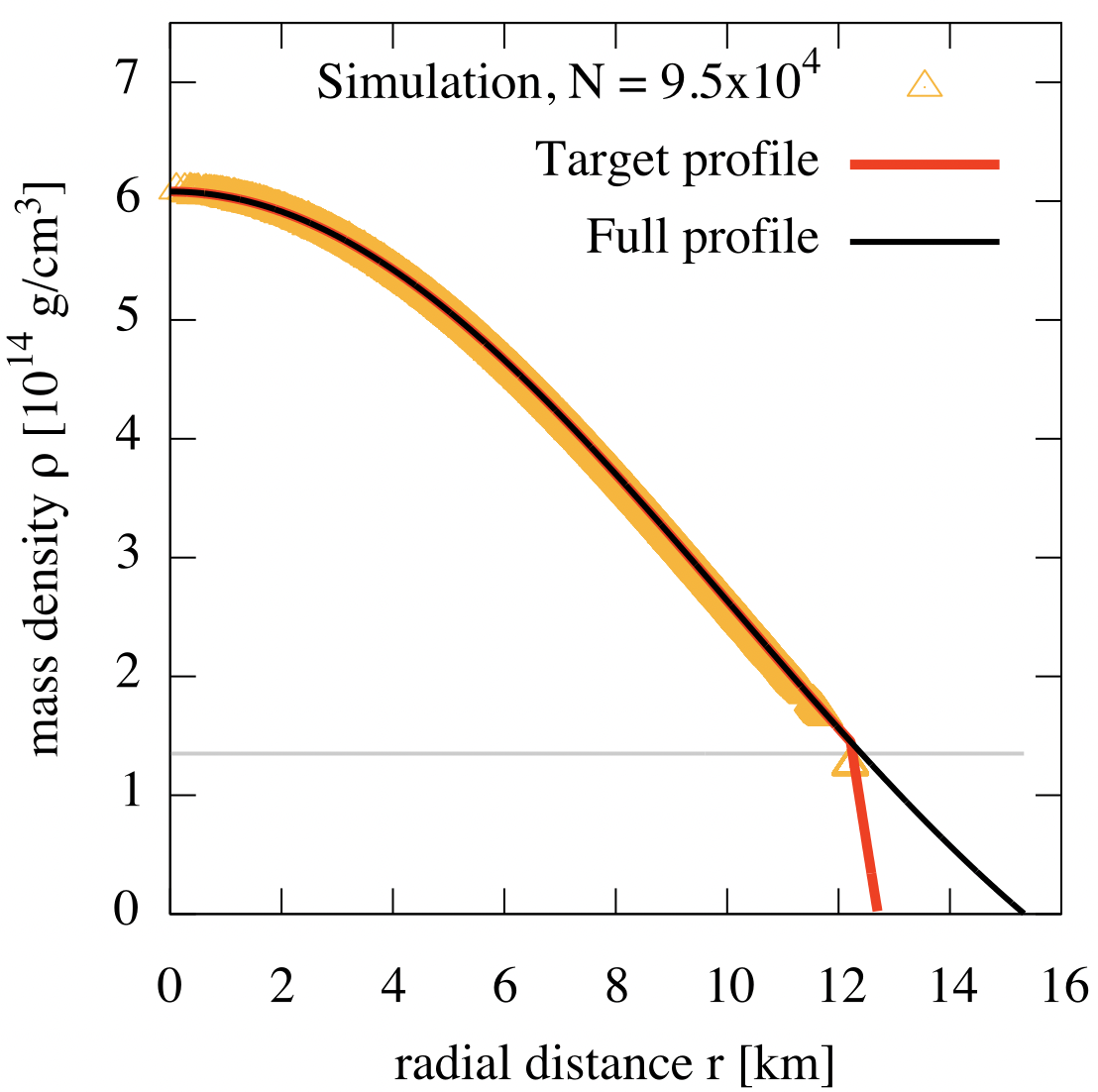}\\
    \vspace{0.4cm}
    \includegraphics[width = 0.35\textwidth]{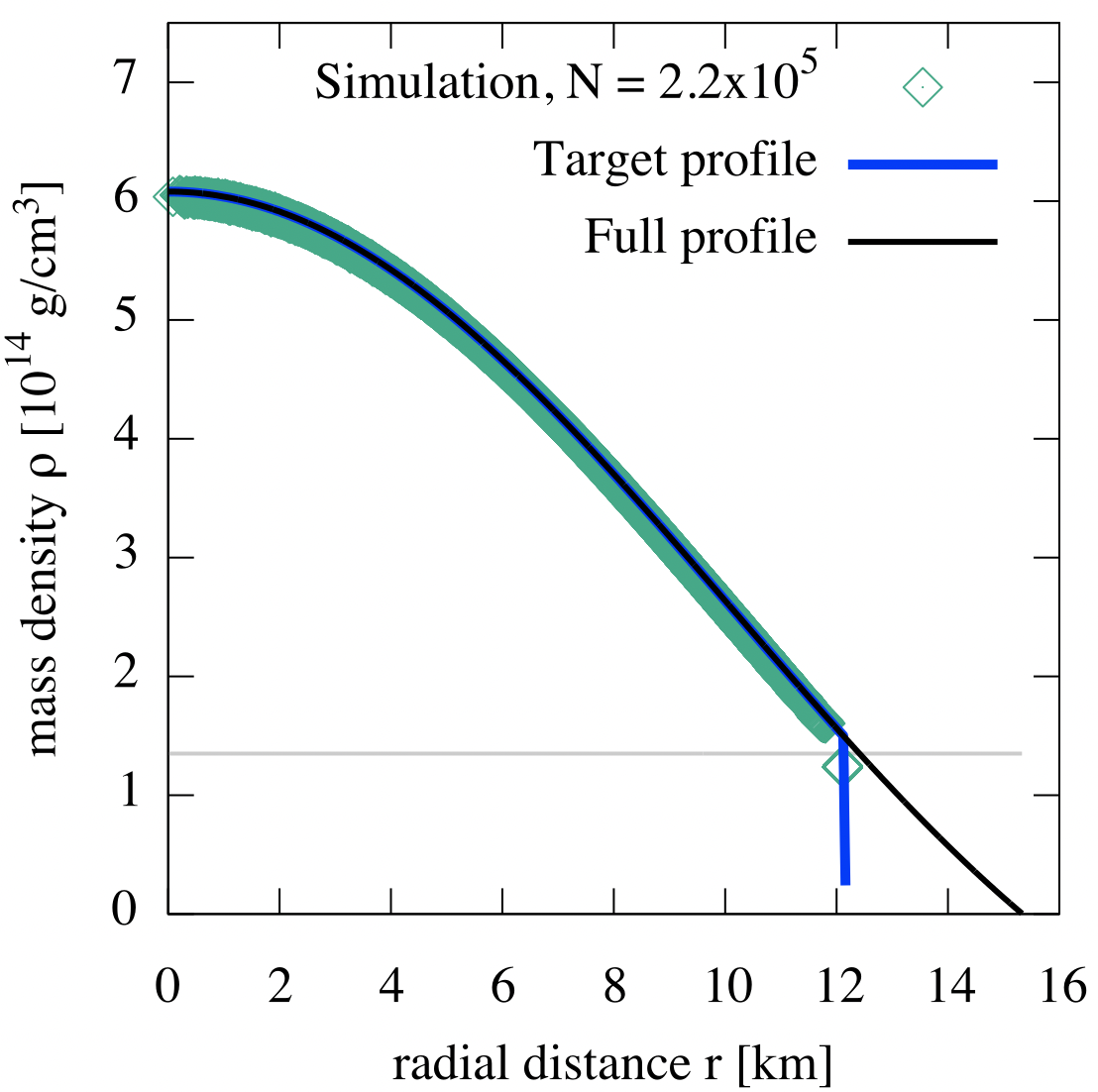}
    \caption{Density profiles used to initialize, relax, and evolve the stars in Fig.~\ref{fig:star_cross}. The profiles are altered from the original profile for a 1.4 solar mass star with the polytropic EoS by cutting them off at around $0.5 \: \rho_0$ (grey line). With that, we ensure that the crust shell is located at similar densities for all three neutron-star setups.}
    \label{fig:profiles}
\end{figure}
The 2D cross-sections for density, crust and core particle distributions in the stars are given in Fig.~\ref{fig:star_cross}.
The original density profile, the modified profile, and resulting particle densities for all three NSTs are shown in Fig.~\ref{fig:profiles}.
In all three setups, the crust is modeled as a thin, single-particle layer on top of the core with a density of $\sim 1.2 \times 10^{14} \: \mathrm{g/cm^3} \: \sim 0.5 \: \rho_{\rm sat}$ and radius $R_c \sim 12.3 \: \mathrm{km}$. 
Due to the relatively high but homogeneous density in the crust, we will use a constant shear modulus of $\mu = 10^{30} \: \mathrm{erg/cc}$.
In this setup, the expected toroidal oscillation frequencies are $f_{P,l} \sim 28.9\,{\rm Hz}$ and $f_{S,l} \sim 16.7\,{\rm Hz}$. \\

FleCSPH posses the capability to use external potentials to move particles to a desired distribution. 
These potentials are usually applied in particle relaxation but can also replace N-body and FMM gravitational force calculations by assigning particle accelerations which are expected from gravitational forces~\citep{Tsao2021}. 
The use of external potentials speeds up NTS simulations and facilitates the relaxation of particles into the desired density profile with a quasi-regular distribution. 
We will describe the relaxation process in more detail in the next section and apply external gravitational potentials throughout the rest of the paper for crustal toroidal oscillations. 

To initialize the latter, we assign velocities according to an $m=0,\:l=2$ mode of the form: 
\begin{align}
    v_x &= - \frac{y}{|\mathbf{r}|} \: M \sin(\theta) \cos(\theta)\,,\\
    v_y &= - \frac{x}{|\mathbf{r}|} \: M \sin(\theta) \cos(\theta)\,,\\
    v_z &= 0, \:\: 
    \theta = \tan^{-1} \left(\frac{x^2 + y^2}{z}\right)\,,
\end{align}
where $M$ is the amplitude of the perturbation.
\subsection{Particle Relaxation}
The relaxation of SPH particles into a desired density profile consists of multiple steps \citep{Kaltenborn2022}. 
The general aim is to reach a particle distribution that is quasi-regular, however, without particles being strictly located on a lattice. 
As has been seen in the previous discussion of standard solid material tests, such placement can lead to lattice imprints and numerical artefacts in the dynamical evolution.
There are different approaches to relax particle configuration and we refer to \cite{Kaltenborn2022} for an overview. 
However, the usage of external potentials to guide particles to a specific distribution has been proven to be an efficient and stable approach in our simulations. 

We begin by placing particles according to an icosahedral distribution \citep{Tegmark1996} with random position perturbations whose amplitude is a small fraction of the smoothing length, here about 15\%.
The neutron-star particles are then relaxed in an external potential according to eq.(\ref{eq::phi_ext}) and the desired density profile \citep{Tsao2021}. 
Here, we use a polytropic EoS with $\Gamma \sim 1$ which ensures a uniform sound speed throughout the star and therefore fast relaxation as the time steps size is usually tied to the largest values of $c_s$.
In this step, the goal is to move particles from their perturbed positions to a distribution that is close to equilibrium. 
The particle density is determined via the summation of particle masses.
By using $\Gamma \sim 1$, we achieve the correct density profile, however, with a pressure that is offset.  
As a next step, we therefore map the the correct pressure using $\Gamma = 2$ onto the particles and perform a second relaxation run. 
This damps out breathing modes and other large-scale dynamics of the star. 
To track the progress of the relaxation, we monitor the total kinetic energy. 
A plateau in the kinetic energy indicates that a relaxed state is reached where the energy is given by velocity noise. 
At this point, we assign particles in the crust a tag which marks them as a solid.

In a third relaxation step we solve the hydrodynamic and constitutive equations for the star and its solid crust.  
Here, the goal is to prepare the neutron star to the mapping of toroidal oscillations by further reducing velocity noise and any global motion that was introduced by the solid crust and its interaction with the core, e.g. decoupling techniques which will be discussed in the next sections. 
Throughout this step, we again monitor the total kinetic energy of the star but also the maximum particle velocities in the crust. 
For the latter, we ensure that the velocity noise is at least one order of magnitude lower than the imposed amplitude $M$ of toroidal oscillations. 
Once the kinetic energy of the star does not show large-scale oscillations and enters a plateau, we stop the relaxation process.

The relaxation steps are often more computationally intensive than the final simulation of crustal motion. 
However, it is crucial to relax particles to a configuration where the velocity and density noise is minimized in order to be able to extract the motion of the solid crust.
\subsection{Crust-Core Coupling}
After the relaxation steps described in the previous section, we can simulate the star with the constitutive model for all particles in the crust. 
The density is now evolved according the to continuity equation. 
\begin{figure}
    \centering
    \includegraphics[width = 0.30\textwidth]{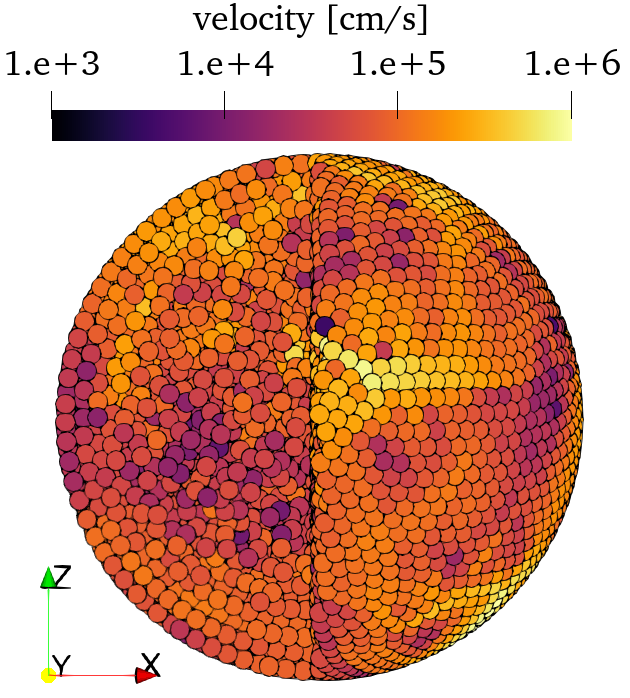}
    \caption{Velocity distribution after relaxation with coupled crust and core where the crust is described as a solid. The maximal particle velocities typically range up to several times $10^6 \: \mathrm{cm/s}$.}
    \label{fig:relax_douple_initial}
\end{figure}
Figure~\ref{fig:relax_douple_initial} shows the velocity distribution in the crust and core of a star with $N = 2.8 \times 10^4$ at $t \sim 0.15 \: {\rm s}$ (after ca. $2.2 \times 10^5$ iterations).
We find that the maximum particle velocities saturate at around $\sim 10^6 \: {\rm cm/s}$.
We therefore we impose a toroidal motion with $M\sim 2 \times 10^7 \: {\rm cm/s}$ to ensure that the resulting particle motion is not dominated by noise.  
\begin{figure}
    \centering
    \includegraphics[width = 
    0.3\textwidth]{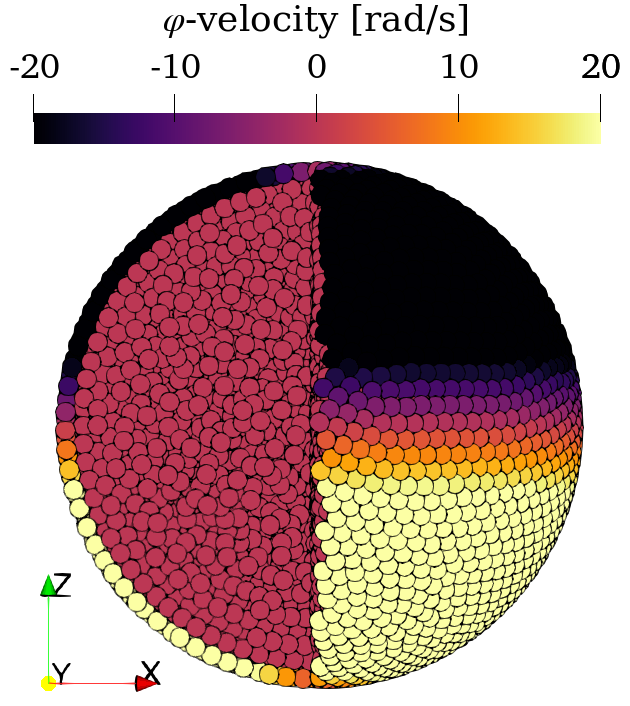}\\
    \vspace{0.3cm}
    \includegraphics[width = 0.3\textwidth]{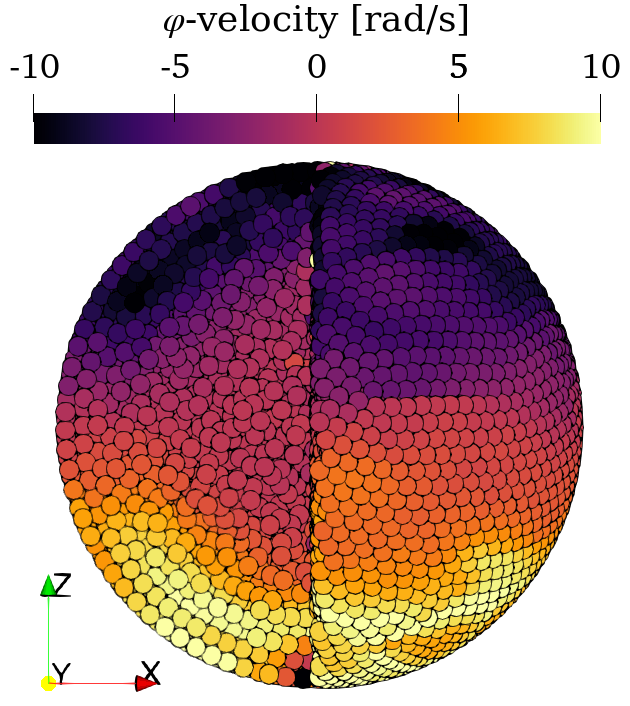}    
    \caption{Top: Initial shear velocity distribution in the $\varphi$-direction in rad/sec, corresponding to $M \sim 4 \times 10^7 \: \mathrm{cm/s}$. Bottom: The velocity distribution after about 1250 iterations, or ca. 0.8 msec. It can be seen that the entire star, i.e. crust and core, is now in motion due to the toroidal velocity having propagated into the fluid core.}
    \label{fig:init_vel}
\end{figure}
The initial $\varphi$-velocities 
\begin{align}
    v_\varphi = \frac{v_x y - v_y x}{x^2 + y^2}
\end{align}
are shown in Fig.~\ref{fig:init_vel}. 
It can be seen that the toroidal velocities are initially only present in the thin crust. 
However, as we evolve the crust and allow its particles to interact with the fluid core neighbors, we find that after less than 1000 iterations (or a fraction of a millisecond), the toroidal velocities propagate from the crust into the NST core. 
This numerical coupling between the crust and the core disturbs the crustal oscillation modes, drains the crust's kinetic energy over time, and quickly damps out any directed large-scale motion.
It becomes very obvious that in order for us to be able to model the crustal toroidal oscillations, we have to find a strategy to remove the numerical coupling between the crust and the core. 
\subsection{Decoupling of Crust and Core}
SPH offers different strategies to address the interaction of different materials at a fluid-solid interface, similar to the CCI (for an overview, see e.g. \cite{Liu2019} and \cite{Yan2016} as well as references therein). 
Our main challenge is to only suppress numerical shear forces while preserving gravitational and nuclear interactions as both ensure the stability of the NST.  
Forces which are due to the numerical viscosity are also important since they keep the simulation stable.
In general, the accelerations of an SPH particle $i$ is a sum of contributions from the nuclear interaction via the EoS pressure, numerical viscosity, material strength, external forces, and gravity: 
\begin{align}
    {\bf a}_i =& - \sum_j \left( {\bf a}_{ij,\mathrm{EoS}} + {\bf a}_{ij,\mathrm{Visc}}  + {\bf a}_{ij,\mathrm{Strength}} \right) \nonumber\\ 
    &+ {\bf a}_{i,\mathrm{Ext}} + {\bf a}_{i,\mathrm{Grav}}\,.
\end{align}
For spherical stars we can assume that (1) the CCI is also spherical, and that (2) gravitational and EoS pressure forces should be strictly normal to the CCI. 
We can therefore decompose accelerations into
\begin{align}
    \mathbf{a}^\parallel_{ij,\mathrm{EoS}} &= \left(\mathbf{a}_{ij,\mathrm{EoS}} \cdot \mathbf{n}_\mathrm{CCI}\right) \: \mathbf{n}_\mathrm{CCI},\nonumber\\
    \mathbf{a}^\perp_{ij,\mathrm{EoS}} &= 
    \mathbf{a}_{ij,\mathrm{EoS}} - \mathbf{a}^\parallel_{ij,\mathrm{EoS}}\,,
    \label{eq::radial_perp_acc}
\end{align}
i.e., components that are parallel ($\parallel$) and perpendicular  ($\perp$) to the normal vector $\mathbf{n}_\mathrm{CCI}$ of the interface. 
When eliminating the perpendicular contribution, we preserve the physical interaction between the crust and the core while removing numerical effects.    
In the case of spherical stars, the normalized radial vector of each particle is automatically parallel to $\mathbf{n}_\mathrm{CCI}$.
For non-spherical or time-dependent deformation, we will need a strategy to determine the normal on the fly.
In the following, we will discuss these different cases. 
\subsection{Radial Vector Decoupling}
Here, we use a particle's radial vector or position ${\bf r}_i$ with respect to the star's center of mass to determine $\mathbf{n}_\mathrm{CCI}$ simply via 
\begin{align}
    \mathbf{n}_{i,\mathrm{CCI}} = \mathbf{r}_i \, |\mathbf{r}_i|^{-1}.
\end{align}
Since the simulated stars and therefore the gravitational external potential are spherically symmetric, the acceleration due to gravitational forces is automatically parallel to $\mathbf{n}_\mathrm{CCI}$.    
The accelerations from the EoS and numerical viscosity, on the other hand, have to be split according to Eq.~(\ref{eq::radial_perp_acc}).
followed by the elimination of the perpendicular component.
\\
\begin{figure}
    \centering
    \includegraphics[width = 0.45\textwidth]{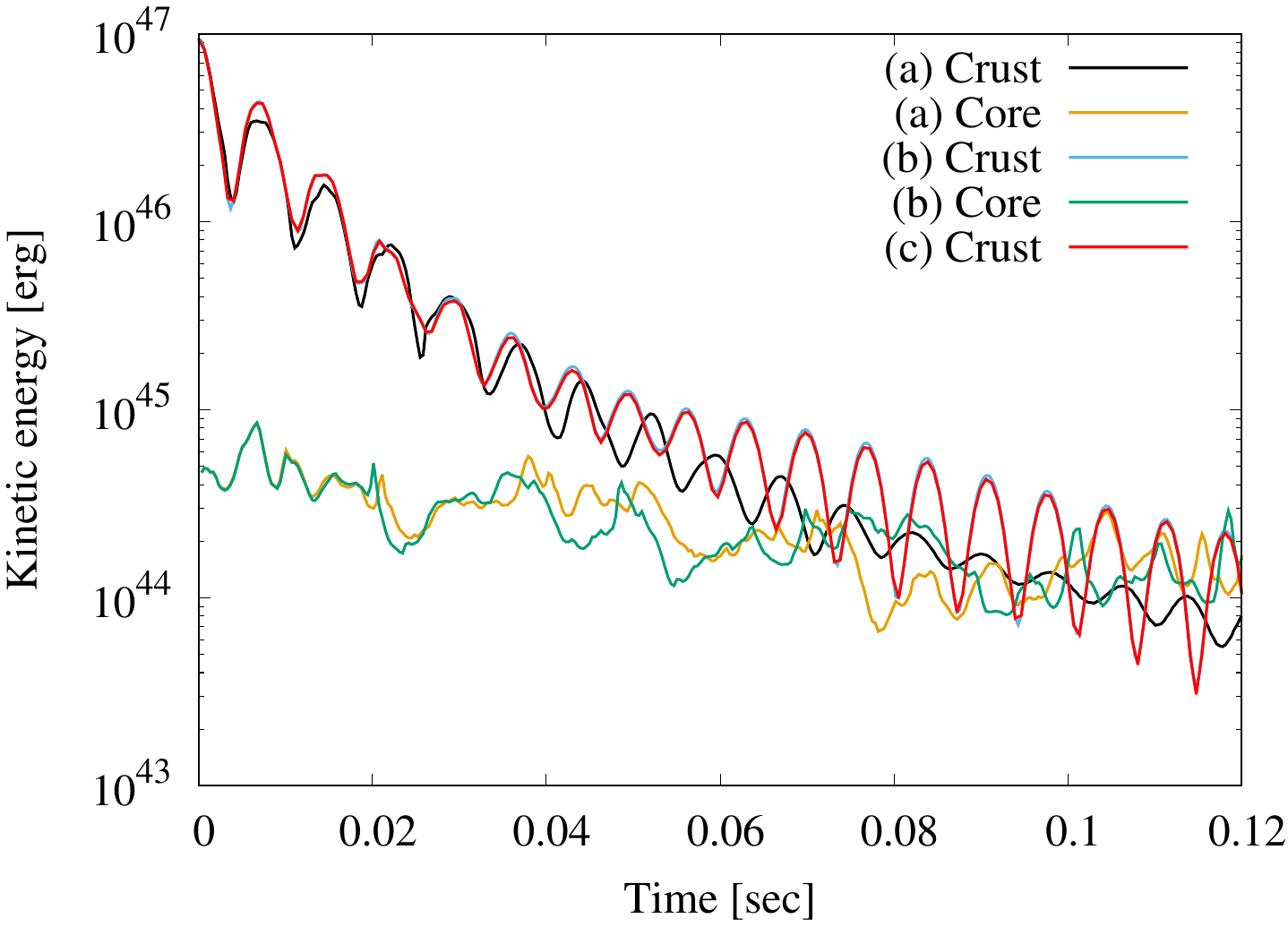}
    \caption{Total particle kinetic energy in the neutron-star crust and core for (a) a decoupled fluid crust and a fluid core, (b) a solid crust and fluid core, and (c) a solid crust on top of a frozen core. The kinetic energy of the latter is not shown as it is zero.}
    \label{fig:crust_core}
\end{figure}
\begin{figure}
    \centering
    \includegraphics[width = 0.3\textwidth]{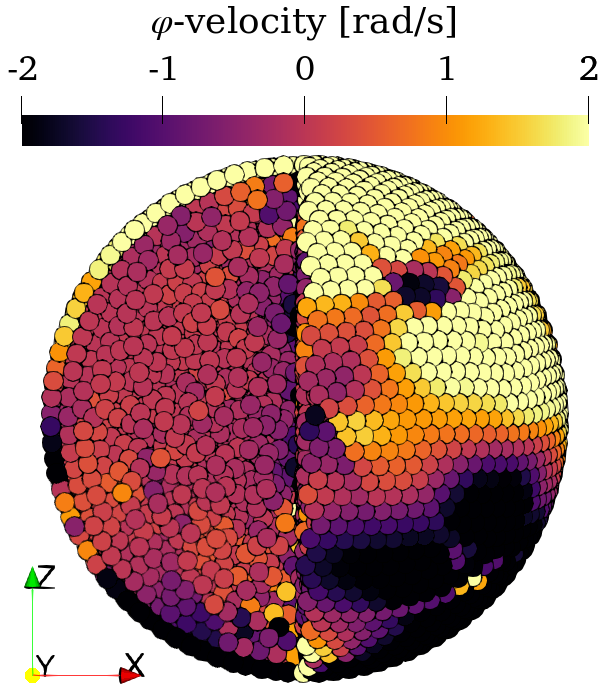}
    \caption{$\varphi$-velocity in the fluid crust and core at 0.035 seconds showing the core velocities being much smaller smaller than the velocities in the crust.}
    \label{fig:fluid_crust}
\end{figure}
Figure~\ref{fig:crust_core} shows the total kinetic energy for the NST crust and core.
We test three different configurations where the star is composed of (a) a fluid crust and core, (b) a solid crust and a fluid core, and (c) a solid crust and a frozen core. 
For the latter, we reset the particle velocities in the core to zero at each iteration. 
The initial toroidal crust velocity has $M \sim 4 \times 10^7 \: \mathrm{cm/s}$. 
For both the solid and fluid crust, we see that the core's kinetic energy stays below the energy of the crust until $t \sim 0.07\:\mathrm{sec}$. 
After that, even though the energies of the crust and core are similar, the crust continues to show an oscillation pattern while the behavior of the core's kinetic energy is irregular.  
This indicates that the motion of both is decoupled and the toroidal motion is, therefore, confined to the crust.
In Fig.~\ref{fig:fluid_crust}, we plot the $\varphi$-velocity distribution of the SPH particles at $t \sim 0.07\,{\rm s}$. 
Here, we can clearly see a distinction between the high particle velocities in the crust versus the generally lower values in core.

\begin{figure}
    \centering
    \includegraphics[width = 0.45\textwidth]{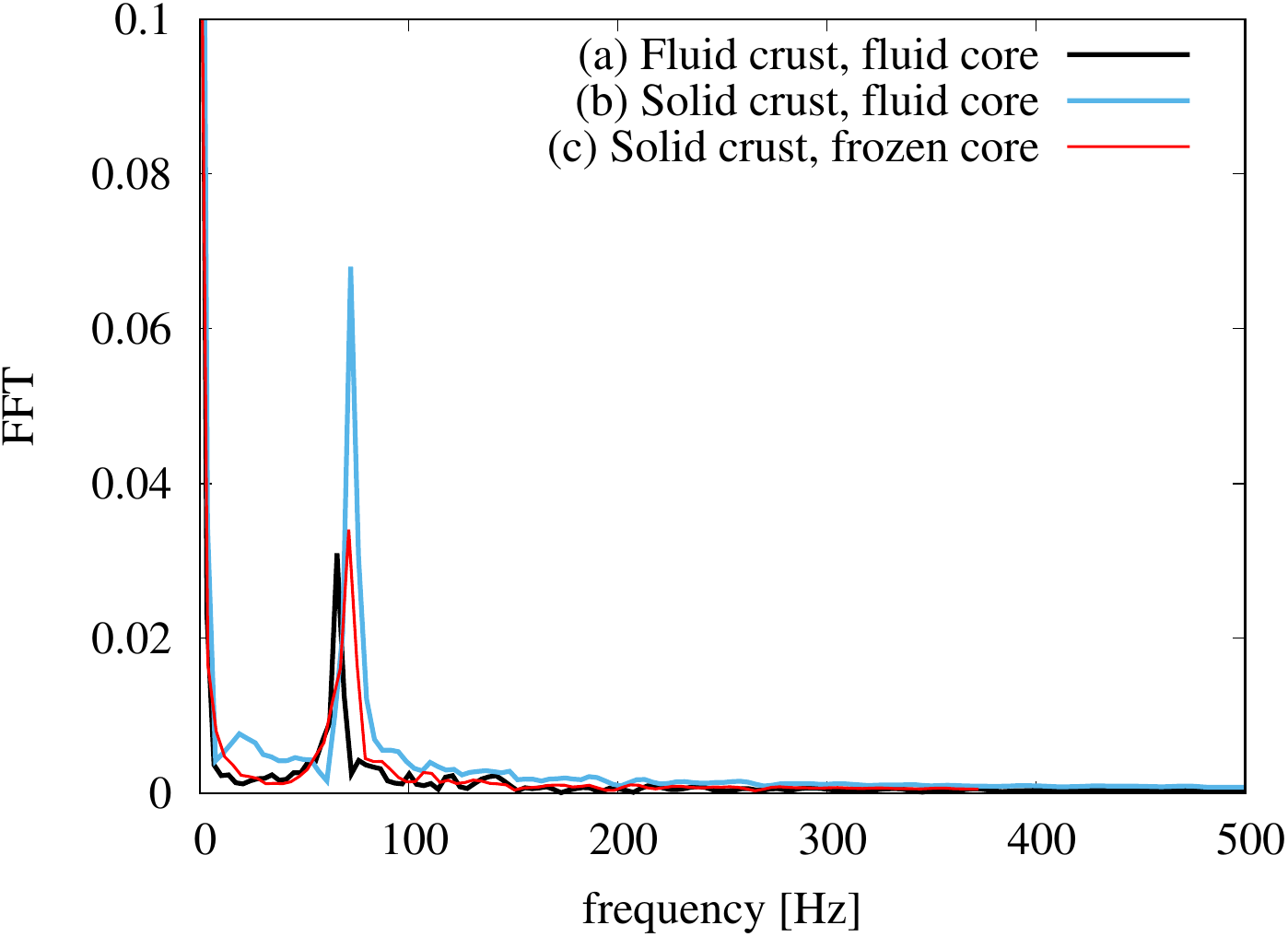}    
    \caption{FFT analysis of crust energy oscillations seen in Fig.\ref{fig:crust_core}.}
    \label{fig:crust_core_FFT}
\end{figure}
To extract the dominant frequencies of the oscillations in the crust, we perform a Fourier analysis of the kinetic energy with a correction its exponential decrease with time.
The results are shown in Fig.~\ref{fig:crust_core_FFT}. 
The low-mode frequencies are at about 64\:Hz for case (a), i.e., the fluid crust on top of the fluid core, and about 74\:Hz for both cases (b) and (c) of a solid crust on top of a fluid and frozen core, respectively.
The oscillations are not modified by the state of the core and, more importantly, are present for a fluid and solid crust.
This implies that the restoring force for the periodic motion is not due to material strength. 
Instead, it is most likely caused by density perturbations during the initial toroidal particle displacement and then sustained by the bulk modulus. 
\begin{figure}
    \centering
    \includegraphics[width = 0.35\textwidth]{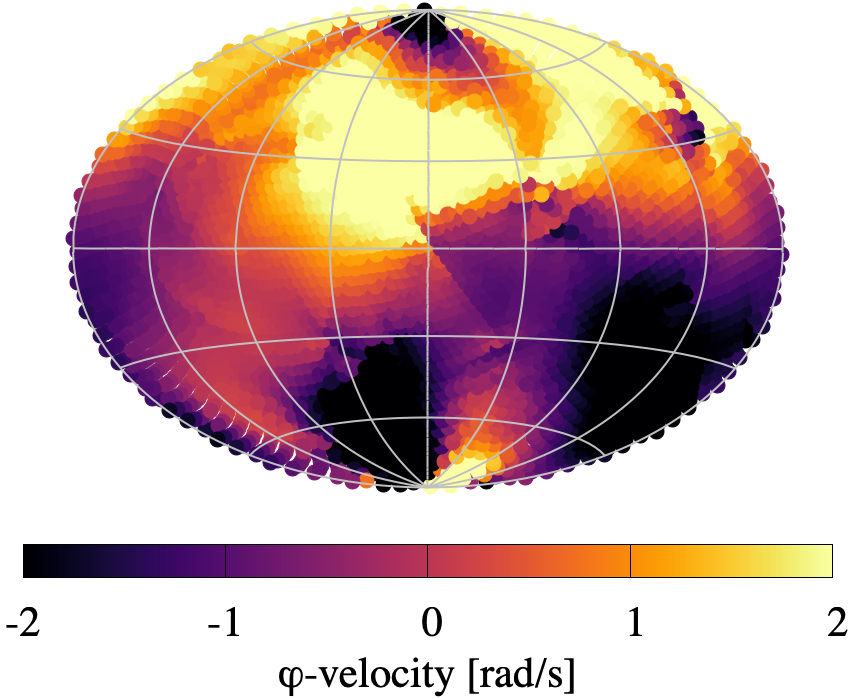}\\
    \vspace{.3cm}
    \includegraphics[width = 0.35\textwidth]{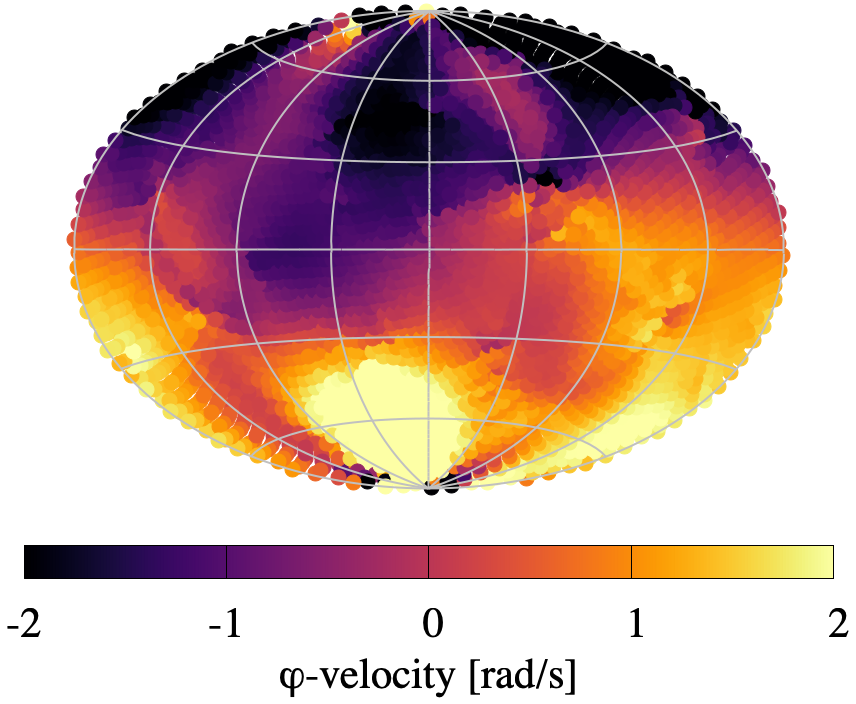}\\
    \vspace{.3cm}
    \includegraphics[width = 0.35\textwidth]{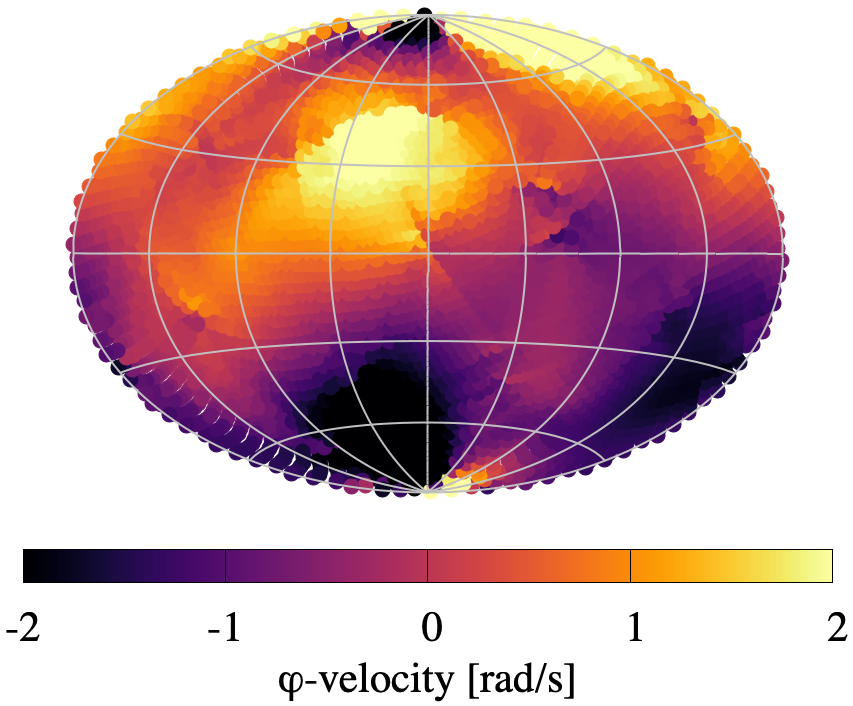}    
    \caption{Oscillations of a fluid crust with a fluid core, showing the $\varphi$-velocity in the crust using the Hammer projection at 0.067, 0.074, and 0.082 seconds.}
    \label{fig:fluid_oscillations}
\end{figure}
Figure \ref{fig:fluid_oscillations} shows the distribution of the $\varphi$-velocity in the fluid crust during one period of the low-frequency oscillation. 
While there is large-scale material motion in the upper vs. the lower hemispheres, the velocity pattern is noisy and certainly not like the one expected for the $m=0,\,l=2$ mode. 
With that we conclude that while the decoupling of the crust and core shear motion is successful, the crustal toroidal oscillations cannot be sustained due to the dominating bulk modulus which converts density perturbations into particle velocity noise.  
Next, we will therefore discuss strategies how to reduce the impact of the bulk modulus and ensure that the shear modulus dictates the toroidal oscillations.  
\subsection{Reduction of EoS Acceleration}
Here, we will attempt to decrease the effect of the bulk modulus on the particle motion in the crust.  
For that, we reduce the corresponding particle EoS accelerations ${\bf a}_{ij,{\rm EoS}}$ by a factor $\kappa$ if both particles, $i$ and $j$, belong to the crust. 
Since the crust's bulk modulus is about three orders of magnitude larger than its shear modulus, we expect that $\kappa$ must be set to similar values in order for the crust's solid material properties to be visible in the simulations. 
We test this by running simulations of crustal oscillations in the $m=0,\,l=2$ mode for $1 \leq \kappa \leq 10^5$. 
NST relaxation simulations with a decoupled crust and core show that the maximal velocity of crust particles can be decreased to several times $10^4\,{\rm cm/s}$. 
With that, we initialize the $m=0,\,l=2$ mode with $M = 3.8 \times 10^5 \: \mathrm{cm/s}$ to reduce density perturbations from particle motion while also ensuring that the initial toroidal velocity is not dominated by noise. 
In addition, we reduce the numerical viscosity for particle accelerations that are perpendicular to ${\bf n}_{\rm CCI}$ to 10\% of $\mathbf{a}^\parallel_{ij,\mathrm{Visc}}$. 
This ensures that the toroidal oscillations are not damped out too quickly. 

\begin{figure}
    \centering
    \includegraphics[width = 0.45\textwidth]{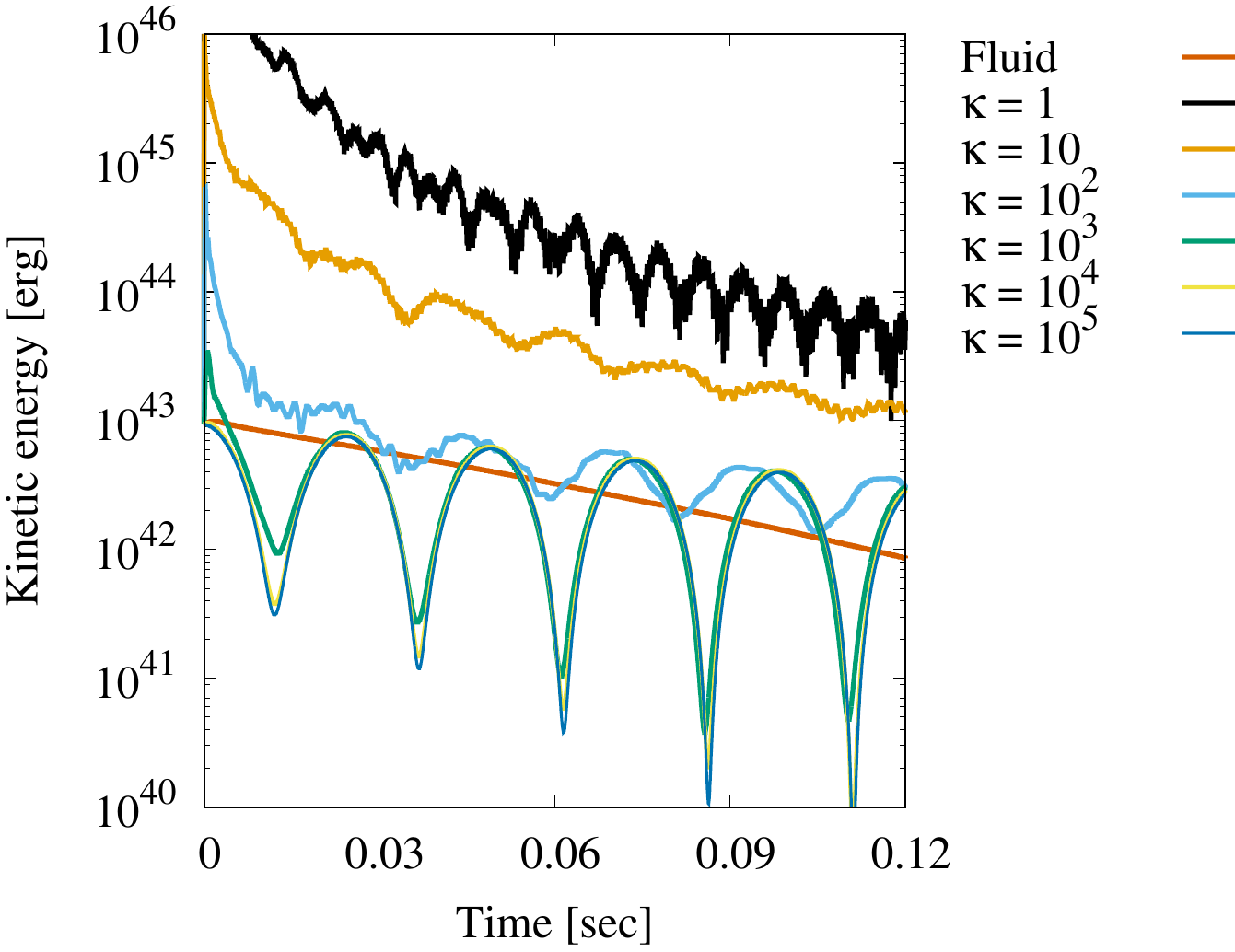}
    \caption{Comparison of oscillation patterns for the NST crust after imposing a $m=0,\,l=2$ toroidal mode with $M = 3.8 \times 10^5 \: \mathrm{cm/s}$. The oscillations are tracked via the crust's kinetic energy. Different reduction factors $\kappa$ are shown as well as a case where the crust is modeled as fluid.}
    \label{fig:osci_pred}
\end{figure}

\begin{figure}
    \centering
    \includegraphics[width = 
    0.3\textwidth]{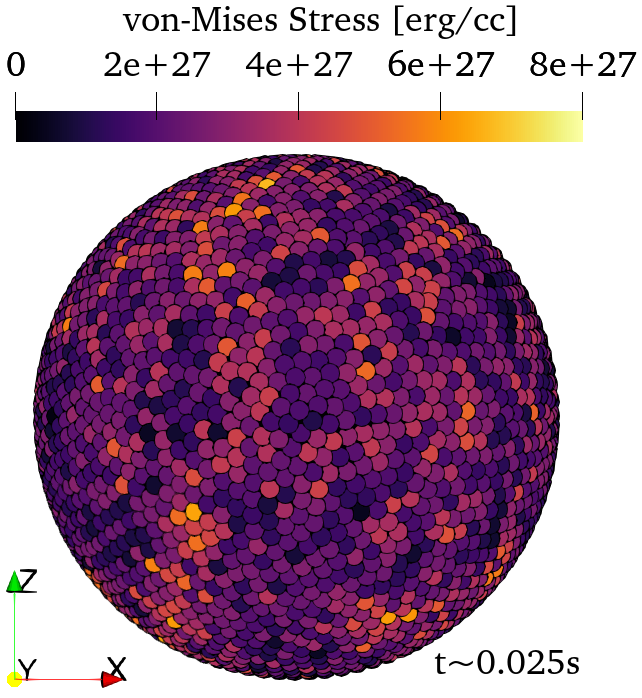}\\
    \vspace{0.3cm}
    \includegraphics[width 
    = 0.3\textwidth]{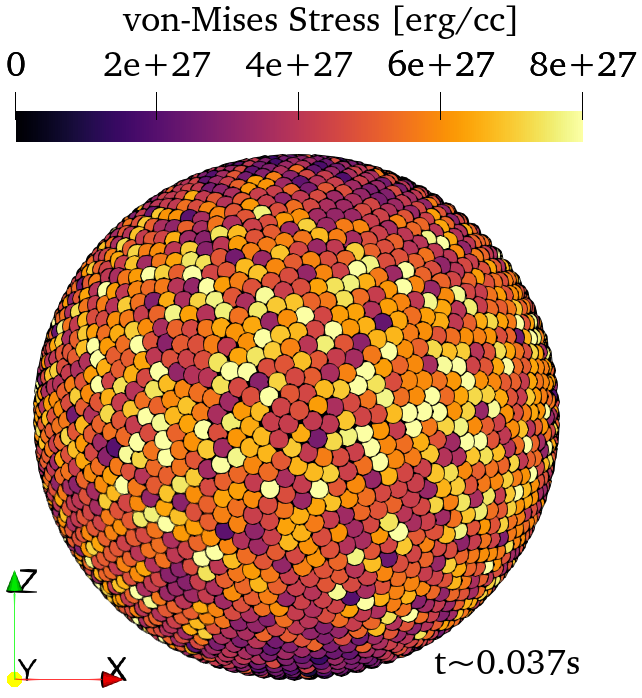}
    \caption{von-Mises stress for $N = 2.8 \times 10^4$ particles, $M = 3.8 \times 10^5 \:\mathrm{cm/s}$, and $\kappa = 10^4$.}
    \label{fig:frozencore_stress_0p0001}
\end{figure}

\begin{figure}
    \centering
    \includegraphics[width = 0.45\textwidth]{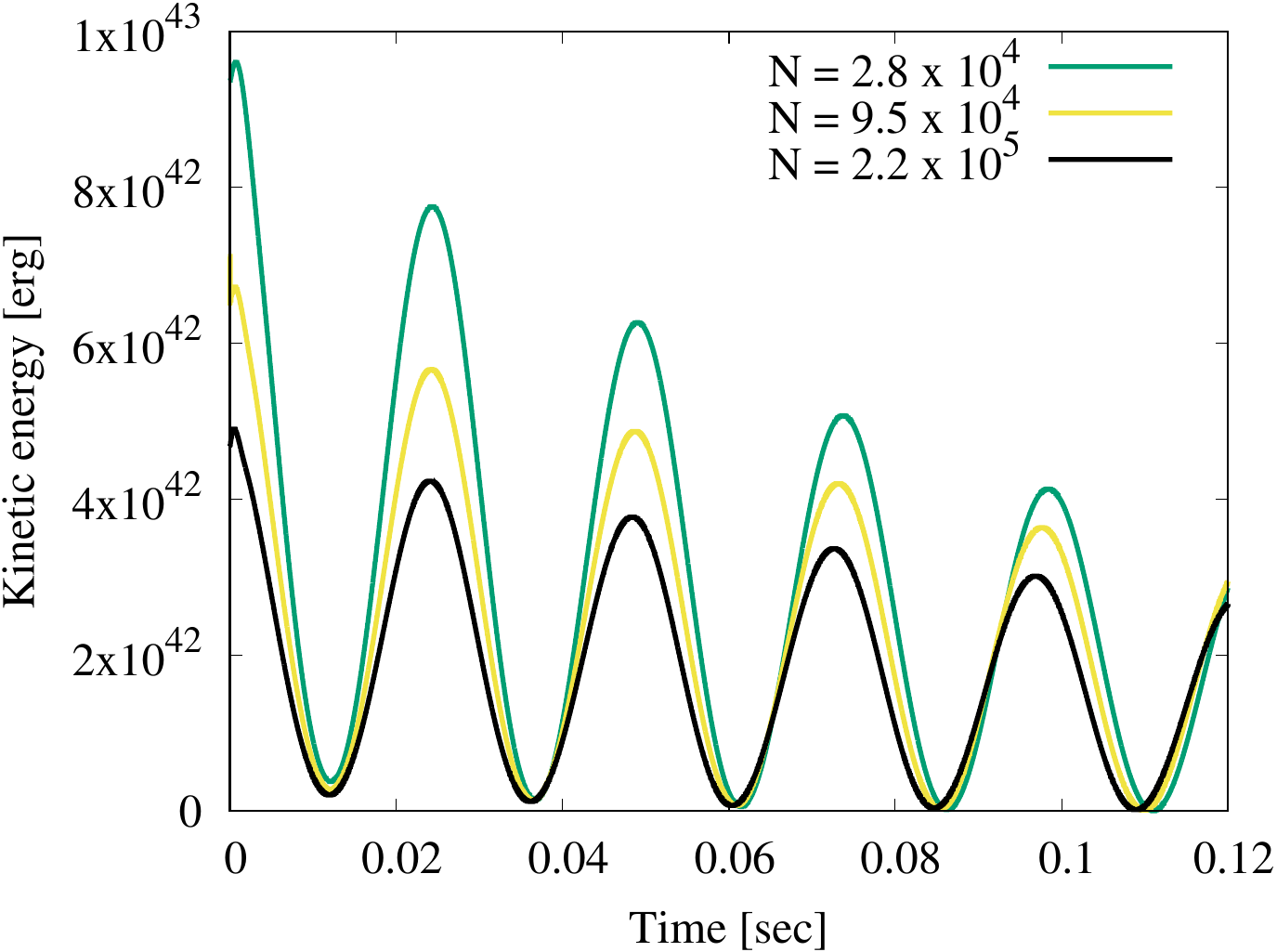}
    \caption{Toroidal oscillations with a frozen core for the three star setups in Fig.~\ref{fig:star_cross}.}
    \label{fig:frozencore_oscillations_N}
\end{figure}

Figure~\ref{fig:osci_pred} shows the resulting kinetic energies of the crust. 
All configurations start out from a star that has been relaxed using $\kappa = 10^5$.
As previously seen in Fig.~\ref{fig:crust_core}, if ${\bf a}_{ij,{\rm EoS}}$ is not reduced, i.e., for $\kappa = 1$, the crust experiences oscillations which are supported by density variations and the bulk modulus. 
Due to the change in $\kappa$ from the relaxation to the toroidal oscillation setup, we also find high-frequency noise in the crust which is visible for $\kappa = 1, \, 10$ and $10^2$.
However, as we increase $\kappa$ to $10^3$, we find a clear emerging oscillation pattern with frequencies of about $20.2\,{\rm Hz}$.
The crustal motion does not change when we further increase $\kappa$ which indicates that the oscillations are not dictated by the crust EoS but by the shear modulus which remains unchanged for all values of $\kappa$.
\begin{figure}
    \centering
    \includegraphics[width = 0.45\textwidth]{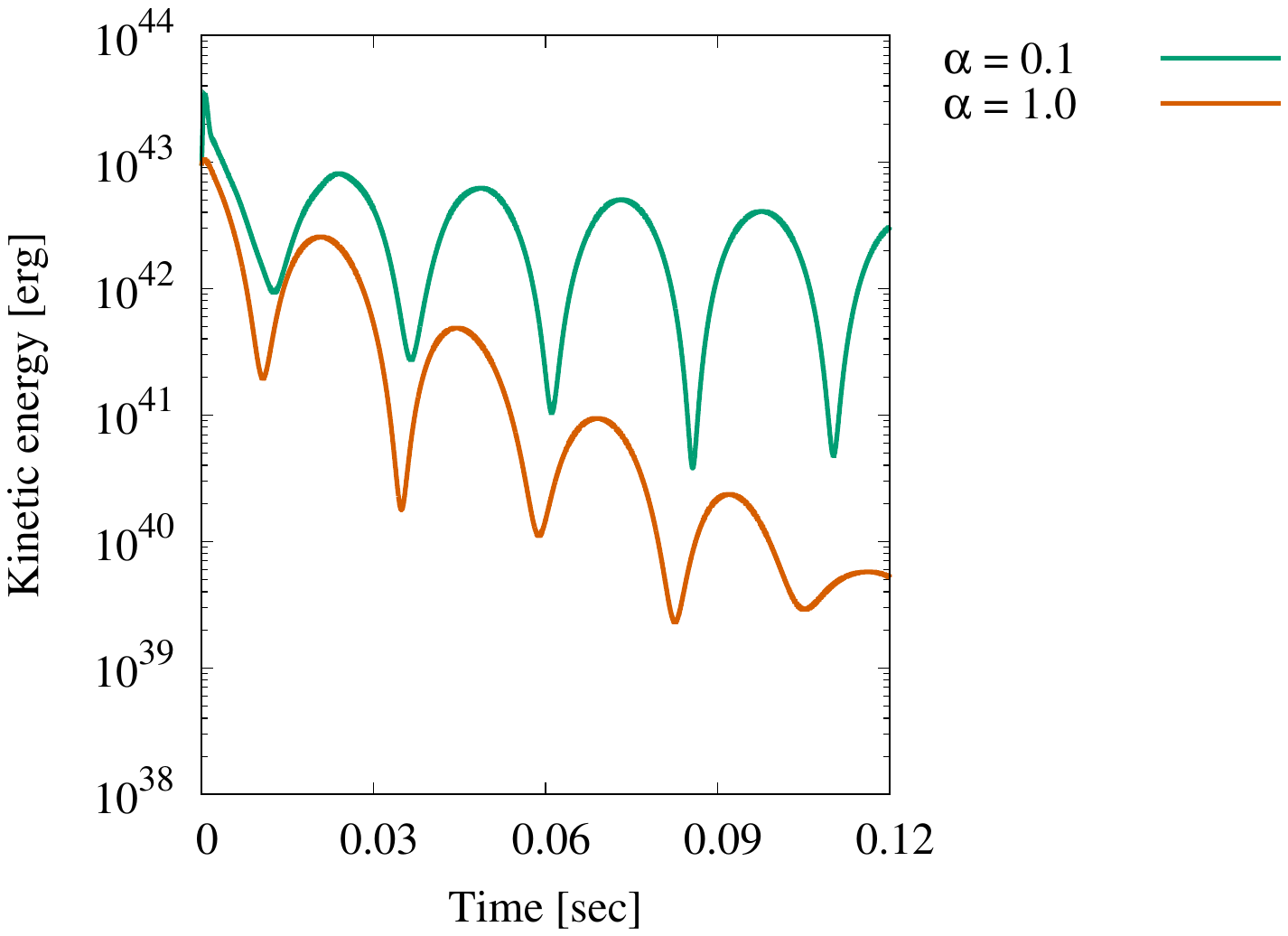}
    \caption{Kinetic energy oscillations for a NST crust initialized as in Fig.\ref{fig:osci_pred} with numerical viscosities using $\alpha = 0.1$ versus $\alpha = 1.0$.}
    \label{fig:osci_visc}
\end{figure}
For comparison, we also model the crust as a fluid with $\kappa = 10^4$. 
We find no oscillation behavior but instead a slow decrease in $\varphi$-velocity and crust kinetic energy. 
This is a behavior that is expected for a fluid and shows again that the crustal oscillations in Fig.~\ref{fig:osci_pred} are due to its shear modulus. 
As seen in Fig.~\ref{fig:osci_visc}, numerical viscosity accelerates the damping of the oscillations and leads to a prompt decrease of the kinetic energy. 
However, it does not largely affect the oscillation frequencies. 
\begin{figure}
    \centering
    \includegraphics[width = 
    0.3\textwidth]{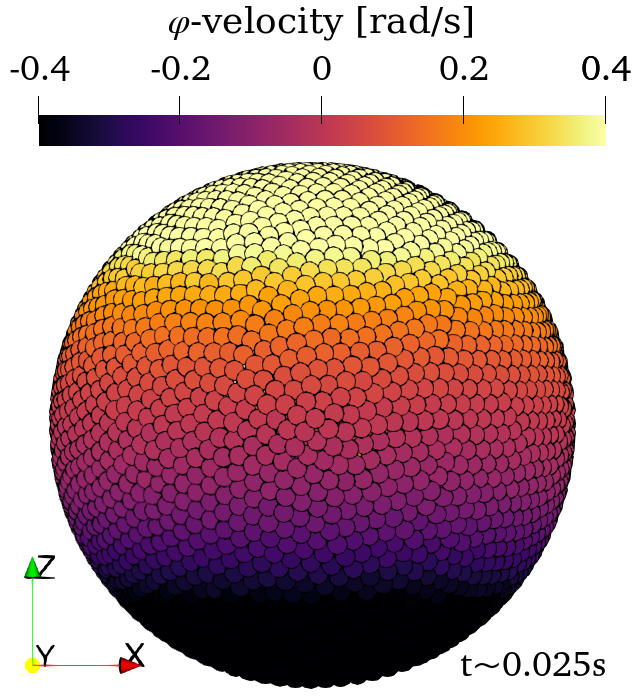}\\
    \vspace{0.3cm}
    \includegraphics[width 
    = 0.3\textwidth]{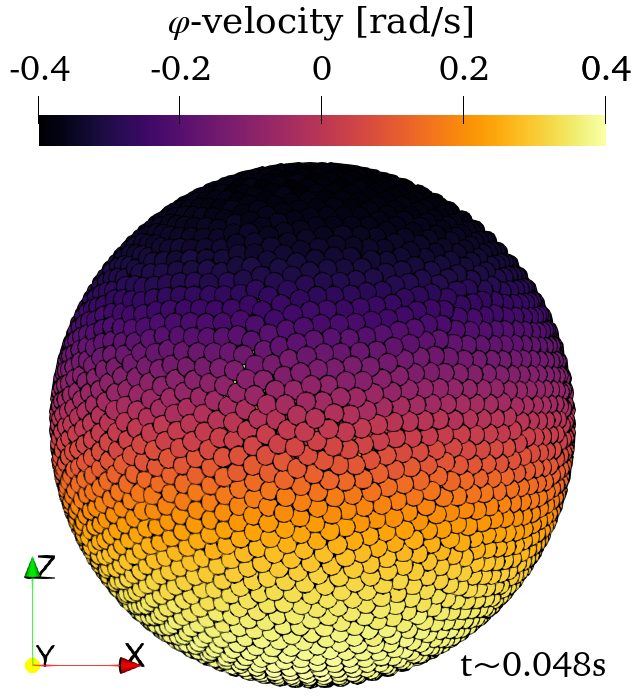}\\
    \vspace{0.3cm}
    \includegraphics[width 
    = 0.3\textwidth]{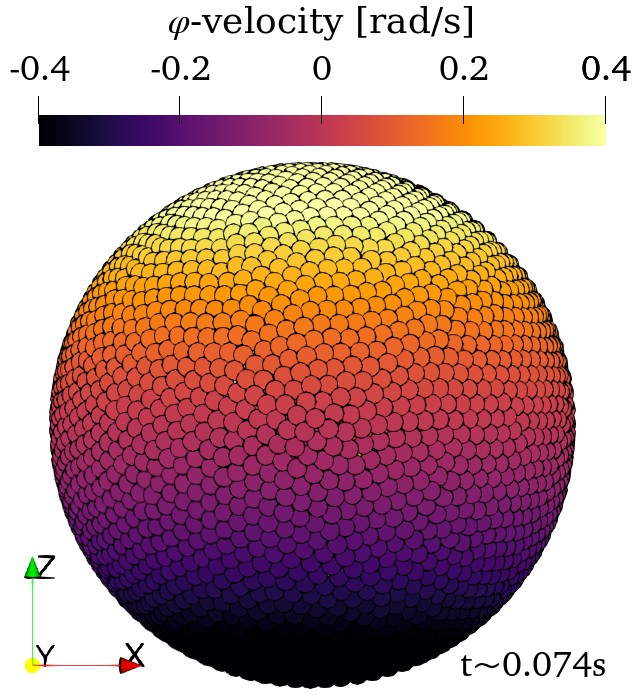}    
    \caption{Toroidal oscillation sequence of the neutron-star crust with $N = 2.8 \times 10^4$ particles, $M = 3.8 \times 10^5 \:\mathrm{cm/s}$, and a frozen core.}
    \label{fig:frozencore_oscillations}
\end{figure}
Figure~\ref{fig:frozencore_oscillations} shows the smooth $\varphi$-velocity distribution during one period of toroidal oscillations as expected for the $m=0,\,l=2$ mode, for $\kappa = 10^4$. 
Figure~\ref{fig:frozencore_stress_0p0001} gives the corresponding distribution of the von-Mises stress in the crust at maximum and minimum kinetic energy. 
The stress reaches values of about $8 \times 10^{27}\,{\rm erg/cm^3}$ for individual particles. 
We test the dependence of the oscillations on $N$ and find only small variation in the oscillation frequency, as shown in Fig.~\ref{fig:frozencore_oscillations_N}.
\\
\begin{figure}
    \centering
    \includegraphics[width = 0.45\textwidth]{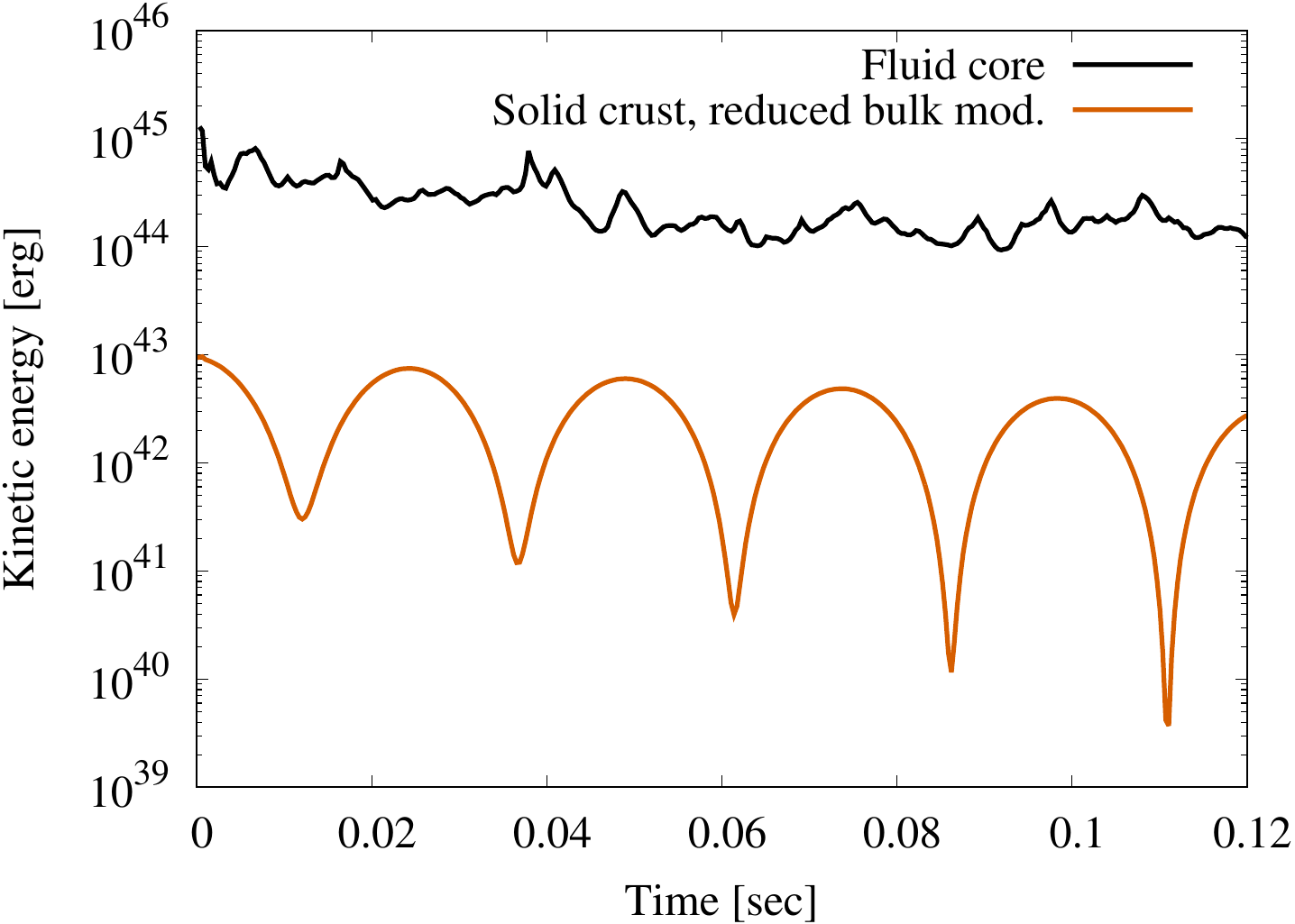}
    \caption{Total particle kinetic energy in the crust and the core for the case of a solid crust with reduced EoS acceleration ($\kappa = 10^4$) that is decoupled from the fluid core. The decoupling is done using the particle positions to determine the normal to the CCI.}
    \label{fig:crust_core_pred}
\end{figure}
\begin{figure}
    \centering
    \includegraphics[width 
    = 0.3\textwidth]{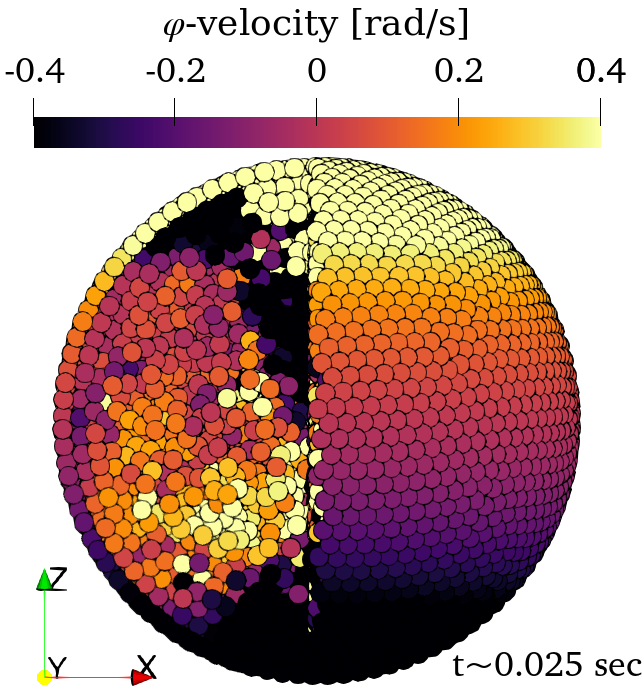}
    \caption{$\varphi$-velocity distribution in the neutron-star crust and core. The crust is decoupled from the core via the particle position. The core is a fluid while the crust is modeled with a solid with a reduced EoS acceleration by $\kappa = 10^4$.}
    \label{fig:fluidcore_oscillations_pred}
\end{figure}
As a next step, we allow the core to move. 
The initial crustal toroidal velocity distribution remains the same as for simulations with a frozen core and we use $\kappa = 10^4$.
We simulate a NST with $N = 2.8 \times 10^4$, applying radial decoupling of the crust and core. 
Figure \ref{fig:crust_core_pred} shows the resulting kinetic energies in the crust and core while Fig.~\ref{fig:fluidcore_oscillations_pred} is a snapshot of the velocity distribution. 
We can clearly see the oscillatory behavior of the crust.
It is identical in frequency to the stars with a frozen core. 
Interestingly, the kinetic energy of the fluid core is much higher than the one of the crust. 
Even considering that we plot the total kinetic energy and the number of particles in the core is about an order of magnitude larger than in the crust, it is still surprising that the decoupling allows the core to have velocity variations that are comparable to the toroidal velocities in the crust, without affecting the oscillatory motion of the latter. 

Overall, we conclude that despite the artificial reduction of ${\bf a}_{ij,{\rm EoS}}$ in the crust, the oscillations seen in Fig.~\ref{fig:frozencore_oscillations} correspond to the $m=0,\,l=2$ toroidal mode. 
The extracted frequencies are neither dependent on resolution nor numerical viscosity. 
The frequencies lie between the predicted values by \cite{Piro:2005jf} and \cite{Samuelsson:2006tt}.

It should be noted that as long as the dynamics of the crust is dominated by its solid material properties, we can expect our model to reproduce the correct behavior.  
However, by reducing the EoS acceleration and thereby the crust's bulk modulus, we made the crust material unphysically compressible. 
Therefore, in the presence of large physical density variations, this method will likely fail. 
\subsection{Weakly-compressible SPH}
In the previous section, we studied toroidal oscillation of the NST crust by artificially reducing the otherwise dominating effect of the nuclear bulk modulus and associated particle accelerations.
This resulted in a very soft crust EoS. 
On the one hand, this allowed the shear modulus to provide the dominating restoring force for crust motion. 
On the other hand, making crust material very compressible is contrary to the behavior of nuclear matter during toroidal oscillations. 
Here, we expect that the material maintains its constant density, similar to an incompressible flow. 
SPH offers different ways to model such flows in engineering problems~\citep{Lind2020, Khayyer2016, Cummins1999, Monaghan1994}.
A frequently chosen approach for weekly compressible flows is to use the Tait-Murnaghan EoS with e.g. $\gamma = 7$. 
Here, we will follow this approach and set $\rho_0$ as the density of the crust layer of our $N = 2.8 \times 10^4$-particle stars with $\rho \sim 1.2374 \times 10^{14} \: \mathrm{g/cm^3}$. 
The sound speed is chosen to be equal to the one expected from a polytropic EoS at the density $\rho_0$
\begin{align}
    c_0 = K \: \Gamma \: \rho_0^{\Gamma-1}\,,
\end{align}
with $\Gamma$ and $K$ as in eq.(\ref{eq::polytropic}).
Finally, we set $P_0$ to zero, although setting it to the corresponding $P(\rho_0)$ of the polytropic EoS would also be an option. 
With this parameter choice, the crust EoS has zero pressure when the particle density equals $\rho_0$, with negative and positive values if $\rho$ drops below or rises above it, respectively. 
The large value of $\gamma$ leads to a fast adjustment of particle densities in the crust to $\rho_0$ and suppresses density perturbations. 
\begin{figure}
    \centering
    \includegraphics[width = 0.45\textwidth]{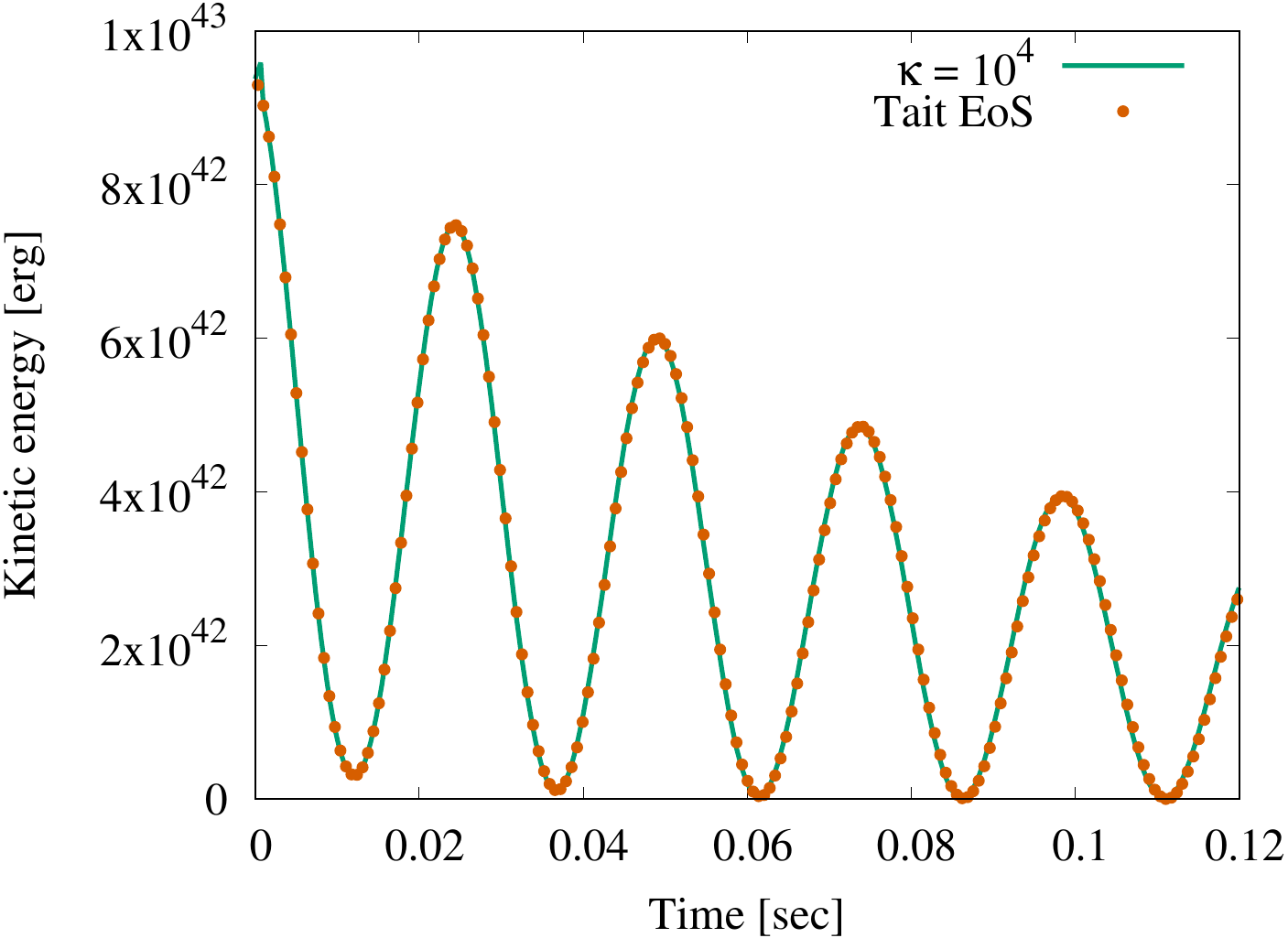}
    \caption{Kinetic energy oscillations for the Tait-Murnaghan EoS crust, decoupled via the particle position from the fluid core. The oscillations follow the previously explored approach to reduce the EoS acceleration of for crust particles by $\kappa = 10^4$.}
    \label{fig:liquid_pred_comp}
\end{figure}
\begin{figure}
    \centering
    \includegraphics[width 
    = 0.3\textwidth]{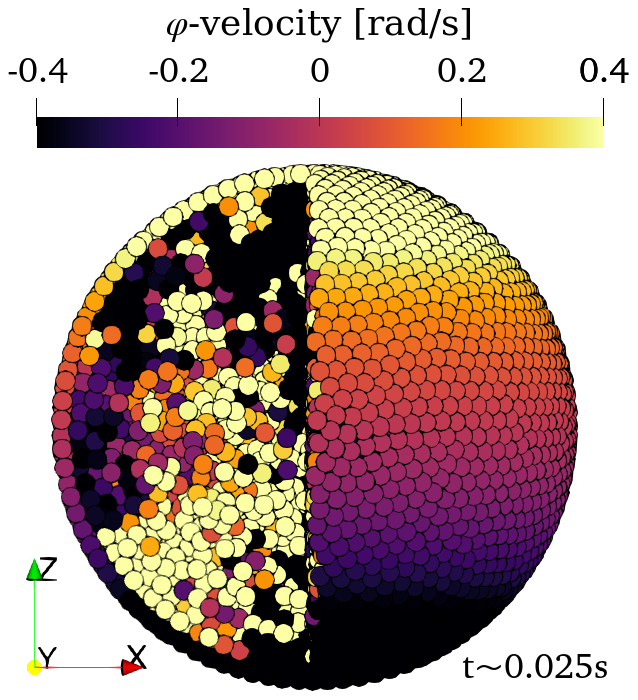}\\
    \vspace{0.3cm}
    \includegraphics[width 
    = 0.3\textwidth]{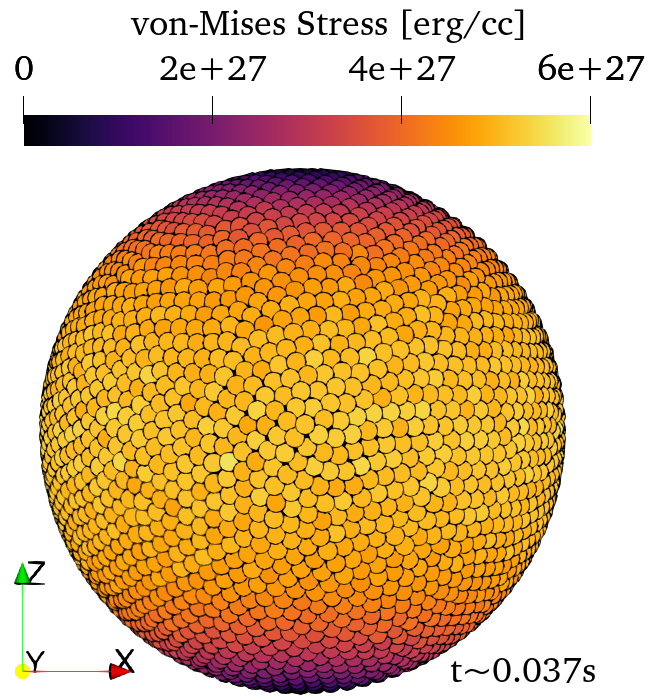}    
    \caption{$\varphi$-velocity and von-Mises stress distribution in the crust and core of a neutron star during toroidal oscillations as in Fig.\ref{fig:frozencore_oscillations} with a fluid core and a Tait EoS crust.}
    \label{fig:fluidcore_oscillations_liquid}
\end{figure}
We note that the Tait-Murnaghan EoS is only used for the interaction between crust particles. 
For the core as well as the interaction between crust and core, we still use the regular Polytropic EoS.
Furthermore, while the presented approach is adequate for dynamics where we expect the nuclear material in the NST to behave as weakly compressible, i.e. for small density variations, the method will probably give incorrect results for physical situations with large density changes.

Figures \ref{fig:fluidcore_oscillations_liquid} and \ref{fig:liquid_pred_comp} show the velocity and von-Mises stress of the Tait-Murnaghan EoS crust on top of a fluid core. 
The crust and the core are decoupled by the particle position. 
As in the previous subsection, we see that despite the generally large velocities in the core, the $m=0,\,l=2$ pattern of $v_\varphi$ is sustained in the crust. 
The resulting frequency of the oscillations and crust motion is nearly identical to the one obtained with a reduced bulk modulus. 
The von-Mises stress has a smooth distribution which can most likely be attributed to the uniform density distribution and lack of corresponding perturbations in the crust. 
Overall, this seems to be a very promising approach.
Further future studies will be required to explore the impact of, e.g., $\gamma$ and $P_0$ in the EoS on e.g. the crust's oscillatory behavior.  
\subsection{Density Gradient Decoupling}
For a neutron star that is evolving and deforming, as e.g., in a binary, we cannot extract the CCI via the radial vector approach since the object is not spherically symmetric. 
The location of the interface and its surface normal has to be determined on the fly in each iteration. 
One possibility is to determine the interface and its normal for each individual particle and its neighbors via a so-called color field~\citep{Mueller2003}.
This approach is often used to calculate fluid surface tension. 
Fluid particles are assigned a tag or color value. 
A smoothed color field can be defined as
\begin{align}
    c_i = \sum_j \frac{m_j}{\rho_j} \: W_{ij}\,.
\end{align}
A vector normal to the interface that is pointing into the fluid can then be calculated from $\mathbf{n}_i = \nabla c_i$.
In principle, the color-field approach can also be adjusted to determine the interface of two materials where particles are used on both sides of the interface. 
However, for a neutron star with small deformations, we expect that ${\bf n}_{i,{\rm CCI}}$ is generally aligned with the density gradient. 
So, instead of a color field that is e.g. associated with crust particles, we will use the density to determine the interface normal. 
First, we calculate 
\begin{align}
    \nabla \rho_i = \sum_j m_j \: \nabla W_{ij}\,.
    \label{eq::gardrho}
\end{align}
For each particle $i$ that lies close to the interface, we determine the CCI surface normal as
\begin{align}
    \mathbf{n}_{i,\mathrm{CCI}} = \left| \nabla \rho_{i,S} \right|^{-1} \: \nabla \rho_{i,S}   \,,
    \label{eq::nFSI} 
\end{align}
where $\nabla \rho_{i,S}$ is the density gradient with additional smoothing 
\begin{align}
    \nabla \rho_{i,S} = \sum_{j} \nabla \rho_j \: \frac{W_{ij}}{W_{ii}}\,.
\end{align}
We found that the latter is necessary for numerical stability. 
Once calculated, we use the $\mathbf{n}_{i,\mathrm{CCI}}$ to decouple the shear motion of the crust and core particles. 

\begin{figure}
    \centering
    \includegraphics[width 
    = 0.3\textwidth]{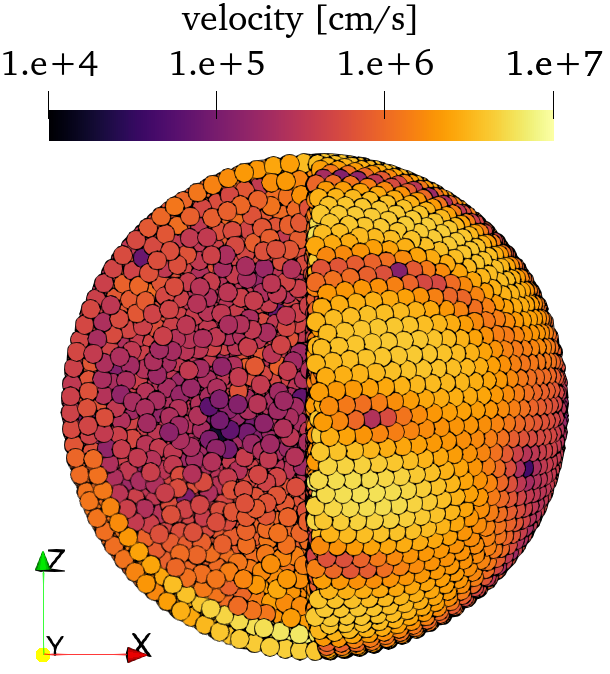}\\
    \vspace{0.3cm}
    \includegraphics[width 
    = 0.3\textwidth]{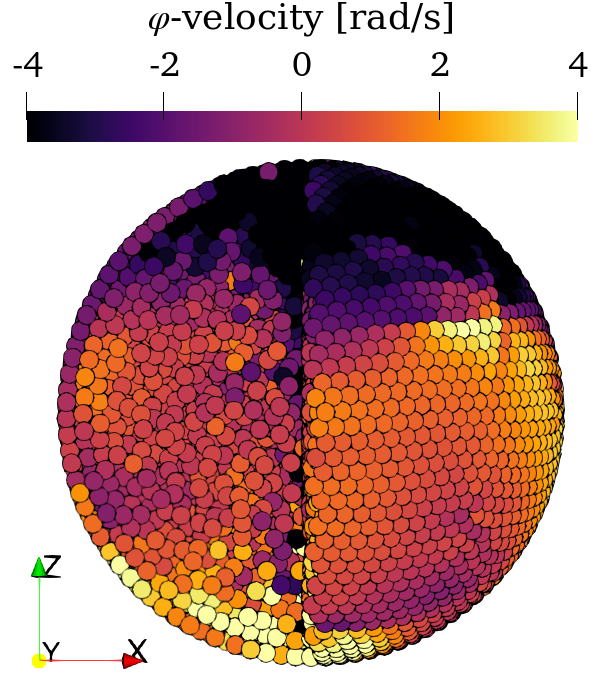}    
    \caption{Velocity distribution in the crust and core of a neutron star after relaxation at time $t \sim 0.05\:\mathrm{sec}$ and after toroidal velocity initialization with a fluid core and a Tait-Murnaghan EoS crust at $t \sim 0.005\:\mathrm{sec}$. The decoupling was done via the density gradient.}
    \label{fig:fluidcore_liquid_color}
\end{figure}

Figure \ref{fig:fluidcore_liquid_color} shows the particle velocities in the core and crust of a NST after relaxation for about 0.05 seconds and after evolving the star with toroidal oscillations using $M \sim 3.8 \times 10^7 \: \mathrm{cm/s}$.
The crust was again described by the Tait-Murnaghan EoS.   
We find that the crust is stable on top of the fluid core, however, as for the decoupled case, the particle velocities are very high.
Although we initialize the toroidal oscillations with a high $M$, the $m=0,\,l=2$ velocity pattern is quickly washed out and replaced by large velocity noise where we can also see a clear coupling between the crust and core. 
We note that our method to determine the density gradients is very simple and more advanced approaches are available \citep{Rosswog2015}.
Although preliminary simulations (which will not be shown here) did not show a large change when using the so-called constant-exact or linear-exact gradients \citep{Rosswog2015}, we will explore these options in the future. 

We suspect that the noise is due to the fact that the CCI is determined independently for each individual particle using a finite number of neighbors and is therefore subject to statistical fluctuations. 
The fact that the crust is stable with a homogeneous density is a promising sign that the deviations of the $\mathbf{n}_{i,\mathrm{CCI}}$ from normals that are determined with the particle radial vectors are small. 
However, a method that suppresses statistical noise is required. 
In the next section we will, therefore, explore the representation of the CCI and its surface normals by spherical harmonics where we use all particles close to the interface for the CCI determination and include low-pass filtering on noise. 
\subsection{Spherical Harmonics Decoupling}
\label{section::ssh}
Here, we will follow the idea of \cite{Weiss2011} and represent the CCI in form of spherical harmonics which are determined from SPH particle positions. 
The spherical harmonics representation will then be used to calculate the interface normal at the position of particles that are close to the CCI. 

\begin{figure}
    \centering
    \includegraphics[width = 0.45\textwidth]{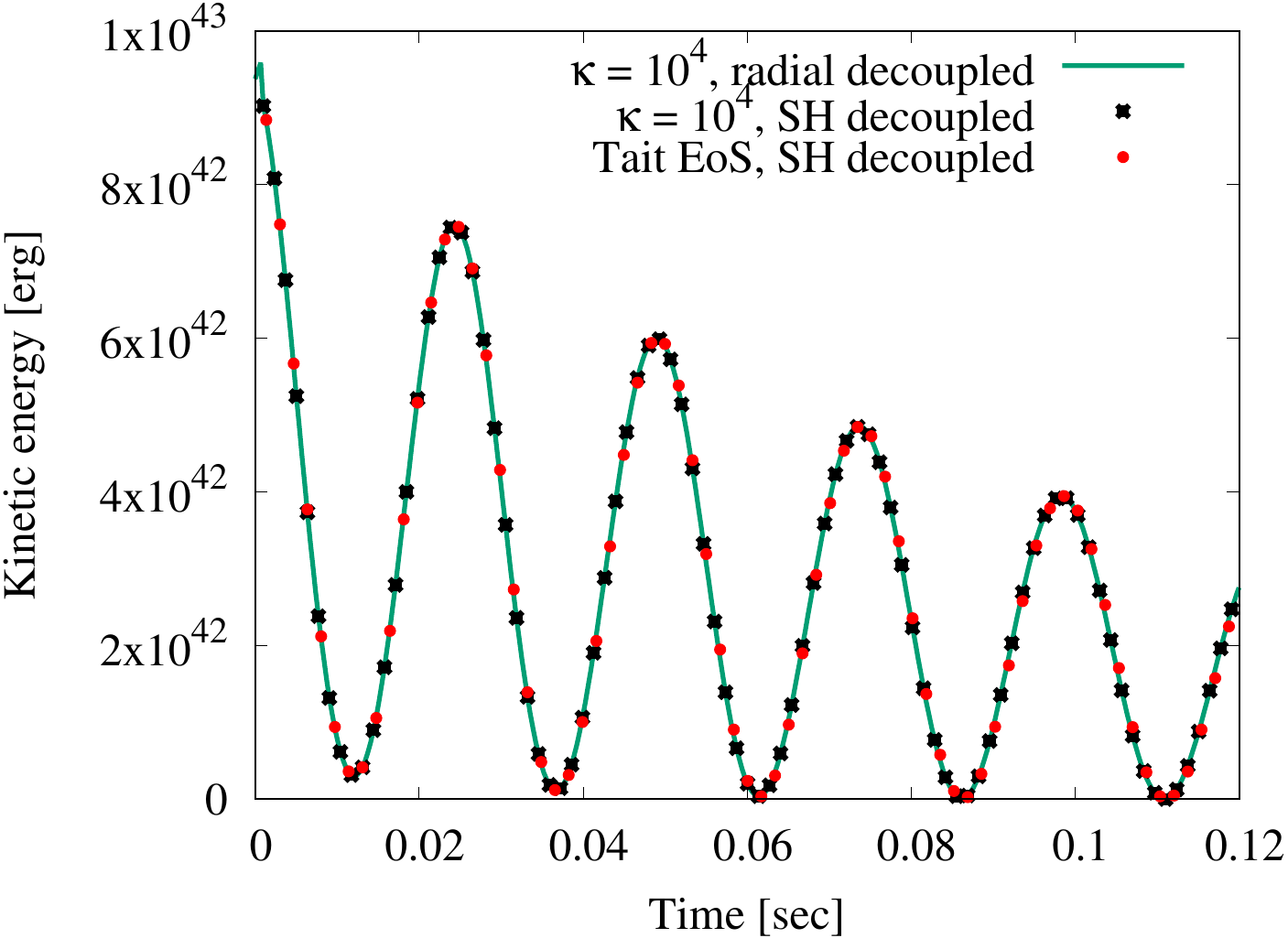}
    \caption{Kinetic energy oscillations for the Tait-Murnaghan EoS crust, decoupled via spherical harmonics decomposition of the CCI from the fluid core. The oscillations agree well with the previously explored approximation where we reduced the EoS pressure acceleration for crust particles by $\kappa = 10^4$.}
    \label{fig:ssh_liquid_pred_comp}
\end{figure}

We assume that the location of the interface is described by the function $R(\theta, \varphi)$ and can expanded in spherical harmonics $Y_{lm}$ with the corresponding coefficients $C_{lm}$: 
\begin{align}
    R (\theta, \varphi) = \sum_{l=0}^\infty \: \sum_{m = -l}^l \: C_{lm} Y_{lm} (\theta, \varphi). 
\end{align}
We divide the star into angular bins $B_{\theta}$ and $B_{\varphi}$ and sort the SPH particles into $B_{\theta} \cap B_{\varphi}$.
The spacing in $\theta$ and $\varphi$ has to be coarse enough for each bin to contain at least one particle. 
Given the number of particles in each bin $N_{\theta \varphi}$, we can calculate the coefficients of the spherical harmonics as
\begin{align}
    C_{lm} &= \sum_{\theta, \varphi} \: \frac{1}{N_{\theta \varphi}} \sum_{i \in B_\theta \cap B_\varphi} p_i \: Y_{lm}(\theta, \varphi) \: \Delta \Omega\,, \\
    \Delta \Omega &= \sin \theta \: \Delta \theta \: \Delta \varphi\,,
\end{align}
with $\theta$ being the central angle of the corresponding bin and $p_i$ given by
\begin{align}
    \left|\textbf{p}_i\right| = \left|\mathbf{r}_i - \Delta a \: \mathbf{n}_{i, \mathrm{CCI}}\right|, \:\:\:
    \Delta a = \frac{\rho_\mathrm{CCI} - \rho_i}{| \nabla \rho_i|}\,.
\end{align}
Here, $\rho_\mathrm{CCI}$ is the density at the CCI and $\mathbf{n}_{i,\mathrm{CCI}}$ is an estimate for the interface normal, determined with the density-gradient method. 
To remove noise, we set coefficients to zero if they are smaller than 0.5\% of $C_{00}$. 
The normal vector $\mathbf{n}_{i,\mathrm{SH}}$ can then be determined via 
\begin{align}
    \mathbf{n}_{i,\mathrm{SH}} &= \left( \sin \theta \cos \varphi + \frac{\cos \theta \cos \varphi}{r} R_\theta + \frac{\sin \varphi}{r \sin \theta} R_\varphi \right) \: \hat{e}_x \nonumber\\
    &+ \left(\sin \theta \sin \varphi + \frac{\cos \theta \sin \varphi}{r} R_\theta - \frac{\cos \varphi}{r \sin \theta} R_\varphi \right) \: \hat{e}_y \nonumber\\
    &+ \left( \cos \theta - \frac{\sin \theta}{r} R_\theta \right)\: \hat{e}_z \,,\\
    R_\theta & = \frac{\partial R}{\partial \theta}, \:\:\: R_\varphi = \frac{\partial R}{\partial \varphi}\,.
\end{align}
In the current calculations, given that the NSTs and the CCI are spherical, we only include $C_{lm}$ for $0 \leq l \leq 2$ and $-l \leq m \leq l$. 
This has to be extended in the future in order to study deformed neutron stars.  
\begin{figure}
    \centering
    \includegraphics[width 
    = 0.3\textwidth]{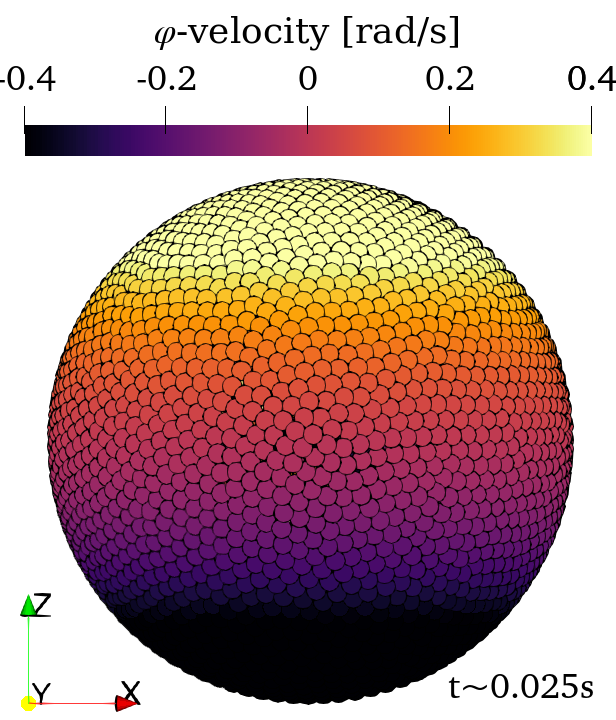}
    \caption{$\varphi$-velocity distribution in the crust of a neutron star during toroidal oscillations as in Fig.\ref{fig:frozencore_oscillations} with a fluid core and a Tait-Murnaghan EoS crust where both are decoupled from each other via the spherical harmonics approach.}
    \label{fig:sh_fluidcore_oscillations_liquid}
\end{figure}
Figure~\ref{fig:sh_fluidcore_oscillations_liquid} shows the distribution of $v_\varphi$ for a Tait-Murnaghan EoS crust that is decoupled from the fluid core via the spherical harmonics interface determination. 
We can see a clear $m=0,\,l=2$ pattern at $t \sim 23~\mathrm{ms}$ while Fig.~\ref{fig:ssh_liquid_pred_comp} gives the evolution of the crust's kinetic energy.
We find a very good agreement with the oscillatory behavior of the model with reduced bulk modulus with $\kappa = 10^4$.
\section{Summary and Outlook}
We have presented the solid material modeling capabilities of the Los Alamos National Laboratory Smoothed Particle Hydrodynamics code FleCSPH using the elastic-perfectly plastic strength model with linear hardening. 
The code has many capabilities that allow it to study neutron stars and compact-star mergers, such as analytic and tabulated equations of state, Newtonian gravity via FMM and fixed relativistic background metric.
With solid material modeling, the code can be applied to study the dynamics of the solid neutron star crust for astrophysically relevant scenarios like crustal oscillations or compact star binaries. 
Here, we test the modeling capability with crustal toroidal oscillations in the fundamental mode and compare the outcome to analytically predicted oscillation frequencies using a polytropic EoS for the fluid core. 
As the behavior of the strength model of the crust is akin to that of gelatin, where the bulk pressure in the crust is three orders of magnitude higher than its shear stress, the particle model of the crust is very sensitive to particle noise.
To suppress the latter, we explore several techniques which attempt to (a) decouple the crustal motion from the core, and (b) minimize the impact of the dominating bulk modulus on the crust dynamics. 
We find that using the Tait-Murhagan or Liquid EoS to describe the crust in a weakly compressible regime, together with a decoupling technique that determines the crust-core interface and its surface normal via spherical harmonics, is a promising approach for non-spherical neutron stars, either isolated or in binaries. 

Our approach demonstrated promising results on a single-layer, thin-shell model of the crust. However, this technique could be extendable to multi-layer configurations, such that the thickness of the crust is also resolved.
While we tested our method in Newtonian setting, there is no reason for the approach not to work in relativistic regime.
With these extensions in place, it should be possible to explore the crust dynamics in the context of binary neutron star mergers, or in single stars with crustal oscillations following a giant flare.
\begin{acknowledgments}
This work was supported by the Advanced Simulation and Computing program (NNSA/DOE) and the Laboratory Directed Research and Development program of Los Alamos National Laboratory (LANL) under project number 20200145ER. The research used resources provided by the LANL Institutional Computing Program and the LANL Darwin testbed. LANL is operated by Triad National Security, LLC, for the National Nuclear Security Administration of the U.S.DOE  (Contract No. 89233218CNA000001). This work is authorized for unlimited release under LA-UR-22-30036. 
\end{acknowledgments}

\appendix
\section{Oscillation Mode Analysis}
Here, we perform a spherical harmonics analysis of the crustal motion after its initialization in the fundamental toroidal mode. 
The setup of the crust is as in Section \ref{section::ssh} with the Tait-Murnaghan EoS. 
Decoupling from the core is performed using interface reconstruction with scalar spherical harmonics. 
The reason for this analysis is the possibility of energy cascading down to lower modes or exciting higher-mode oscillations during the simulation. 
Both options could make the observed crustal movement a combination of different modes and not the simple $l=2,m=0$ motion, as assumed.    
For our analysis, we follow \cite{Dahlen1998} and decompose the crustal velocity field $\mathbf{u}$ via vector spherical harmonics : 
\begin{align}
\mathbf{u} = \sum_l \sum_m U_{lm} \mathbf{P}_{lm} + V_{lm} \mathbf{B}_{lm} + W_{lm} \mathbf{C}_{lm} 
\end{align}
where 
\begin{align}
\mathbf{P}_{lm} &= \hat{\mathbf{r}} Y_{lm} \nonumber\\
\mathbf{B}_{lm} &= \frac{1}{\sqrt{l (l + 1)}} \nabla Y_{lm} \nonumber\\
\mathbf{C}_{lm} &= \frac{1}{\sqrt{l (l + 1)}} -\hat{\mathbf{r}} \times \nabla Y_{lm} 
\end{align}
with $\nabla = \hat{\mathbf{\theta}} \: \partial \theta + \hat{\mathbf{\phi}} \frac{1}{\sin \theta} \: \partial \phi$.
The coefficients of the spherical harmonics decomposition are obtained from
\begin{align}
U_{lm} &= \int \left( \mathbf{P}_{lm} \cdot \mathbf{u} \right) d\Omega \nonumber\\
V_{lm} &= \int \left( \mathbf{B}_{lm} \cdot \mathbf{u} \right) d\Omega \nonumber\\
W_{lm} &= \int \left( \mathbf{C}_{lm} \cdot \mathbf{u} \right) d\Omega 
\end{align}
The expression $U_{lm} \mathbf{P}_{lm} + V_{lm} \mathbf{B}_{lm}$ describes spheroidal motion which has radial and tangential components while $W_{lm} \mathbf{C}_{lm}$ represents purely toroidal displacements. Since we initialize the crustal motion in the fundamental toroidal $l=2,m=0$ mode, we expect the $W_{20}$ coefficient to be dominant in the above decomposition. 

For the numerical determination of the spherical harmonics coefficients, we apply the same approach of particle binning as described in Section \ref{section::ssh}. 
\begin{figure}
    \centering
    \includegraphics[width = 0.45\textwidth]{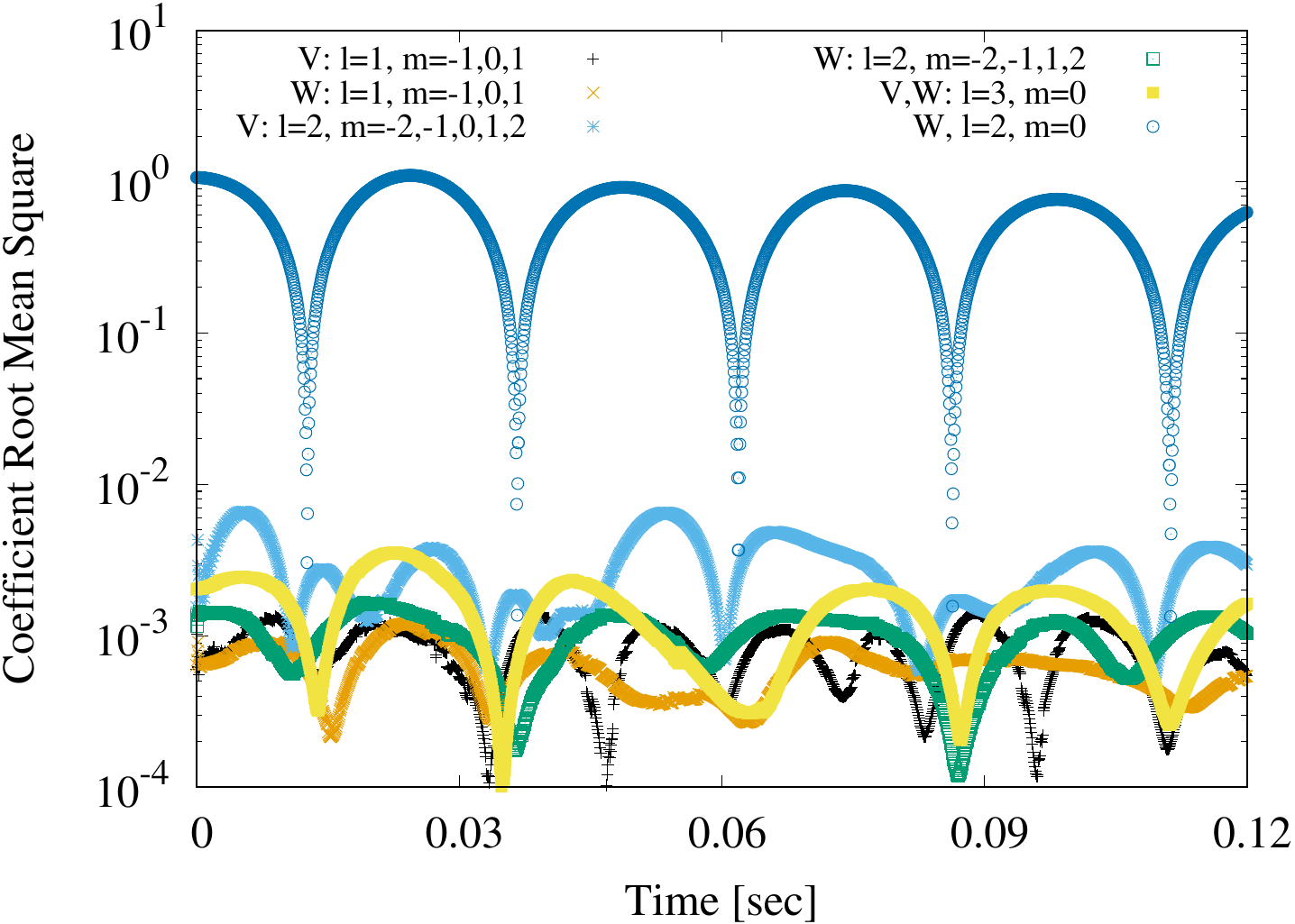}
    \caption{Root mean squares of spherical harmonics coefficients $V_{lm}$ and $W_{lm}$ during the simulated toroidal oscillations of the neutron star crust.}
    \label{fig:modes}
\end{figure}
Furthermore, we are only interested in the tangential and toroidal motion and only consider particles in the crust. 
Figure \ref{fig:modes} shows the resulting root mean squares of $V_{lm}$ and $W_{lm}$ which are determined by
\begin{align}
    \sqrt{\frac{1}{n_V} \: \sum_m V_{lm}^2} \:\:\: \mathrm{and} \:\:\:  \sqrt{\frac{1}{n_W} \: \sum_m W_{lm}^2}
\end{align}
for given values of $l$ and $m$ with $n_V$ and $n_W$ being the number of considered coefficients. 
We consider possible oscillations with $l=1,2$ and the corresponding values of $m$ as well as the $l=3$, $m=0$ mode. 
As expected, $W_{20}$ dominates the vector spherical harmonic decomposition while the root mean squares of higher and lower modes are at least a factor of 100 smaller. 
Furthermore, these modes seem to be present from the very beginning of the simulation with no noticeable growth in amplitude with time. 
With that, the corresponding particle motion is initialized at the start of the simulations and most likely a numerical artefact of particle placement, numerical viscosity, or finite particle number. 
In any case, the impact of these modes on the toroidal motion of the crust with $l=2, m=0$ is negligibly small.
This confirms that our extracted oscillation frequency of the crust indeed corresponds to the fundamental toroidal mode. 

\bibliographystyle{aasjournal}
\bibliography{reference}
\end{document}